\documentclass [a4paper, 11pt]{article}
\pdfoutput=1
\usepackage{jheppub1, amsmath, amsfonts, amssymb, amssymb,amsxtra, mathrsfs, makeidx, amsthm, stmaryrd, shuffle, slashed,multirow}

\usepackage[latin1]{inputenc}

\usepackage{graphicx}
\usepackage{float}

\usepackage{color}
\definecolor{dgreen}{rgb}{0,0.70,0.30}
\definecolor{gold}{rgb}{0.85,.66,0}
\definecolor{purple}{rgb}{1.0,0.3,0.6}

\usepackage{tikz}

\input ulem.sty

\restylefloat{figure}

\def\beq{\begin{equation}}
\def\eeq{\end{equation}}

\def\Re{{\rm Re\,}}
\def\Im{{\rm Im\,}}

\newcommand{\vecb}{\left(\begin{array}{c}}
\newcommand{\vece}{\end{array}\right)}
\newcommand{\ccb}{\left(\begin{array}{cc}}
\newcommand{\cce}{\end{array}\right)}
\newcommand{\cccb}{\left(\begin{array}{ccc}}
\newcommand{\ccce}{\end{array}\right)}
\newcommand{\ccccb}{\left(\begin{array}{cccc}}
\newcommand{\cccce}{\end{array}\right)}
\newcommand{\cccccb}{\left(\begin{array}{ccccc}}
\newcommand{\ccccce}{\end{array}\right)}

\newcommand{\RR}{\mathbb R}
\newcommand{\R}{\mathbb R}

\newcommand{\NN}{\mathbb N}
\newcommand{\ZZ}{\mathbb Z}
\newcommand{\Z}{\mathbb Z}
\newcommand{\QQ}{\mathbb Q}

\allowdisplaybreaks[1]

\numberwithin{equation}{section}

 \newcommand{\bea}{\begin{eqnarray}}
 \newcommand{\eea}{\end{eqnarray}}

\def\half{{\scriptstyle {1\over 2}}}
\def\threeh{{\scriptstyle {3 \over 2}}}
\def\fiveh{{\scriptstyle {5 \over 2}}}

\def\t2{\tau_2}
\def\IZ{\relax\ifmmode\mathchoice {\hbox{\cmss Z\kern-.4em Z}}
{\hbox{\cmss Z\kern-.4em Z}}
{\lower.9pt\hbox{\cmsss Z\kern-.4em Z}}
{\lower1.2pt\hbox{\cmsss Z\kern-.4em Z}}
\else{\cmss Z\kern-.4em Z}\fi}

\def\g{\gamma}

\def\c1{{\chi^1}}

\def\N4{{\cal N}=4}

\def\nn{\nonumber}

\def\IR{\relax{\rm I\kern-.18em R}}
\def\IZ{Z}

\def\x0{{x_0}}
\def\y0{{y_0}}
\def\p{\partial}
\def\Im{{\rm Im\,}}
\def\Re{{\rm Re\,}}

\newcommand{\calA}{{\mathcal A}}

\newcommand{\calD}{{\mathcal D}}
\newcommand{\calE}{{\mathcal E}}

\DeclareMathAlphabet{\mathpzc}{OT1}{pzc}{m}{it}
\newcommand\ap{\alpha'}
\newcommand{\te}{\textrm}

\newcommand{\pa}{\partial}

\newcommand{\al}{\alpha}
\newcommand{\be}{\beta}
\newcommand{\ga}{\gamma}
\newcommand{\de}{\delta}
\newcommand{\ep}{\epsilon}

\newcommand{\si}{\sigma}

\newcommand{\Si}{\Sigma}

\newcommand{\Om}{\Omega}

\newcommand{\eq}{=} 
\newcommand{\co}{\ , \ \ \ \ \ \ }
\newcommand{\dd}{\mathrm{d}}

\newcommand{\Tt}{\tilde{T}}
\newcommand{\Vt}{\tilde{V}}

\newcommand{\lat}{\Xi}
\newcommand{\cov}{D}

\title{Multiparticle one-loop amplitudes and S-duality in closed superstring theory}
\author[a]{Michael B. Green,}
\author[a]{Carlos R. Mafra,}
\author[a,b]{Oliver Schlotterer}

\affiliation[a]{Department of Applied Mathematics and
Theoretical Physics, \\
Wilberforce Road, Cambridge CB3 0WA, UK}
\affiliation[b]{Max--Planck--Institut f\"ur Gravitationsphysik, \\
Albert--Einstein--Institut, 14476 Golm, Germany}
\emailAdd{M.B.Green@damtp.cam.ac.uk}
\emailAdd{C.R.Mafra@damtp.cam.ac.uk}
\emailAdd{oliver.schlotterer@aei.mpg.de}

\abstract{Explicit expressions for one-loop  five supergraviton scattering amplitudes in both type II superstring theories are determined by making use of the  pure spinor formalism. The  type IIB
amplitude  can be expressed in terms of a doubling of ten-dimensional super Yang--Mills tree amplitude, while the type IIA amplitude has additional pieces that cannot be expressed in that manner. We evaluate the coefficients of terms in the analytic part of the low energy expansion of the amplitude, which correspond to a series of terms in an effective action of the schematic form $\cov^{2k}R^5$  for $0\le k \le 5$ (where $R$ is the Riemann curvature).  Comparison with earlier analyses of the
tree amplitudes and of the four-particle  one-loop amplitude leads to an interesting extension of the action of
$SL(2,\Z)$ S-duality on the  moduli-dependent coefficients in the  type IIB theory.   We also investigate  closed-string five-particle amplitudes that violate conservation of the $U(1)$ R-symmetry charge -- processes that are forbidden in supergravity. The coefficients of their low energy expansion are shown to agree with S-duality systematics. A less detailed analysis is also given of the six-point
function, resulting in the vanishing of the analytic parts of the $R^6$ and $\cov^4 R^6$ interactions in the ten-dimensional effective action, but not in lower dimensions.}

\begin{document}

\setcounter{tocdepth}{2}
\maketitle


\section{Introduction and overview}

\subsection{Introduction}
\label{introduction}

Over the past few years the study of scattering amplitudes in supersymmetric quantum field theory and string theory
has led to a stimulating interplay between physical ideas and structures of mathematical interest. These beautiful relationships strongly constrain theories with maximal supersymmetry, which can be viewed as descending from ten-dimensional
open or closed-string theories.  This has led to the discovery of a number of intriguing properties of both
perturbative and non-perturbative aspects of such theories in various backgrounds and in a variety of dimensions.  In
particular, the relationship between perturbative and non-perturbative properties of string theory embodied in its
duality symmetries has stimulated many fruitful research avenues.

The objectives of this paper are twofold.  The first is to investigate properties of closed-string one-loop amplitudes
in type II superstring theory with more than four external particles, which have not previously been studied
systematically.  Here we will mainly discuss five-particle amplitudes, with some additional results for six-particle
amplitudes. 
This will make use of the general rules for constructing open-string one-loop amplitudes that were
derived in \cite{Mafra:2012kh} by use of the pure spinor formalism \cite{Berkovits:2000fe,MPS}, combined with the constraints imposed by BRST
invariance. In the type IIB case we will see that the kinematic structure of the five-particle amplitude is similar to that of the tree-level amplitude of \cite{Mafra:2011nv, Mafra:2011nw, Schlotterer:2012ny}, as
well as with the structure of four-particle tree and loop amplitudes.  We will also present a detailed analysis of properties of the low energy expansion of these amplitudes.

The second objective is to make use of these perturbative results to extend the understanding of the exact,
non-perturbative, structure of terms in the low energy expansion of the amplitude. These are strongly constrained by
S-duality, which relates strong and weak coupling. Such higher dimensional interactions have coefficients that are functions of the scalar fields, or moduli.
The ten-dimensional type IIB theory provides the simplest
nontrivial example, with duality group $SL(2,\ZZ)$ \cite{Hull:1994ys}.  
Compactification to $D=10-d$  dimensions  on a $d$-torus results in a theory with maximal supersymmetry with a rank-$(d+1)$ duality group that is an arithmetic subgroup of the real split version of  $E_{d+1}$  \cite{Hull:1994ys}\footnote{For $d=1, \dots, 7$ compact dimensions, these are specific discrete subgroups of the global supergravity groups \cite{Cremmer:1978ds, Cremmer:1979up, Julia:1981, Julia:1988}, $SL(2)$, $SL(2)\times SL(3)$, $SL(5)$, $Spin(5,5)$, $E_6$,
$E_7$, $E_8$.  The two distinct $d=0$ ($D=10$) theories are type IIA and IIB, with S-duality groups $1$ and $SL(2)$,
respectively.}.

Past results concerning $SL(2,\Z)$ duality of four-particle scattering have led to expressions for the exact dependence on the moduli of the coefficients of certain BPS interactions that arise at the first three orders in the low energy expansion of the effective action  beyond classical supergravity   \cite{Green:1997tv,Green:1998by, Sinha:2002zr,Green:2005ba}.  Four-particle processes conserve the $U(1)$ R-charge of type IIB supergravity, but it is not generally conserved in $N$-particle string theory amplitudes with $N \ge 5$.   The modular properties of the coefficients of a number of $U(1)$-violating effective interactions at the same order as $R^4$ in the low energy expansion were considered in \cite{Green:1997me, Green:1999qt, Berkovits:1998ex, Basu:2008cf, Boels:2012zr, Basu:2013goa}.   
   In the following we will show how this structure matches with our results for the five-particle amplitude and we will
find interesting extra structure that is not seen in the four-point function
and which might lead to further insights into the exact moduli dependence.

\subsection{Overview of paper and brief summary of results}
\label{overview}

We will devote section \ref{background}  to a review of background material that is relevant to the results of the subsequent sections.
Section \ref{trees} will review the construction of closed-superstring tree amplitudes from open-string amplitudes by
an extension of the Kawai--Lewellen--Tye (KLT) procedure \cite{Kawai:1985xq}. The open-string $N$-point tree amplitudes
were constructed in \cite{Mafra:2011nv,Mafra:2011nw} making use of the manifestly supersymmetric pure spinor formalism
\cite{Berkovits:2000fe}, which packages all processes related by linearised supersymmetry into a single
expression.  Any colour-ordered open-string tree amplitude is expressed by a simple extension of massless supersymmetric
Yang--Mills (YM) amplitudes in the schematic form valid for any multiplicity $N$
\beq
\calA_{\rm{tree}} = F_{\te{tree}}(s_{ij})\, A_{YM}\, .
\label{openamp}
\eeq
In this expression, $\calA_{\rm{tree}}$ and $A_{YM}$ are $(N-3)!$ component vectors of superstring- and YM tree amplitudes, respectively,
referring to a basis of independent colour-ordered amplitudes \cite{BjerrumBohr:2009rd,Stieberger:2009hq,Bern:2008qj}. The quantity $F_{\te{tree}}(s_{ij})$
is a $(N-3)!\times (N-3)!$ matrix that depends on external on-shell momenta $k_i$ through the dimensionless Mandelstam invariants, $s_{ij}=\frac{\ap}{4}(k_i+k_j)^2$,
in a manner that is determined by generalised Selberg integrals of $N-3$ colour-ordered vertex operators on the
boundary of a disk world-sheet.  These amplitudes and their expansions in powers of the Mandelstam invariants
$s_{ij}$, or equivalently, their expansions in powers of $\alpha'$, were studied in detail in
\cite{Schlotterer:2012ny, Broedel:2013tta, Broedel:2013aza}. The coefficients of terms in this low energy expansion exhibit a fascinating pattern of
kinematic factors involving nontrivial multi-zeta-values (MZVs), which will be reviewed in section \ref{motivmzv}. We
will also review the structure of closed-string multiparticle tree-level amplitudes that were also studied in
\cite{Schlotterer:2012ny} and shown to have the structure
\beq 
{\cal M}_{\te{tree}} = A^{t}_{YM} \, {\cal S}_{\rm{tree}}(s_{ij}) \, \tilde A_{YM}\, .
\label{closedamp}
\eeq
Here, ${\cal S}_{\rm{tree}}$ is again an $(N-3)!\times (N-3)!$ matrix acting in the space spanned by a basis of colour-ordered YM tree amplitudes $A_{YM}, \tilde A_{YM}$ associated with the left- and right-movers, respectively.
This matrix is a function of the Mandelstam invariants and is given by an integral over $N-3$ vertex insertion points in a spherical world-sheet. It is again
of interest to study the ``low energy'' expansion of the amplitude, which involves an expansion of ${\cal S}_{\rm{tree}}(s_{ij})$ in powers of Mandelstam
invariants
\beq
{\cal S}_{\rm{tree}}(s_{ij})   = S_0 \sum_{\{k_i\} \in (2\NN+1)^\times}   (MZV_{ \{k_i \} })  \times (M_{k_1} M_{k_2} M_{k_3} \ldots + \cdots) \,,
\label{treestructure}
\eeq 
The factors $M_{k_i}$ are $(N-3)!\times (N-3)!$ matrices with entries that are homogeneous polynomials in Mandelstam
invariants of odd degree $k_i$ which already enter the open-string $\ap$ corrections (\ref{openamp}) along with $\zeta_{k_i}$. The notation $(MZV_{ \{k_i \} })$ indicates (linear combinations of) MZV products whose overall weight $w=\sum_i k_i$ matches the order in $\ap$.  Non-trivial MZVs (i.e., MZVs that are not polynomial in ordinary Riemann-zeta values) do not occur until order $(\alpha')^{11}$ in this expansion, whereas they first occur at order $(\alpha')^8$ for the open string.
 The leading low energy term $S_0$ (i.e., $k_i=0$ $\forall i$) proportional to $N-3$ powers of Mandelstam invariants is known as the momentum kernel \cite{BjerrumBohr:2010ta, BjerrumBohr:2010yc, BjerrumBohr:2010hn}, and
reproduces the supergravity tree. Higher order terms, on the other hand, are proportional to monomials in the $M_{k_i}$'s  and represent stringy corrections. These matrices in the $\ap$ expansion were
determined explicitly at multiplicity $N=5$ in \cite{Schlotterer:2012ny} and later in \cite{Boels:2013jua}, and a systematic derivation of their form at general multiplicity $N$ is given in \cite{Broedel:2013tta, Broedel:2013aza}.  The precise form of the expansion in \eqref{treestructure} will be given up to order $(\alpha')^{11}$ in \eqref{het1,33}.

The structure of the open-string one-loop multiparticle amplitudes $ \calA_{\te{1-loop}}$ was established in \cite{Mafra:2012kh}, as will be reviewed in section \ref{oneloopreview}.  The construction again made use of the pure spinor formalism and led to
 amplitudes of the form 
 \beq \calA_{\te{1-loop}} = F_{\te{1-loop}}(s_{ij})\, A_{YM}\,,
 \label{openloop} \eeq
 where the matrix function of the invariants, $F_{\te{1-loop}}(s_{ij})$, is given in terms of integrals
 over vertex positions on either boundary of an annular world-sheet. These integrals, which generalise the Selberg integrals of the tree-level term, have not yet been systematically analysed in their general low energy behaviour, even
 for small values of $N$. The vector of $ A_{YM}$ still refers to the independent tree-level amplitudes in YM field theory.

A major focus of this paper is the construction of $N$-particle closed-string one-loop amplitudes and, in the type IIB case, how their relationship to tree amplitudes is constrained by S-duality.  Since this is a non-perturbative symmetry and therefore constrains the structure of the theory over the whole of moduli space.    Up to now almost all
the work on closed-string loop amplitudes has been restricted to four-particle ($N=4$) scattering except for \cite{Richards:2008jg, BjerrumBohr:2008vc, BjerrumBohr:2008ji}.   Even though this is a particularly
special case, it has provided interesting input for analysing the constraints of S-duality, as will be reviewed in section \ref{dualityreview}. There we will describe the manner in which the $SL(2,\Z)$ S-duality group acts on the low energy expansion of the four-particle closed-string scattering amplitude
in the ten-dimensional type IIB theory.  This requires the terms in the low energy expansion of the four graviton amplitude to have coefficients that are
modular functions of the complex scalar field, $\Omega$.  For example, the lowest order correction to the classical supergravity tree amplitude is a term of
order ${\alpha'}^3 \, R^4$ (where $R$ is the Weyl curvature) and its coefficient is a particular $SL(2,\Z)$ Eisenstein series.  The
dependence on $\Omega$ of the coefficients of the higher derivative terms of order $\cov^4\, R^4$ and $\cov^6\, R^4$ will also be discussed.

The construction of closed-string one-loop amplitudes in the pure spinor formalism is the subject of section
\ref{purespinor}.  Certain technical problems that arise with the composite $b$ ghost are alleviated by imposing BRST
invariance as a restriction on the form of the amplitude. Although these amplitudes can again be viewed, in a certain
sense, as a doubling of the open-string amplitudes, they incorporate an important new feature.  In contrast to tree-level $N$ particle amplitudes and the one-loop $N=4$ amplitude, the  $N>4$ one-loop amplitudes involve contractions between left and right moving world-sheet fields.  This happens both through  OPE contractions and through integration by parts and leads to new classes of terms.  As a result, with present
methods the explicit construction of the loop amplitudes becomes very complicated as $N$ increases and we will limit
the discussion in section \ref{purespinor} to the case $N=5$.

There is a qualitative distinction between the structure of the type IIA and IIB amplitudes.  
 We will find that the type IIB amplitude can once again be expressed
in terms of a doubling of the YM tree amplitude in the form
\beq
N=5\ : \ \ \ \ \ \ {\cal M}_{\te{1-loop}} = A^{t}_{YM} \, {\cal S}_{\te{1-loop}}(s_{ij}) \, \tilde A_{YM}\,.
\label{closedfiveamp}
\eeq
 This provides the first nontrivial indication that the polarisation dependence in closed-string loop amplitudes is captured by bilinears of YM trees.  
 It seems plausible that the structure in
(\ref{closedfiveamp}) extends to higher numbers of loops and possibly to higher $N$.   Furthermore,  since the low-energy limit of closed-string theory reproduces maximal supergravity, these comments should also apply to loop amplitudes in (maximal) supergravity.
The structure of \eqref{closedfiveamp} will be shown to apply not only to five-particle amplitudes  that conserve the $U(1)$ R--symmetry charge of classical supergravity, but also those that do not.  The dependence on the charge violation is encoded in  the coefficient function, ${\cal S}_{\te{1-loop}}(s_{ij})$.

We will also see that the type IIA five-particle amplitude contains  extra terms that cannot be expressed in the form \eqref{closedfiveamp}, as will be discussed in section \ref{type2a}.  These include parity-violating components (terms with a single  $\epsilon$ tensor), such as the amplitude with a Neveu--Schwarz/Neveu--Schwarz antisymmetric potential and four gravitons, which reproduces the familiar $BR^4$ interaction. While  the type IIA and IIB theories are distinct in $D=10$ dimensions, they are equivalent upon toroidal compactification to dimension $D=10-d$ in which the Yang--Mills theory is non-chiral.   Therefore, the compactified version of \eqref{closedfiveamp} applies to the scattering of massless states in either of these theories.

The expression ${\cal S}_{\te{1-loop}}(s_{ij})$ in (\ref{closedfiveamp}) involves integrals of the vertex positions over a toroidal world-sheet of complex
structure $\tau$, which is also to be integrated with an $SL(2,\ZZ)$-invariant measure.  This makes it very difficult to analyse the full amplitude, but a great
deal of information about its low energy expansion can be obtained, as we will see in section \ref{lowtorus}.   This is an expansion in powers of the world-sheet scalar Green function and its derivatives.  At least to the order we consider in this paper the amplitude can be separated into a non-analytic piece that contains thresholds and an analytic piece that can be expanded in powers of the Mandelstam
invariants.  The terms that arise at a given order in $\ap$ are
world-sheet Feynman diagrams with free propagators joining the external vertex positions.  To the order that we will reach in this paper most of these diagrams were
evaluated in \cite{Green:2008uj} in studying the low energy expansion of the one-loop four-particle amplitude. Furthermore, following section 5 of \cite{Green:2008uj} it
is straightforward to generalise the discussion to compactifications of the amplitude to lower dimensions on tori. 

In this paper we will concentrate on the analytic part of the loop amplitude, although the interplay of the analytic and non-analytic parts  is significant in determining properties of the amplitude. 
The analysis  in sections \ref{lowtorus} and \ref{sec:sdual} will determine the detailed structure of the expansion of ${\cal S}_{\te{1-loop}}$ in a form that can be compared
with the tree amplitude structure given in \eqref{treestructure},  
\beq
S_{\te{1-loop}}(s_{ij}) = S_0 \sum_{\{k_i\} \in (2\NN+1)^{\times} \atop{\{l_i\} \in (\NN+6)^{\times}}}    \lat^{(d)}_{\{ k_i,l_i, \ldots\} } \, \times  (M_{k_1} M_{k_2}  \ldots \\ M'_{l_1} M'_{l_2} \ldots + \cdots ) \,,
\label{loopstructure}
\eeq
where $M_{k_i}$ and $M_{l_i}'$ are $2\times 2$ matrices that depend on the Mandelstam invariants and $S_0$ again describes the field theory limit of (\ref{treestructure}).  A
striking feature of this set of matrices is the augmentation of tree-level matrices $M_{k_i}$ of odd degree in $s_{ij}$ by additional matrices $M_{l_i}'$ of
(non necessarily odd) degrees $ \geq 7$. The  quantities $\lat^{(d)}_{\{ k_i,l_i,\ldots\} }$ in \eqref{loopstructure} are determined by a sum of world-sheet Feynman diagrams with $w = \sum_i (k_i + l_i) -3$ propagators and represent the 
coefficients of terms such as the supersymmetric completion of $(\ap)^{w+3}\, \cov^{2w-2} R^5$ in the low energy effective action. The diagrams that contribute to the $\lat^{(d)}_{\{ k_i,l_i,\ldots\}
}$'s up to order $w=6$ in the four-point function were evaluated in \cite{Green:1999pv, Green:2008uj} and resulted in polynomials in Riemann-zeta values in $D=9,10$
space-time dimensions. A slightly extended set of diagrams enters in the calculation of the five-point function, the first novel example showing up at order
$(\ap)^8$. One of the major impediments to obtaining results at higher order in $\ap$ is the difficulty in calculating these diagrams.  A very interesting issue
concerning the structure of the diagrams beyond the order at which they have so far been evaluated is at what order nontrivial MZVs arise (recalling that  they arise in the factor
$(MZV_{\{ k_i\} })$ in \eqref{treestructure} for $w\ge 11$).

The results in section \ref{lowtorus} on the low energy expansion of type IIB loop amplitudes and the parity-conserving components of the type IIA loop amplitudes are restricted to parts of the torus integrals that are analytic in the Mandelstam invariants. Hence, they concern the local part of the one-loop effective action of the type II theories. General features of these results are shown in tables \ref{operators} and \ref{uoneviolating}. 
Table \ref{operators} indicates the dimension of the space of kinematic invariants that arise in the expansion of the tree and
one-loop graviton amplitudes with $N=4$, $5$ and $6$ at each order in $\alpha'$.  The precise form of the rank $2k+4N$
tensors that contract with the derivatives and curvature tensors in terms of the form $\cov^{2k} R^N$ can in principle be
extracted from the amplitudes and the polynomial structure of the matrices $M_k, M'_l$. Novel techniques to bypass the tedious procedure and to obtain superstring effective actions without excessive diagram bookkeeping are described in \cite{Barreiro:2012aw}.
 
We will find that the pattern of kinematic invariants that arises for $N>4$ is more elaborate. Generically, curvature interactions at order $(\ap)^{k+3}$ are organised in sequence of the form $\cov^{2k-2l} R^{4+l}$ with $l=0,1,2,\ldots,k$, i.e. which start with a four curvature term. However, the five-particle one-loop amplitude generates additional  $\cov^{2k-2}R^5$ interactions  at $k \geq 4$ that do not have any four-point ancestor, $\cov^{2k}R^4$.  At tree level, the first example of a sequence of $N \geq 5$ interactions  that is not related to an $N=4$ interaction occurs at weight eleven and has  the schematic form $(\ap)^{11} \, \{\cov^{14} R^5,\cov^{12}R^6,\ldots\}$ (which is not shown in table \ref{operators} but see \cite{Schlotterer:2012ny}).  Strikingly, the coefficient of this sequence involves a triple zeta value, $\zeta_{3,3,5}$, that cannot be reduced to a monomial in Riemann zeta values.  We will find that the one-loop results contain additional kinematic structures at lower orders $7,8,9$ in the ${\alpha'}$ expansion, as indicated in table~\ref{operators}.   There we see that there are two distinct
contributions at order $(\ap)^7\, \cov^6 \, R^5$.  One of these belongs to the same family of invariants $(\ap)^7 \, \cov^{8-2l} \, R^{4+l}$ that arises at tree level and the
second is a new contribution that does not match any tree-level result. The number of independent contributions increases further to four at order  $(\ap)^{9}\, \cov^{10} \, R^5$ as shown in the last column of table \ref{operators}.  This pattern has important consequences for the implementation
of $SL(2,\Z)$ duality, as we will see in section \ref{sec:s}.  

\begin{table}[t]
\begin{center}
\begin{tabular}{|r|c|c|c|c|c|c|c|c|c|}\hline
 & $n=0$ &$1$ &$2$ &$3$ &$4$ &$5$ &$6$ & \\
\hline
\multirow{2}{*}{$(\alpha')^{n+3}D^{2n}R^4$}
& $1$ & $0$ & $1$ & $1$ & $1$ & $1$ & $2$ & tree     \\
& $1$ & $0$ & $\uwave{1}$ & $1$ & $\uwave{1}$ & $1$ & $2$ & 1-loop   \\\hline
\multirow{2}{*}{$(\alpha')^{n+3}D^{2n-2}R^5$}
& $0$ & $0$ & $1$ & $1$ & $1$ & $1$ & $2$ & tree    \\
& $0$ & $0$  & $\uwave{1}$ & $1$ & $\uwave{2}$ & $2$ & $4$  & 1-loop  \\\hline
\multirow{2}{*}{$(\alpha')^{n+3}D^{2n-4}R^6$}
& $0$ & $0$ & $1$ & $1$ & $1$ & $1$ & $2$ & tree    \\
& $0$ & $0$  & $\uwave{\le 1}$ & $\le 2$ & $\uwave{\le 4}$ & $??$ & $??$   & 1-loop  \\
\hline
\end{tabular}
\end{center}
\vskip-10pt
\caption{\small   A schematic list of the number of independent kinematic invariants  that contribute at each order in the $\ap$ expansion of $N \leq 6$ graviton amplitudes.
The underlined entries vanish in the $D=10$ case, but not in lower dimensions.  The detailed pattern of contraction of indices between the curvatures
and derivatives is encoded in the $A_{YM}, \tilde A_{YM}$ and $M_{k_i}, M'_{l_i}$, where the expressions include the other process related by supersymmetry that conserve the $U(1)$ charge.
The inequalities and the $??$ in the last row indicate our presently incomplete knowledge of the six-particle amplitude.}
\label{operators}
\end{table}

The terms in table \ref{operators} that are underlined with a wavy line have coefficients that vanish in ten dimensions, but not 
after toroidal compactification to lower dimensions.  The prototype for such terms is the $\cov^4\, R^4$ interaction,
which was shown to vanish at one loop in ten dimensions in \cite{Green:1999pv} but not in nine dimensions
\cite{Green:2008uj}.  More generally, this interaction is nonzero in any dimension less than ten.   


\begin{table}[t]
\[
\begin{array}{|c|c||c|c|} \hline
\te{boson} &q  &\te{fermion} &q
 \\ \hline \hline
\ \te{(anti-)holomorphic axio-dilaton} \  &\ \pm 2 \
&\te{dilatino} &\ \pm 3/2 \
 \\\hline
 \te{(anti-)holomorphic three-form} &\pm 1 
 &\ \te{gravitino} \ &\pm 1/2
  \\ \hline  %
\te{graviton and five-form} &0 &&
\\ \hline 
\end{array}
\]
\vskip-12pt
\caption{\small States of the massless type IIB supermultiplet and their R symmetry charges $q$.}
\label{u1over}
\end{table}

There are many other amplitudes describing scattering
of other component massless fields that are related to the graviton amplitudes by supersymmetry. Any interaction that conserves the $U(1)$ R-symmetry charge
$q$ of type IIB supergravity follows the same pattern.  Table \ref{u1over} reviews the $q$ charges of the states in the type IIB supergravity multiplet.
Four-particle amplitudes conserve the $U(1)$ charge of the type IIB theory, but this may be violated in five-particle
scattering.  In fact, as noted in \cite{Green:1999qt}, the maximal $U(1)$ charge violation, $q$, of an
$N$-particle amplitude is 
\beq 
q = \pm (2N -8)\,,
\label{maxuone}
\eeq
so when $N=5$ the only $U(1)$-violating processes are ones for which $q=\pm 2$.  These $U(1)$-violating  five-particle processes are also discussed in sections 3 and 4, leading to results for the degeneracies of the kinematic invariants 
displayed in table \ref{uoneviolating} using the example of derivatives acting on $G^2\, R^3$ interactions.  Here $G$ is the complex combination of
Neveu--Schwarz/Neveu--Schwarz (NSNS) and Ramond/Ramond (RR) three-form field strengths that carries unit $U(1)$
charge\footnote{We should emphasise that these $G^2 R^3$ operators are understood to involve two alike combinations of field strengths rather than complex conjugate pairs $G \bar G$.}. As we will see, the coefficients in the low energy expansion of the the five-particle one-loop amplitude  (\ref{loopstructure}) in the $q=\pm 2$ sector  are related to those in the $q=0$ sector in a simple manner. 

\begin{table}[htdp]
\begin{center}
\begin{tabular}{|r|c|c|c|c|c|c|c|c|c|}\hline
 & $n=0$ &$1$ &$2$ &$3$ &$4$ &$5$ &$6$ & \\
\hline
\multirow{2}{*}{$(\alpha')^{n+3}D^{2n}G^2R^3$}
& $1$ & $0$ & $1$ & $1$ & $1$ & $1$ & $2$ & tree     \\
& $1$ & $0$ & $\uwave{1}$ & $1$ & $\uwave{2}$ & $2$ & $5$ & 1-loop   \\\hline
\multirow{2}{*}{$(\alpha')^{n+3}D^{2n-2}G^2R^4$}
& $0$ & $0$ & $1$ & $1$ & $1$ & $1$ & $2$ & tree    \\
& $0$ & $0$  & $\uwave{\le 1}$ & $\le 2$ & $\uwave{\le 4}$ & $??$ & $??$          & 1-loop  \\
\hline
\end{tabular}
\end{center}
\vskip-10pt
\caption{\small The degeneracy of kinematic invariants that violate the conservation of $U(1)$ by $2$ units. This applies to all the
interactions of the same dimension as $\cov^{2k} \, G^2\, R^3$ and $\cov^{2k} \, G^2\, R^4$. Once again, the underlined entries vanish in the $D=10$ case, but not in lower dimensions.}
\label{uoneviolating}
\end{table}

Some implications of these results for S-duality are described in section \ref{dualityreview} and \ref{sec:sdual}.  In
particular, we will see that the results concerning the modular functions of the modulus $\Omega$ that implement type IIB S-duality in the
four-particle amplitude extend to the five-particle and higher point analogues in a simple manner.  However, as mentioned above,  the explicit one-loop
amplitudes for $N$-particle amplitudes with $N>4$ have additional pieces that enter at the order
$(\alpha')^{7}\cov^8R^4$ that are not present in the $N=4$ case.  Therefore, there must be modular invariant coefficients for these kinematic factors that begin at one loop and have no tree-level pieces.  Interestingly, this is the order at which an
understanding of the modular invariance of the four-particle amplitude is extremely limited and there is still
much to be understood.   We will also consider the implications of S-duality for the $U(1)$-violating five-particle interactions listed in table \ref{uoneviolating}.   In these cases the coefficient functions  are expected to be modular forms of nontrivial weight that are related to those of the $U(1)$-conserving graviton interactions.  We will see that certain modular functions described in section \ref{dualityreview} tie in with the $\ap$ expansion in the $U(1)$-violating sector displayed in section \ref{sec:u1vio}.   Furthermore, we offer an explanation for the pattern of $q=\pm 2$ and $q=0$ coefficients referred to in the previous paragraph. 

Some general properties of the six-particle amplitude will be considered in section \ref{sixpoint} although our analysis is not yet complete
due to technical obstacles that will be described.  The extent to which we have analysed the world-sheet integrals gives rise to the information listed in the last column of table \ref{operators}. At present, we can only give upper bounds on the number of independent $\cov^{2k} R^6$ operators after classifying the lattice sums appearing in the low energy expansion of the underlying torus integrals. The main six-point result is the vanishing of the $R^6$ and $\cov^4 R^6$ interactions in $D=10$ dimensions (see table  \ref{operators}) as well as the related  $U(1)$-violating operators of the same dimension (see e.g. table \ref{uoneviolating}).

The paper concludes with a brief summary and some comments in section \ref{summary}.

Several technical issues are left for the appendices. The first appendix \ref{compconv} helps to compare some of the present results with the literature. Appendix \ref{superspacecalc} contains a derivation of a an
important pentagon numerator identity using pure spinor superspace in \ref{super1}, and a proof that the low energy
limit of the five-particle amplitude is totally symmetric in \ref{super2}.  The diagrams that enter the low energy
expansion of the five-particle amplitude are exhibited in detail in appendix \ref{morediag}. A third appendix \ref{app:w4} summarizes the analytic parts of the five-point torus integrals to order $O(\ap^7)$. Finally, appendix \ref{anotherapp} is devoted to the key ingredients in the momentum expansion of the six-particle amplitude.


\section{Review of background material}
\label{background}

\subsection{The structure of superstring tree amplitudes}
\label{trees}

Tree level amplitudes involving any number of massless open-string states have been computed in
\cite{Mafra:2011nv,Mafra:2011nw} based on pure spinor cohomology methods \cite{Mafra:2010ir,Mafra:2010jq}. The colour
stripped amplitude
was found to be
\beq
{\cal A}_{\te{tree}}(1,\si(2,3,\ldots,N-2),N-1,N) \eq  \sum_{\pi \in S_{N-3}} A^{\pi}_{YM} \, F_{\te{tree}}^{\si,\pi}(s_{ij}) \ .
\label{tree}
\eeq
Remarkably, all the polarisation dependence of the superstring amplitude (\ref{tree}) is encoded in the (super-)YM field theory subamplitudes $A^\pi_{YM}$. The objects $F_{\te{tree}}^{\si,\pi}(s_{ij})$ originate in world-sheet integrals over the disk boundary, encode the string theory modifications to the field theory amplitude and can be mathematically classified as generalised Euler or Selberg integrals. Both ingredients on the right-hand side of (\ref{tree}) appear in their $(N-3)!$ element basis \cite{Bern:2008qj}
\begin{align}
A^{\pi}_{YM}  &:= A_{YM}(1,\pi(2,3,\ldots,N-2),N-1,N)
\label{het1,27}
\\
F_{\te{tree}}^{\si,\pi}(s_{ij})  &:= 4^{N-3}  \! \! \! \! \! \! \! \! \! \! \! \! \! \! \! \! \! \! \! \! \! \! \! \int \limits_{\phantom{z_{\si(i)}< z_{\si(i+1)}}z_{\si(i)}< z_{\si(i+1)}} \! \! \! \! \! \! \! \! \! \! \! \! \! \! \! \! \! \! \! \! \! \! \! \! \dd z_2 \ldots \int \dd z_{N-2} \  \prod_{i<j}^{N-1} |z_{ij}|^{4s_{ij}} \notag \\
& \ \ \ \ \ \ \ \ \ \ \ \ \times  \ \pi \left\{ \, \prod_{k=2}^{\lfloor N/2 \rfloor} \, \sum_{m=1}^{k-1} \frac{ s_{mk}}{z_{mk}} \,  \prod_{k=\lfloor N/2 \rfloor+1}^{N-2} \, \sum_{n=k+1}^{N-1} \frac{ s_{kn}}{z_{kn}} \, \right\} \ ,
\label{het1,27a}
\end{align}
and we have also restricted the string subamplitudes on the left-hand side to their $(N-3)!$  dimensional basis \cite{BjerrumBohr:2009rd,Stieberger:2009hq}. The conformal Killing group $SL(2,\RR)$ of the disk topology has been used to fix three world-sheet positions in (\ref{het1,27a}) to the values $(z_1,z_{N-1},z_N)=(0,1,\infty)$. The $S_{N-3}$ permutation $\pi$ acts on the labels $2,3,\ldots,N-2$ of the variables\footnote{Our definition of $s_{ij}$ incorporates an extra factor of $\frac{1}{4}$ compared to many references such as \cite{Mafra:2011nv,Mafra:2011nw,Schlotterer:2012ny}. This facilitates the discussion of closed strings and introduces the unusual factors of $4$ into the open string statement (\ref{het1,27a}).} 
\beq
z_{ij} :=  z_i\,-\,z_j \ , \ \ \ \ \ \ 
s_{ij} :=  \frac{\ap}{4}\, (k_i+k_j)^2
\label{manddef}
\eeq
in the curly bracket whereas $\pi(1,{N-1}) = (1,N-1)$ are unaffected. The vectors $k_i$ denote external on-shell momenta.

Tree level correlation functions of closed-string states factorise into the left- and right moving correlators, so the open-string result (\ref{tree}) yields the following closed-string amplitude
\beq
{\cal M}_{\te{tree}} = \sum_{\si,\pi \in S_{N-3}} A_{YM}^\si \, {\cal S}_{\te{tree}}^{\si,\pi}(s_{ij}) \, \tilde A_{YM}^\pi \ .
\label{het1,30}
\eeq
The integrals over the spherical closed-string world-sheet at genus zero form an $(N-3)! \times (N-3)!$ matrix
\begin{align}
{\cal S}_{\te{tree}}^{\si,\pi}(s_{ij}) :=  \int \dd^2 z_2 \ldots \int \dd^2 z_{N-2} \ \prod_{i<j}^{N-1} &|z_{ij}|^{2s_{ij}}  \  \si \left\{ \, \prod_{k=2}^{\lfloor N/2 \rfloor} \, \sum_{m=1}^{k-1} \frac{ s_{mk}}{z_{mk}} \,  \prod_{k=\lfloor N/2 \rfloor+1}^{N-2} \, \sum_{n=k+1}^{N-1} \frac{ s_{kn}}{z_{kn}} \, \right\} \notag \\
 &\times \ \pi \left\{ \, \prod_{k=2}^{\lfloor N/2 \rfloor} \, \sum_{m=1}^{k-1} \frac{ s_{mk}}{ \bar z_{mk}} \,  \prod_{k=\lfloor N/2 \rfloor+1}^{N-2} \, \sum_{n=k+1}^{N-1} \frac{ s_{kn}}{ \bar z_{kn}} \, \right\} 
\label{het1,29}
\end{align}
which describes the string theory extension of the  supergravity tree amplitude.

\subsection{Low energy expansion of superstring tree amplitudes}
\label{motivmzv}

The $(\ap)^w$ order in the low energy expansion of massless superstring tree amplitude involves multiple zeta values (MZVs) \cite{Mafra:2011nw,Stieberger:2009rr,Brown1,Brown2,Terasoma},
\begin{equation}
\zeta_{n_1,\ldots,n_r} :=
\sum\limits_{0<k_1<\ldots<k_r}\ \prod\limits_{l=1}^r k_l^{-n_l}\,,\quad n_l\in \NN\ ,\ n_r\geq2\ ,
  \label{00,4}
\end{equation}
of weight $w=\sum_{j=1}^r n_j$. The number $r$ of arguments in (\ref{00,4}) is referred to as the depth of a MZV. The systematics of their appearance was analysed in
\cite{Schlotterer:2012ny} and extended to the closed-string sector through the KLT relations \cite{Kawai:1985xq}. Let
$M_{2k+1}$ denote the $\zeta_{2k+1}$ coefficient\footnote{The coefficient of primitive zeta values $\zeta_{2k+1}$ depend on the $\QQ$ basis chosen for
MZVs of weight $2k+1$, see e.g. \cite{Blumlein:2009cf} for a minimal depth basis at weights $\leq 22$ which was also used in \cite{Schlotterer:2012ny} at weights $\leq 16$.} of the matrix (\ref{het1,27a}) of open-string $\ap$
corrections,\footnote{The $M_{2k+1}$ matrices from \cite{Schlotterer:2012ny, Broedel:2013tta, Broedel:2013aza} are written in open string conventions with $s_{ij}^{\te{open}}:= \ap (k_i+k_j)^2$. The present closed string discussion preserves the functional dependence $M_{2k+1}(s_{ij}^{\te{open}}) \mapsto M_{2k+1}(s_{ij})$ to implement the ``rescaling'' of $\ap$ in the translation between open and closed string results (see in particular \cite{Kawai:1985xq}).}
\beq
M_{2k+1} := F_{\te{tree}}\big(\tfrac{s_{ij}}{4}\big) \, \Big|_{  \zeta_{2k+1} } \ ,
\label{defM}
\eeq
its entries are degree $2k+1$ polynomials in the dimensionless Mandelstam variables (\ref{manddef}) to be determined through the methods of \cite{Broedel:2013tta, Broedel:2013aza}. Moreover, let
$S_0$ denote the field theory limit of the closed-string integral matrix (\ref{het1,29}), then the $\ap$-expansion of (\ref{het1,29}) has the structure 
\begin{align}
{\cal S}_{\te{tree}} &= S_0 \, \big( \, 1 + 2\zeta_3 \, M_3 + 2\zeta_5 \, M_5 + 2\zeta_3^2 \, M_3^2 + 2\zeta_7 \, M_7
+ 2\zeta_3 \zeta_5 \, \{ M_3,M_5\} + 2\zeta_9 \, M_9 \notag \\
&\quad{} + \tfrac{4}{3} \, \zeta_3^3 \, M_3^3  + 2\zeta_5^2 \, M_5^2  + 2\zeta_3 \zeta_7\, \{ M_3,M_7\}  + 2\zeta_{11} \, M_{11} + \zeta_3^2 \zeta_5 \, \{M_3,\{M_3,M_5\} \} \notag\\
&\quad{} + 2 \, \big( \tfrac{1}{5}\zeta_{3,3,5} - \tfrac{4}{35} \zeta_2^3 \zeta_5
+ \tfrac{6}{25} \zeta_2^2 \zeta_7 + 9 \zeta_2 \zeta_9 \big) \, [M_3,[M_3,M_5]]  + \cdots \,\big)
\label{het1,33}
\end{align}
The coefficients in this expansion are products of Riemann zeta values up to weight $11$, where the first irreducible MZV arises (whereas in the low energy expansion of the open-string tree \eqref{tree} the first irreducible MZV arises at weight $8$).
The MZV content of the string corrections can be better understood by lifting the $\zeta_{n_1,n_2,\ldots,n_r} \in \RR$  to their motivic version $\zeta^{\mathfrak{m}}_{n_1,n_2,\ldots,n_r}$
\cite{BrownDecomp,Schlotterer:2012ny}. The latter are endowed with a Hopf algebra structure which can be used to make
the basis more transparent: References \cite{BrownDecomp,Schlotterer:2012ny} describe a Hopf algebra isomorphism
$\phi$ which maps motivic MZVs to noncommutative polynomials in cogenerators $f_3,f_5,f_7,\ldots$
supplemented by a commutative element $f_2 = \phi ( \zeta^{\mathfrak{m}}_2 )$. The latter turns out to be absent in the motivic version ${\cal
S}^{\mathfrak{m}}_{\te{tree}}$ of ${\cal S}_{\te{tree}}$:
\begin{align}
{\cal S}_{\te{tree}}^{\mathfrak{m}} &= S_0 \, \phi^{-1} \big( \, 1 \ +\ 2f_3 \, M_3 \ +\ 2f_5 \, M_5 \ +\ 4 f_3^2 \, M_3^2 \ +\ 2f_7 \, M_7 \ +\ 2f_3 \shuffle f_5 \, \{ M_3,M_5\} \notag \\
& \ \ \ \ +\ 2f_9 \, M_9 \ +\ 8\, f_3^3 \, M_3^3 \ +\ 4f_5^2 \, M_5^2 \ +\ 2f_3 \shuffle f_7\, \{ M_3 ,M_7\} \ +\ 2f_{11} \, M_{11}  \notag \\
& \ \ \ \ +\ f_3 \shuffle f_3 \shuffle f_5 \, \{M_3,\{M_3,M_5\} \}\ +\ 2 \, f_5 f_3^2 \, [M_3,[M_3,M_5]] \ + \ \ldots\, \big)
\label{het1,37} \\
&= S_0 \, \phi^{-1} \bigg( \, \sum_{p=0}^{\infty} \, \sum_{ i_1,i_2,\ldots,i_p \atop {\in 2\NN+1}} \! \! M_{i_1}\,   \ldots \, M_{i_p} \, \sum_{k=0}^p \, f_{i_1}\,  \ldots\, f_{i_k} \shuffle f_{i_p} \, f_{i_{p-1}} \, \ldots \, f_{i_{k+1}} \, \bigg)
\label{het1,38}
\end{align}
The $\shuffle$ symbol denotes the commutative shuffle product\footnote{The shuffle product on non-commutative words in $f_{i_j}$ with $i_j \in 2\NN+1$ is defined by
\begin{align}
&f_2^p \, (f_{i_1} \, f_{i_2}\, \ldots\, f_{i_r}) \, \shuffle \, f_2^q \, (f_{i_{r+1}} \, f_{i_{r+2}}\, \ldots\, f_{i_{r+s}}) \eq f_2^{p+q} \sum_{\si \in \Si(r,s)} f_{i_{\si(1)}} \, f_{i_{\si(2)}} \, \ldots \, f_{i_{\si(r+s)}} \notag \\
&\Si(r,s) \eq \big\{ \, \si \in S_{r+s} : \ \si^{-1}(1) < \si^{-1}(2)<\ldots < \si^{-1}(r) \ \te{and} \ \si^{-1}(r+1) <\ldots < \si^{-1}(r+s) \, \big\}
\end{align}} 
on the non-commutative words in $f_{2k+1}$. The simplicity of the $\phi$ image of ${\cal S}_{\te{tree}}^{\mathfrak{m}}$ can firstly be seen at weight 11 where
the awkward coefficient of $[M_3,[M_3,M_5]]$ in (\ref{het1,33}) is mapped to $\phi(\tfrac{1}{5}\zeta^{\mathfrak{m}}_{3,3,5} - \tfrac{4}{35}
(\zeta^{\mathfrak{m}}_2)^3 \zeta^{\mathfrak{m}}_5 + \tfrac{6}{25} (\zeta^{\mathfrak{m}}_2)^2 \zeta^{\mathfrak{m}}_7 + 9 \zeta^{\mathfrak{m}}_2
\zeta^{\mathfrak{m}}_9) = f_5 f_3^2$. The absence of the commutative element $f_2$ in (\ref{het1,38}) reflects the cancellation of $\zeta_{2k}$ at low weights
of the KLT relations. The projection of the noncommutative words in $f_{2k+1}$ to $\sum_{k=0}^p f_{i_1} f_{i_2} \ldots f_{i_k} \shuffle f_{i_p} f_{i_{p-1}}
\ldots f_{i_{k+1}}$ imposes a selection rule\footnote{We should stress that the form of (\ref{het1,38}) is independent on the choice of $\QQ$ basis for MZVs
even though the polynomial structure of matrices $M_{w}$ at $w \leq 11$ is subject to possible redefinitions by (possibly nested) commutators in $M_{i<w}$.
This ambiguity in $M_w$ is compensated by a reshuffling of $f_w$ in the $\phi$ images at weight $w$.} on the MZVs of depth $r \geq 2$. As one can see from
(\ref{het1,37}), depth two MZVs at weight $w=8,10$ such as $\zeta^{\mathfrak{m}}_{3,5}$ and $\zeta^{\mathfrak{m}}_{3,7}$ do not enter gravity tree
amplitudes, and the first instance $\zeta^{\mathfrak{m}}_{3,3,5}$ of a depth $r>1$ MZV occurs at weight $w=11$.

\subsubsection{Four- and five-particle examples of closed-string tree amplitudes}

The S-duality connection between the four- and five-particle closed-string amplitudes at tree level and one loop are the main topic of this work.
This motivates us to explicitly spell out the $N=4,5$ version of the ingredients in (\ref{het1,30}): Multiplicity $N=4$ gives rise to scalars
\beq
N= 4 \ : \ \ \  A_{YM} = A_{ YM}(1,2,3,4) \ , \ \ \ S_0 = \frac{ \pi s_{12}s_{14}}{s_{13}}
\ , \ \ \ M_w = -\frac{1}{w}( s_{12}^w + s_{13}^w + s_{14}^w)
\label{Stree4pt}
\eeq
such that all the commutators among $M_{2k+1}$ and therefore all the MZVs of depth $\geq 2$ cancel. 

The $N=5$ point amplitude, on the other hand, is built from two component kinematic vectors $A_{ YM}$ and
$2\times 2$ matrices $S_0$, $M_{2k+1}$
\beq
N=5 \ : \ \ \ A_{ YM} = \begin{pmatrix} A_{YM}(1,2,3,4,5)\\ A_{YM}(1,3,2,4,5)\end{pmatrix} \ , \ \ \  S_0 = \frac{ \pi^2 }{s_{14} \, s_{25} \, s_{35}} \, \ccb \si_{11} & \si_{12} \\ \si_{12} & \si_{22} \cce
\label{YMvec}
\eeq
with entries $\si_{12} =- s_{12} s_{34} s_{13} s_{24} (s_{45}+s_{51})$ as well as $\si_{22}  =- s_{13} s_{24} (s_{12} s_{23} s_{45} + \te{cyclic}(12345))$ and $\si_{11} =\si_{22} \big|_{2\leftrightarrow 3}$. The simplest example of the matrices $M_{2k+1}$ reads
\beq
N=5 \ : \ \ \ M_3 = \ccb m_{11} &m_{12} \\ m_{21} &m_{22} \cce \ , \ \ \ \begin{array}{l}
m_{11}  =  s_3 [ - s_1 (s_1+2s_2+s_3)+s_3s_4+s_4^2 ]+s_1s_5 (s_1+s_5) \\
  m_{12} = - s_{13} s_{24} (s_1+s_2+s_3+s_4+s_5)
  \end{array}
  \label{Stree5pt}
\eeq
(with $m_{21} = m_{12} \big|_{2\leftrightarrow 3}$ and
$m_{22} = m_{11} \big|_{2\leftrightarrow 3}$ as well as $s_i := s_{i,i+1}$ subject to $s_{5}= s_{15}$). There are several avenues to determine their explicit form in more general cases, either by exploiting the representation of $F_{\te{tree}}$ (\ref{het1,27a}) in terms of (multiple Gaussian) hypergeometric functions (different techniques were used in \cite{Schlotterer:2012ny}, \cite{Oprisa:2005wu, Stieberger:2006bh, Stieberger:2006te} and \cite{Boels:2013jua}), or by use of  polylogarithm integration \cite{Broedel:2013tta}, or by use of the Drinfeld associator \cite{Broedel:2013aza}\footnote{See \cite{drin1, drin2} for mathematical background to the Drinfeld associator, and \cite{Drummond:2013vz} for its first connection with superstring amplitudes.}. The explicit expressions for $M_3,M_5,M_7$ and $M_9$ in the $N=5$ case are contained in the auxiliary files to this paper and
to \cite{Boels:2013jua}, as well as on the website \cite{WWW}, where matrices for higher multiplicity are available.

\subsection{One-loop amplitudes of the open superstring}
\label{oneloopreview}

The structure of one loop amplitudes among massless open-string states has been analysed in \cite{Mafra:2012kh}. Their BRST invariant part unaffected by the
hexagon anomaly \cite{Green:1984sg,Green:1984qs} was found to again boil down to linear combinations of YM tree subamplitudes
\beq
{\cal A}_{\te{1-loop}}(1,\Si(2,3,\ldots,N-2,N-1,N)) \eq  \sum_{\pi \in S_{N-3}} A^{\pi}_{YM} \, F^{\Si,\pi}_{\te{1-loop}}(s_{ij}) \ .
\label{1loop}
\eeq
We introduce a collective notation $\Si$ for both planar and non-planar arrangements of vertex operators along the boundary of the cylinder or Moebius strip
world-sheet. It governs the integration range for the vertex operator positions and the modular parameter of the genus one Riemann surface in
$F^{\Si,\pi}_{\te{1-loop}}(s_{ij})$. We will later on make use of the five-point correlator (\ref{corre}) underlying (\ref{1loop}) whose double copy furnishes
a subsector of one loop amplitudes of the closed string.

In contrast to the tree-level results (\ref{tree}) and (\ref{het1,30}) which are completely universal with respect to the number of space-time dimensions and
supercharges preserved, the structure of one loop amplitudes (\ref{1loop}) crucially depends on maximal supersymmetry.

\subsection{Some S-duality constraints on Type IIB amplitudes}
\label{dualityreview}
 
Ten-dimensional closed-string perturbation theory is an expansion around a limit in moduli space in which the type IIA  string coupling, $g_A=e^{\varphi_A}\to 0$, or the type IIB string coupling,  $g_B=e^{\varphi_B}\to 0$,  (where $\varphi_{A,B}$ are the dilatons of the type II theories).  The complete dependence of the amplitude on the moduli involves the nonperturbative completion of the perturbative amplitude. The only modulus in the type IIA theory is $e^{\varphi_A}$  and the duality group is trivial.  The type IIB theory depends on the complex scalar, $\Omega =\Omega_1 + i \Omega_2 =  C^{(0)} + ie^{\varphi_B}$, which transforms in the standard nonlinear manner under $SL(2,\Z)$,
\beq
\Omega\to \frac{a \Omega + b}{c\Omega + d}\,,
\label{dualscalar}
\eeq
where $a,b,c,d \in \Z$ and $ad-bc=1$.  The fields in the IIB supergravity supermultiplet transform under $SL(2,\Z)$ with weights that correspond to the $U(1)$ charges shown in table \ref{u1over}, where a field $\Phi_q$ of charge $q$ transforms as 
\beq
\Phi_{q} \to \left( \frac{c\Omega + d}{c\bar\Omega +d}\right)^{q/2}\,\Phi_{q} \,.
\label{uonetrans}
\eeq
Invariance of the type IIB theory and its effective action under $SL(2,\Z)$ transformations of the scattering states severely constrains the structure of the amplitudes. When combined with the constraints of maximal supersymmetry, $SL(2,\Z)$ invariance determines the precise dependence on the modulus, $\Omega$, of terms of low order in the $\alpha'$ expansion.   In other words, it determines the precise non-perturbative behaviour of these low order terms. 

\subsubsection{$U(1)$-conserving amplitudes}
\label{uoneconserve}

In the case of four-particle scattering, the $U(1)$ charge is conserved and the analytic part of the (both perturbatively and non-perturbatively completed) amplitude can be written in component form as 
\beq
{\cal M}(\Om) = s_{12}^2 s_{13}s_{14} \,A_{YM}(1,2,3,4) \, \tilde A_{YM}(1,2,4,3) \, T(s_{ij};\Omega)\,,
\label{fullfour}
\eeq
where the totally symmetric kinematic factor $s_{12}^2 s_{13}s_{14} A_{YM}(1,2,3,4)  \tilde A_{YM}(1,2,4,3)$ reproduces the standard $t_8 t_8$ tensor \cite{Green:1987mn} in the four graviton component. The symmetry of (\ref{fullfour}) in the external states implies that the low energy expansion of the analytic part of the scalar function\footnote{The interplay of the analytic and non-analytic parts of the amplitude is discussed in \cite{Green:2008uj}.} is a symmetric function of powers of the Mandelstam variables and has the form 
\begin{align}
T(s_{ij};\Omega)  \, \Big|_{\te{analytic}} = &\frac{1}{s_{12}s_{13}s_{14}}\left(\Omega_2^2 + M_3\, \Omega_2^{\frac{1}{2}}\,\calE_3 + M_5\,
 \Omega_2^{-\frac{1}{2}}\, \calE_5 + M_3^2\,\Omega_2^{-1}\, \calE_{3,3} + M_7\, \Omega_2^{-\frac{3}{2}}\,\calE_7 \right.  \notag \\ 
&\left. + \{M_3,M_5\}\,\Omega_2^{- 2}\, \calE_{\{3,5\}}+ M_3^3\, \Omega_2^{-\frac{5}{2}}\,\calE_{3,3,3}+ M_9 \,\Omega_2^{-\frac{5}{2}}\, \calE_9+ \ldots\right)\,,
\label{symmamp}
\end{align}
with four-point string corrections $M_w$ defined by (\ref{Stree4pt}).

The coefficients $\calE_{\ldots}(\Omega)$ are  $SL(2,\Z)$-invariant functions in the $D=10$ type IIB theory (and the explicit powers of $\Omega_2$ are absent after transforming from the string frame to the Einstein frame).    More generally, after toroidal compactification to $D$ dimensions on a $(10-D$)-torus the type IIA and type IIB theories are identified and the  $\calE_{\ldots}$'s are functions of the moduli space associated with the S-duality group.  The first term in the above expansion gives classical tree-level supergravity when substituted into \eqref{fullfour}. The above notation differs somewhat from earlier conventions in the literature. Appendix \ref{compconv} gathers a couple of conversion rules to compare the subsequent statements with references such as \cite{Green:2008uj}.

The kinematic factor $s_{12}^2 s_{13}s_{14} A_{YM}(1,2,3,4) \tilde A_{YM}(1,2,4,3)$ in the amplitude \eqref{fullfour} is completely determined by supersymmetry but the challenge is to determine the dynamical quantity $T(s_{ij};\Omega)$.  Although its exact form is not known, there are some interesting results concerning the first few terms in its low energy expansion, \eqref{symmamp}.
 The first term in the expansion beyond the supergravity amplitude is given by terms of order $R^4$, which are $1/2$-BPS interactions that may be expressed as
 integrals over 16 Grassmann coordinates, see \cite{Howe:1983sra} for an on-shell linearized superspace description of type IIB supergravity.
The next terms are those of order $D^4 R^4$, which are $1/4$-BPS interactions that may be expressed as integrals over 24 superspace Grassmann
coordinates.  These have $\Omega$-dependent coefficients \cite{Green:1997tv, Green:1998by, Sinha:2002zr}
\beq
\calE_{3}(\Omega) = E_{\threeh}(\Omega)\,,\qquad \calE_{5}(\Omega) = E_{\fiveh}(\Omega)\,,
\label{rfour}
\eeq  
where $E_s(\Omega)$ is an $SL(2,\Z)$ Eisenstein series, which satisfies the Laplace eigenvalue equation
\beq
\Delta_\Omega \, E_s(\Omega) = s(s-1)\, E_s(\Omega)
\label{eisendef}
\eeq
with respect to the modulus, $\Delta_{\Omega} := \Omega^2 \partial_\Omega \partial_{\bar \Omega}$. The unique $SL(2,\Z)$-invariant  solution of this equation that is power behaved in the weak coupling limit, $\Omega_2\to \infty$,  is\footnote{The normalisation convention is the one used in \cite{Green:2008uj}
appendix A.1.} the non-holomorphic Eisenstein series
\beq
E_s(\Omega) = \sum_{(m,n)\ne (0,0)}\frac{\Omega_2^s}{|m+n\Omega|^{2s}}\,.
\label{eisendef2}
 \eeq
with Fourier expansion 
 \beq
E_s(\Omega) = \sum_{N\ne 0} {\cal F}_N(\Omega_2)\, e^{2i\pi N\Omega_1}\,.
\label{eisendef3}
 \eeq
The non-zero modes ${\cal F}_{N\neq 0}(\Omega_2)$ contain the effects of D-instantons, with exponentially suppressed asymptotic behaviour at weak coupling ($\Omega_2 \to \infty$). The zero modes, on the other hand, are a sum of two power behaved terms $\Omega_2^{s}=g_s^{-s}$ and $\Omega_2^{1-s}=g_s^{s-1}$ which correspond to particular terms in string perturbation theory:
 \beq
 {\cal F}_0(\Omega_2) =  2\zeta_{2s}\, \Omega_2^s  +  \frac{ 2 \pi^{1/2}  \Gamma(s-1/2)}{\Gamma(s)} \,\zeta_{2s-1} \, \Omega_2^{1-s}\,,
 \label{zeromode}
 \eeq
In transforming from the string frame to the Einstein frame, the $R^4$ interaction is multiplied a factor of $\Omega_2^{-1/2}$ so the two perturbative terms
in the coefficient $E_{3/2}(\Omega)$ correspond to a tree-level piece proportional to $\Omega_2^{3/2}$, and a one-loop piece proportional to
$\Omega_2^{-1/2}$.  Similarly the $D^4 R^4$ interaction picks up a factor of $\Omega_2^{1/2}$ in transforming from string frame to Einstein frame so the two
perturbative terms in the coefficient $E_{5/2}(\Omega)$ correspond to a tree-level piece proportional to $\Omega_2^{5/2}$, and a two-loop piece proportional
to $\Omega_2^{-3/2}$.  These results explain why the low energy expansion of the ten-dimensional one-loop four-particle amplitude has a $R^4$ term but does
not have a $D^4 R^4$ part. Furthermore, the coefficient of the two-loop $D^4R^4$ interaction predicted by these arguments has been checked by
explicit amplitude calculations in \cite{D'Hoker:2005ht,Gomez:2010ad}. The generalisation of these results to lower dimensional theories with maximal supersymmetry obtained by toroidal compactification
\cite{Green:2008uj} involve combinations of Eisenstein series for higher-rank duality groups, which are functions of more moduli \cite{Green:2010wi,
Green:2010kv} (see also \cite{Pioline:2010kb}) .  An important general feature is that in dimensions $D<10$ the $D^4 R^4$ coefficient, $\calE_{5}$, does have
a one-loop contribution (as is indicated by the wavy underlining of the $\cov^4R^4$ one-loop term in table \ref{operators}).

The coefficient of the $1/8$-BPS terms of order $D^6 R^4$ in the low energy expansion, $\calE_{3,3}$,  is not an Eisenstein series but is expected to be a solution of the inhomogeneous Laplace equation\footnote{Note that the function ${\cal E}_{3,3}(\Omega)$ in the present notation is related to the $D^6 R^4$ coefficient ${\cal E}_{(0,1)}(\Omega)$ in \cite{Green:2008uj} by ${\cal E}_{3,3}(\Omega)= 3 {\cal E}_{(0,1)}(\Omega)$.}
\bea
(\Delta_\Omega  - 12)\, {\cal E}_{3,3}(\Omega) = - 3\big[ {\cal E}_3(\Om) \big]^2\,
\label{inhomo}
\eea
(which was motivated by M-theory considerations in \cite{Green:2005ba}).  The solution to this equation has a constant term (the zero Fourier mode in $\Omega_1$) of the form
\beq
\int_0^1 \dd \Omega_1 \ \calE_{3,3}(\Omega) = 2 \zeta_3^2 \Omega_2^{3} +4  \zeta_2\zeta_3\Omega_2  + {24\over 5}\zeta_2^2  \Omega_2^{-1} +
{4\over 9}\zeta_6 \Omega_2^{-3} +O(\exp(-4\pi\Omega_2))\, ,
\label{ethreehthreeh}
\eeq
which contains four terms power behaved in $\Omega_2$ that correspond to tree-level, one-loop, two-loop and three-loop string theory
contributions,\footnote{The string frame amplitude is obtained by multiplying $\calE_{3,3}$ by $\Omega_2^{-1}$.} together with an infinite sum of
D-instanton/anti D-instanton contributions.  The ratio of the tree-level and one-loop contributions agrees with the explicit string perturbation theory
calculations (and the overall normalisation has been chosen to be consistent with a tree-level amplitude normalised to $1/s_{12}s_{13}s_{14}$).

\subsubsection{$U(1)$-violating amplitudes}
\label{uonenonconserve}
Amplitudes with $N>4$ particles generally do not conserve the $U(1)$ charge, with the violation of the charge  bounded by $|q| \leq (2N -8)$ (see  \eqref{maxuone}).  The moduli-dependent coefficients of interactions that violate $U(1)$ must transform as modular forms of non-zero holomorphic and anti-holomorphic weights in order to ensure that the complete interaction, including its moduli dependence, is invariant under $SL(2,\Z)$.  Recall that the transformation  under \eqref{dualscalar} of a  modular form $f^{(w,w')}(\Omega)$  (with holomorphic and anti-holomorphic weights $(w,w')$)  is given by $f^{(w,w')} (\Omega) \to  (c   \Omega+ d)^w (c \bar \Omega+ d)^{w'} \,  f^{(w,w')} (\Omega)$
so a form with weights $(w,-w)$ transforms with a phase, 
\beq
 f^{(w,-w)}(\Omega)  \to
 f^{(w,-w)} \left(\frac {a\Omega + b}{c \Omega +d} \right) =  \left(\frac{c\Omega+d}{c \bar \Omega+ d}\right)^w \,  f^{(w,-w)}(\Omega)  \,,
\label{mofindef}
\eeq
which corresponds to a charge $q=2w$. 

A modular form of weight $(w+1,-w-1)$, or  $q=2w+2$,  can be obtained from one that transforms with weight $(w,-w)$ by applying a covariant derivative
\beq
{\cal D} f^{(w,-w)} = \Omega_2\left(i\frac{\partial}{\partial \Omega} + \frac{w}{2}\right)f^{(w,-w)} =: f^{(w+1,-w-1)}\,,
\label{covderiv}
\eeq
while the charge is lowered by the operator $\bar {\cal D}$ defined by 
$\bar {\cal D} f^{(w,-w)} = \Omega_2\left( - i\partial/\partial \bar\Omega + w/2\right)  f^{(w-1,-w+1)} $.

The coefficients of the leading interactions in the low energy expansion of amplitudes that violate $U(1)$ charge conservation were related by nonlinear supersymmetry to  $(\ap)^3  {\cal E}_3(\Omega)\, R^4$ in  \cite{Green:1998by} and have the form arrived at by arguments based on M-theory duality in  \cite{Green:1997me}.     An example is the $q=-2$ five-particle amplitude with two $G$'s and three gravitons.  The lowest-order interaction in the low energy expansion is $(\ap)^3   \, G^2R^3$, which has a coefficient  $\calE^{(-1,1)}_3 (\Omega)$ that is a $(1,-1)$-form given by
\beq
\calE_3^{(1,-1)}(\Omega) := {\cal D}   {\cal E}_3 (\Omega) =  \sum_{(m,n)\ne (0,0)} \frac{\Omega_2^s }{|m+n\Omega|^{2s}}
\left(\frac{m+n\bar\Omega} {m+n\Omega}\right)\,.
\label{uonecharge}
\eeq
The zero mode of this expression is easily obtained by acting with ${\cal D}$ on the $ {\cal E}_3$ zero mode in \eqref{zeromode}, giving
\beq
{\cal D}{\cal F}_0(\Omega_2) = s\,\zeta_{2s} \Omega_2^s  - \frac{ \pi^{1/2} \Gamma(s-1/2) }{\Gamma(s-1) } \, \zeta_{2s-1}\Omega_2^{1-s} \,,
\label{pertcharge}
\eeq  
which again has two perturbative contributions that have different coefficients from those of the $R^4$ interaction.  The explicit tree-level and one-loop calculations   in later sections will provide further information concerning $U(1)$-violating processes.
Some aspects of the modular forms associated with higher dimension $U(1)$-conserving and $U(1)$-violating interactions have been considered in \cite{Berkovits:1998ex, Basu:2008cf, Boels:2012zr}.

\section{Type II one-loop amplitudes in pure spinor superspace}
\label{purespinor}

The prescription for computing $N$-point superstring amplitudes at one-loop using the
minimal pure spinor formalism is given by \cite{MPS}
\begin{equation}\label{prescription}
{\cal M}_{\te{1-loop}} \sim \int \dd^2\tau\, \langle |(\mu, b)\prod_{P=2}^{10} Z_{B_P}
Z_J \prod_{I=1}^{11} Y_{C_I}|^2 \, V_1\tilde V_1(0)\prod_{j=2}^N \int\, \dd^2z_j U^j\tilde U^j(z_j,\bar z_j)\rangle.
\end{equation}
The massless closed-string states are represented by double copies of the pure spinor vertex operators $V$ and $U$ for the ten-dimensional ${\cal N}=1$ SYM
gauge multiplet \cite{MPS}. Moreover, $\mu$ is the Beltrami differential, $\tau$ is the Teichm\"uller parameter of the genus one Riemann surface, and
the
angle brackets $\langle {\ldots} \rangle$ denote the path integral discussed in detail in \cite{MPS}.
Finally, $Z_{B_P}$, $Z_J$, $Y_{C_I}$ are picture-changing operators and $b$ is the b-ghost whose schematic
form is given by \cite{MPS, odaY}
\begin{align}
b & = (\Pi d + N\p\theta + J\p\theta)\,d\,\delta(N) + (w\p\lambda + J\p N + N\p J + N\p N)\delta(N) \cr
&+ (N\Pi + J\Pi + \p\Pi + d^2)(\Pi\delta(N) + d^2\delta'(N))\cr
&+ (Nd + Jd)(\p\theta\delta(N) + d\Pi\delta'(N) + d^3\delta''(N))\cr
&+ (N^2 + JN + J^2)(d\p\theta\delta'(N) + \Pi^2\delta'(N) + \Pi d^2 \delta''(N) + d^4 \delta'''(N))
\label{bghost}
\end{align}
where $\delta'(x) = {\p\over\p x}\delta(x)$ and the variables on the right-hand side are conformal fields of the world-sheet theory of the pure spinor
formalism \cite{MPS}. In particular, $\lambda^\al$ is a bosonic ghost subject to the pure spinor constraint $\lambda^\alpha \ga^m_{\alpha\beta} \lambda^\beta=0$
when contracting the $16\times 16$ Pauli matrices $\ga^m_{\alpha\beta}$ of $SO(1,9)$.

Using the above prescription to compute amplitudes involving more than four external strings
can be rather challenging, mostly due to the complicated nature of the b-ghost and
the picture-changing operators. Fortunately there are some shortcuts which
can be taken to simplify this task.

As explained in the four-point computation of \cite{MPS}, the systematics of how the $d_\alpha$ zero-modes -- 16 Weyl spinor components -- are saturated can be exploited to
bypass many complicated features of the b-ghost. Only a few terms of \eqref{bghost} actually
give non-vanishing contributions and Lorentz invariance uniquely fixes the result
of their path integral up to an overall constant. For higher-point {\it open} superstring amplitudes similar arguments were used 
in \cite{Mafra:2012kh} to perform the integration over
the $d_\alpha$ zero modes while BRST invariance of the resulting pure spinor superspace expressions
fixed their relative coefficients.

The four-point amplitude of \cite{MPS}\ did not involve any OPEs among the vertices (except through
the standard Koba--Nielsen factor) since otherwise the $d_\alpha$ zero modes would not be saturated. But
starting at $N=5$ points there are
non-vanishing contributions featuring at most $N-4$ OPE contractions such as
\beq
d_\alpha(z_i)\theta^\beta(z_j)   \rightarrow \frac{\delta^\beta_\alpha}{z_{ij}} \, , \, \ \  \Pi_m(z_i) x^n(z_j,\bar z_j) \rightarrow  - \, \frac{ \delta_m^{n} }{z_{ij}} \, , \, \ \ N^{mn}(z_i) \lambda^{\al}(z_j) \rightarrow   \frac{ (\ga^{mn})^\al{}_{\be} \lambda^{\be} }{2 \, z_{ij}} \label{opesone} \,.
\eeq
The singularities of the genus one correlator caused by primary fields $d_\alpha, \Pi_m$ and $N^{mn}$ of conformal weight one enter through the torus Green function\footnote{Note that $\langle  x(z_i,\bar z_i) \, x (z_j,\bar z_j) \rangle = -\frac{ \ap }{2} \big( \ln \left| \frac{ \theta_1(z_{ij}) }{\theta_1'(0)} \right|^2 - \frac{2\pi }{\tau_2} \big[ \Im(z_{ij}) \big]^2 \big)$. 
}
\begin{equation}\label{defeta}
X_{ij} := s_{ij} {\p \ln \chi_{ij}  \over \p z_i} \co
\ln \chi_{ij} :=  \ln \left| \frac{ \theta_1(z_{ij}) }{\theta_1'(0)} \right|^2 - \frac{2\pi }{\tau_2} \big[ \Im(z_{ij}) \big]^2 + C(\tau,\bar \tau) \,.
\end{equation}
Its zero modes $C(\tau,\bar \tau) := 2  \ln | \sqrt{2\pi} \eta(\tau)|^2$ involving the Dedekind eta function $\eta(\tau)$ drop out of the scattering
amplitude and have thus been subtracted in (\ref{defeta}) for later convenience \cite{Green:1999pv, Green:2008uj}.
The definition of $X_{ij}$ including the Mandelstam variable (\ref{manddef}) is motivated by
integration by parts (see section \ref{intbypart} and \cite{Mafra:2012kh}).

The novelty appearing in
higher-point computations of {\it closed} string amplitudes stems from terms involving the contraction between left- and
right-movers -- either directly using the OPE (involving the volume $\tau_2 := \te{Im}(\tau)$ of the torus)
\begin{equation}\label{PiPibar}
\Pi^m(z)\bar\Pi_n(\bar z) = \delta^{m}_{n} \,\pi \left(  \frac{1}{\tau_2} - \delta^2(z,\bar z)\right).
\end{equation}
or indirectly through integration by parts, as will be explained below. In the following, we will refer to the right-hand side of the OPE (\ref{PiPibar}) as  
\beq \Om := \pi \left(   \delta^2(z,\bar z) - \frac{1}{\tau_2}\right) = \partial \bar \partial \ln \chi \label{defomega}
\eeq
and drop any reference to the argument $z$ because the $\de^2(z,\bar z)$ does not
contribute
in presence of the Koba--Nielsen factor (\ref{KN}).\footnote{This can be seen upon analytic continuation to the kinematic region
where $s_{ij}>0$ \cite{Richards:2008jg}.}

\subsection{The structure of the five-point closed-string correlator}
\label{fivepointamp}

The computation of the closed-string five-point correlator can be separated into three parts:
\begin{itemize}
\item [a)] purely holomorphic-square\footnote{Throughout this work, the term ``holomorphic'' square refers to products of expressions from the left- and right moving sector of the closed string, with $z_i$ and $\tau$ dependencies which are complex conjugate to each other.} terms multiplying
world-sheet functions $X_{ij}\tilde X_{kl}$;
\item [b)] holomorphic-square terms proportional to $\Omega$ generated
from integration by parts;
\item [c)] left/right mixing terms arising from the OPE \eqref{PiPibar}. 
\end{itemize}
These contributions will be labelled $K^{(a)}$, $K^{(b)}$ and $K^{(c)}$, respectively.  
BRST invariance will be used to obtain their relative weights in the final
answer,
\beq
{\cal M}_{\te{1-loop}} = 2 \int \frac{ \dd^2 \tau}{\tau_2^5} \int  \dd^2 z_2 \, \ldots \, \dd^2 z_5 \ {\cal I}(s_{ij})  \ \bigl( K^{(a)} +
K^{(b)} +  K^{(c)} \bigr)\, .
\label{defKs}
\eeq
The overall normalisation $-2$ can be obtained by unitarity.
In \eqref{defKs} we introduced the following shorthand for the Koba--Nielsen factor,
\beq\label{KN}
{\cal I}(s_{ij}) := \Big\langle\prod_{i=1}^N  e^{i k_i \cdot x(z_i,\bar z_i) } \Big\rangle \eq
\prod_{i<j}^N e^{-k_i \cdot k_j \langle x(z_i,\bar z_i) x(z_j,\bar z_j) \rangle } \eq
\prod_{i<j}^N \chi_{ij}^{s_{ij}}\,,
\end{equation}
with $\chi_{ij}$ given by the exponential of (\ref{defeta}). The contributions to $K^{(a)}$ and $K^{(b)}$ can be obtained from the holomorphic square of the open-string results of \cite{Mafra:2012kh}, where it was shown that
the left-moving CFT correlator can be written as\footnote{The
$(i,j,k)$ notation is explained in \cite{Mafra:2012kh} and should not be confused with vector indices.}
\begin{align}
\Big( X_{12} \, \frac{T_{12} \, T_3^i \, T_4^j \, T_5^k}{s_{12}}
  + (2\leftrightarrow 3,4,5) \Big)  +  \Big( X_{23} \, \frac{T_{1} \, T_{23}^i \, T_4^j \, T_5^k}{s_{23}}  +  ( 23 \leftrightarrow 24,25,34,35,45)  \Big)
\label{corre}
\end{align}
in terms of $X_{ij}$ defined by (\ref{defeta}). The objects $T_1 := V_1$ and $T_{12}$ denote BRST building blocks which were introduced in \cite{Mafra:2010ir,Mafra:2010jq}
to compactly represent tree-level kinematic factors in pure spinor superspace. Their relatives $T_3^i$ and $T_{23}^i$ are specific to one-loop open superstring
kinematics \cite{Mafra:2012kh}. Both $T_{12\ldots p}$ and $T_{23\ldots p}^i$ can be though of as the single pole residue of iterated OPEs
among the pure spinor vertex operators $V$ and $U$. The efficiency of these superfields in streamlining amplitude computations stems from their
covariant BRST variations \cite{Mafra:2010ir,Mafra:2010jq}.

The holomorphic square of \eqref{corre} generates 100 integrals but, as we shall see in the next section,
only $37$ are independent under integration by parts. In the open-string amplitude these manipulations lead to
BRST-closed kinematic factors\footnote{The precise form of the BRST building blocks $T_{12},T_1,T_3^i$ and $T_{23}^i$ is not needed for the purpose of this work once the BRST invariant (\ref{Qinv}) is expressed in terms of YM trees via (\ref{CtoYM}).}
\beq
C_{1,23} := \frac{T_{12} \, T_3^i \, T_4^j \, T_5^k}{s_{12}}  +  \frac{T_{31} \, T_2^i \, T_4^j \, T_5^k}{s_{13}} +  \frac{T_{1} \, T_{23}^i \, T_4^j \, T_5^k}{s_{23}}
\label{Qinv}
\eeq
for the left-moving superfields. In the closed-string case, it will be shown in the next subsection that reducing the
integrals to a basis leads to a manifestly BRST-closed piece $K^{(a)}$ composed from the holomorphic square of \eqref{Qinv} together with a correction
$K^{(b)}$ proportional to the function $\Omega$ of \eqref{defomega} which is not by itself BRST-closed.

\subsection{Integration by parts and $K^{(b)}$}
\label{intbypart}

In view of the definition (\ref{KN}) of the Koba--Nielsen factor, the vanishing of total world-sheet derivatives under the $z_i$ integrals leads to identities among the $X_{ij}$ defined in (\ref{defeta}) in open-string amplitudes. In the $N=5$ point context, for example \cite{Mafra:2012kh},
\beq\label{exampOne}
0 = \int \dd z_2 \ \partial_2 \, {\cal I}(s_{ij}) = \int \dd z_2 \ (X_{21} + X_{23} + X_{24} + X_{25}) \, {\cal I}(s_{ij})\,.
\end{equation}
For closed-string amplitudes, however, the presence of a right-moving $\tilde X_{ij}$ interferes with integration by parts performed
on the left-moving variables because
\beq
\tilde\p_i X_{ij} = s_{ij}\,\Omega,
\eeq
which follows from \eqref{defeta} and \eqref{defomega}.
As an example, the following five-point identity
\beq\label{exampTwo}
0 = \int \dd^2 z_2 \ \p_2 \, \big( {\cal I}(s_{ij})\, \tilde X_{2j} \big) =  \int \dd^2 z_2 \ \big((X_{21} + X_{23} + X_{24} + X_{25})  \, \tilde X_{2j} + s_{2j} \Omega \big) {\cal I}(s_{ij}) ,
\end{equation}
generates a fake left/right-mixing term proportional to $s_{2j}\Omega$.
Whenever there's no chance of confusion the Koba--Nielsen factor and the integration sign will be
omitted from now on, so the relations \eqref{exampOne} and \eqref{exampTwo} will be denoted
\begin{equation}\label{ibpcompact}
X_{12} = \sum_{k=3}^5 X_{2k}, \qquad X_{12}\tilde X_{2j} = \sum_{k=3}^5 X_{2k}\tilde X_{2j} + s_{2j}\Omega.
\end{equation}


In order to express the holomorphic square of the open-string CFT correlator \eqref{corre} in terms of a minimal set of integrals, it will
be sufficient to consider four prototype integration by parts  identities from which all others follow
through relabelling,
\begin{align}
X_{12} \, \tilde X_{23} &= (X_{23}+X_{24}+X_{25}) \, \tilde X_{23} +  s_{23} \, \Omega\cr
X_{12} \, \tilde X_{13} &= (X_{23}+X_{24}+X_{25}) \, (\tilde X_{32}+ \tilde X_{34}+ \tilde X_{35})  -  s_{23} \, \Omega\cr
X_{12} \, \tilde X_{12} &= (X_{23}+X_{24}+X_{25}) \, (\tilde X_{23}+\tilde X_{24}+ \tilde X_{25})  -  2s_{12} \, \Omega\cr
X_{12} \, \tilde X_{34} &= (X_{23}+X_{24}+X_{25}) \, \tilde X_{34}\,. \label{IBPs}
\end{align}
After some algebra one finds that the square of (\ref{corre}) yields
\begin{align}
\label{parta}
K^{(a)} &= \sum_{2\leq i<j \atop 2 \leq k<l}^5 \left(X_{ij}\tilde X_{kl} + \delta_{ik}\delta_{jl}\, s_{ij}\Omega\right)\, C_{1,ij} \tilde C_{1,kl}\\
K^{(b)} &= -  \Omega \, \Bigl(\frac{ T_{12} T_3^i T_4^j T_5^k \, \Tt_{12} \Tt_3^i \Tt_4^j \Tt_5^k}{s_{12}}  +  (2 \leftrightarrow 3,4,5) \notag \\
& \quad\qquad{} +  \frac{ T_{1} T_{23}^i T_4^j T_5^k \, \Tt_{1} \Tt_{23}^i \Tt_4^j \Tt_5^k}{s_{23}}  +  (23 \leftrightarrow 24,25,34,35,45) \Big)\,,
\label{partb}
\end{align}
i.e.\ the contributions to a) and b) comprise $37$ integrals and vary as follows under the pure spinor BRST operator $Q$:
\beq
QK^{(a)} = 0\,,
\eeq
\begin{align}
\label{QKb}
QK^{(b)} & =
- \Omega\, V_1 V_2 T_3^i T_4^j T_5^k \big[\Tt_{12} \Tt_3^i \Tt_4^j \Tt_5^k - \Vt_1 \Tt_{23}^i \Tt_4^j \Tt_5^k
 - \Vt_1 \Tt_{24}^i \Tt_3^j \Tt_5^k -\Vt_1 \Tt_{25}^i \Tt_3^j \Tt_4^k\big]\cr
&\quad{} + (2\leftrightarrow 3,4,5)\,.
\end{align}

Using the explicit superfield
representation of the BRST blocks of \cite{Mafra:2012kh} it is not difficult to check that both $s_{pq} C_{1,ij} \tilde C_{1,kl}$
and $|T_{12} T_3^i T_4^j T_5^k|^2/s_{12}$ have the same dimension as  $k^{10} A_{YM} \tilde A_{YM}$, i.e. their five-graviton components have dimensions of $R^5/k^2$.

As will become clear later when considering the $\ap$ expansions of the integrals, it is natural to include the term $s_{ij}\Omega$ together with the diagonal contributions $X_{ij}\tilde X_{ij}$ when
writing \eqref{parta} since the resulting integral over $(X_{ij}\tilde X_{ij} +s_{ij} \Om)$ has no leading low energy contribution or kinematic poles.
Therefore only
$K^{(b)}$ can contribute to the ${\cal O}(\ap^3)$ factorisation channels and once it is
combined with $K^{(c)}$ to form a BRST-closed quantity, the resulting kinematic factor agrees with 
the holomorphic square of
SYM tree amplitudes in the precise combination dictated by the KLT formula at order $(\ap)^3$.

\subsection{Interactions between left and right-movers and $K^{(c)}$}
\label{Kcsection}

The contributions discussed in the last subsections are those in which the pure spinor variables in the holomorphic and anti-holomorphic sectors
are treated separately, as in the open-string case. However,  there are other
ways to saturate the sixteen $d_\alpha$ zero-modes in both the left- and the right moving sectors of the pure spinor formalism,  which also involve an OPE contraction
between the two sectors.

When both the left- and right-moving b-ghosts contribute through terms of the form $d^4\delta'(N)$
one can also saturate all $d$ zero modes if a left-moving $\Pi^m(z_i)$ contracts with a right-moving $\bar \Pi^n(\bar z_j)$ in the external vertices.
Similar arguments as in the open-string calculations of \cite{Mafra:2012kh} can be used to integrate the $d_\alpha$ zero modes, giving  a left-moving contribution of 
\beq\label{LRmix}
V_1 A^m_2 T^i_3 T^j_4 T^k_5 + (2\leftrightarrow 3,4,5)\,,
\end{equation}
with a similar expression for the right-movers. The vector indices $m$ of the $A^m$ superfields of ten dimensional ${\cal N}=1$ SYM are contracted between the two sides.
Both $A^m$ and its spinorial field strength $W^{\alpha}$ (see (\ref{Rmdef}) and subsequent equations) enters the (left-moving) integrated vertex operator $U$.

The b-ghost \eqref{bghost} admits other possibilities involving left/right mixing.
For example, there can be a contraction between a $\bar \Pi^m \tilde d^2$ from the right-moving b-ghost with a $\Pi^n A_n$ from
a left-moving vertex or vice-versa. This can be achieved with a left-moving $b$ contribution of $d^4\delta'(N)$ together with a right-moving $\tilde b$
proportional to $\bar \Pi^m \tilde d^2$. The left-moving $d_\alpha$ zero-mode integration gives rise to the same
contribution $V_1 A^2_m T^i_3 T^j_4 T^k_5 + (2\leftrightarrow 3,4,5)$ as discussed above, but now four $\bar d_\alpha$ are required from
the right-moving vertices. Their contribution is given by $\tilde V_1 \tilde W^m_{2,3,4,5} + (2\leftrightarrow 3,4,5)$, where
\begin{align}
W^m_{2,3,4,5} := \frac{1}{12}\,\big[
   &(\lambda\g^n W^2)(\lambda\g^p W^3)(W^4\g^{mnp}W^5) + (\lambda\g^n W^2)(\lambda\g^p W^4)(W^3\g^{mnp}W^5)\cr
 + &(\lambda\g^n W^2)(\lambda\g^p W^5)(W^3\g^{mnp}W^4) + (\lambda\g^n W^3)(\lambda\g^p W^4)(W^2\g^{mnp}W^5)\cr
 + &(\lambda\g^n W^3)(\lambda\g^p W^5)(W^2\g^{mnp}W^4) + (\lambda\g^n W^4)(\lambda\g^p W^5)(W^2\g^{mnp}W^3)\big]\,.
 \label{Rmdef}
\end{align}
To see this note that group theory considerations imply that only one vector can be constructed
using two pure spinors $\lambda^\alpha$ and four $W^\beta$, therefore \eqref{Rmdef} is the unique such vector which is symmetric in the
labels $2,3,4$ and $5$. The normalisation was chosen for later convenience.

Another kind of interaction between left- and right-movers appears when there is a $\Pi^m\bar\Pi^n$ contraction between the
b-ghosts themselves through the schematic term $(\Pi^m d^2) (\bar\Pi^n \bar d^2)$. In this case both sides of external vertices
contribute with factors of $V_1 W^m_{2,3,4,5}$.

\subsection{BRST invariance}
\label{brstsymm}

After obtaining the different CFT contributions one needs to assemble the parts in order to obtain
a BRST-invariant result in pure spinor superspace. Our claim is that the correct amplitude is obtained up to an overall coefficient after this
step is completed.

The purely holomorphic square part $K^{(a)}$ in \eqref{parta} is written in terms of $C_{1,ij}$ and therefore manifestly BRST-invariant. The relative coefficient between $K^{(b)}$ in \eqref{partb} and $K^{(c)}$ from section \ref{Kcsection} is uniquely fixed by demanding that they combine to form a BRST invariant quantity. Since $K^{(b)}$ does not involve contractions between the left- and right-moving fields the only way that the
purely left/right mixing terms from $K^{(c)}$ can cancel the BRST variation of $K^{(b)}$ is
if $QK^{(c)}$ is also holomorphically factorised. 

Noting that $QA^m = k^m V + (\lambda\gamma^m W)$ as well as $QT^i_2 T^j_3 T^k_4 = 0$ and
\beq
\label{QWm}
QW^m_{2,3,4,5} = - (\lambda \g^m W^2) T^i_3 T^j_4 T^k_5 - (2\leftrightarrow 3,4,5) \ ,
\eeq
one sees that the combination
\beq\label{Tmdef}
T^m_{2,3,4,5} := A^m_2 T^i_3 T^j_4 T^k_5 + A^m_3 T^i_2 T^j_4 T^k_5 + A^m_4 T^i_2 T^j_3 T^k_5 +
A^m_5 T^i_2 T^j_3 T^k_4 + W^m_{2,3,4,5}
\end{equation}
has the property that its BRST variation contains the vector index $m$ only in momenta $k^m$,
\beq\label{QTm}
QT^m_{2,3,4,5} =
k^m_2 V_2 T_3^i T_4^j T_5^k + (2\leftrightarrow 3,4,5)\,.
\end{equation}
This observation justifies the normalisation of (\ref{Rmdef}) and suggests that the left/right mixing terms in $K^{(c)}$ are given by
\begin{equation}\label{partc}
K^{(c)} = - \Omega\,  V_1 T^m_{2,3,4,5} \tilde V_1 \tilde T^m_{2,3,4,5}\,,
\end{equation}
because
\beq
\label{QKc}
QK^{(c)} =\Omega\, V_1 V_2 T^i_3 T^j_4 T^k_5 \; k^2_m \tilde V_1 \tilde T^{m}_{2,3,4,5} + (2\leftrightarrow 3,4,5)
\eeq
is holomorphically factorised and therefore has a chance of cancelling the BRST variation of $K^{(b)}$.
Indeed one can show that 
\beq
\label{KbplusKc}
K^{(b)} + K^{(c)}= - \Omega\,C_{1,2,3,4,5}^m \tilde C_{1,2,3,4,5}^m
\eeq
where\footnote{The $m$ superscripts in $C_{1,2,3,4,5}^m \tilde C_{1,2,3,4,5}^m$ are meant to be symbolic reminders
that some of the terms therein involve vector index contractions between left- and right-movers. One cannot view $C_{1,2,3,4,5}^m$ as
a separate vector of its own right.}
\begin{align}
C_{1,2,3,4,5}^m \, \tilde C_{1,2,3,4,5}^m& := V_1\, T_{2,3,4,5}^m \, \tilde V_1 \, \tilde T_{2,3,4,5}^m
+  \Bigl( \frac{ T_{12} T_3^i T_4^j T_5^k \, \Tt_{12} \Tt_3^i \Tt_4^j \Tt_5^k}{s_{12}}  +  (2 \leftrightarrow 3,4,5) \Bigr)\notag \\
& \qquad{} +  \Bigl( \frac{ T_{1} T_{23}^i T_4^j T_5^k \, \Tt_{1} \Tt_{23}^i \Tt_4^j \Tt_5^k}{s_{23}}  +
(23 \leftrightarrow 24,25,34,35,45)\Bigr)\,.
\label{Cmdef}
\end{align}
is BRST-closed and therefore fixes the relative normalisation among the different CFT contributions
to the closed-string amplitude in \eqref{defKs}.

To verify that \eqref{Cmdef} is BRST closed one uses \eqref{QKb} and \eqref{QKc} to obtain
\beq
\label{QCm}
QC_{1,2,3,4,5}^m \, \tilde C_{1,2,3,4,5}^m = V_1 V_2 T^i_3 T^j_4 T^k_5 \, \tilde J_{1|2|345} 
+  (2 \leftrightarrow 3,4,5)\,,
\eeq
where the right-moving kinematic factor
\beq
\label{Jacdef}
\tilde J_{1|2|345} := \Tt_{21} \Tt_3^i \Tt_4^j \Tt_5^k + \Vt_1 \Tt_{23}^i \Tt_4^j \Tt_5^k + \Vt_1 \Tt_{24}^i \Tt_3^j
\Tt_5^k + \Vt_1 \Tt_{25}^i \Tt_3^j \Tt_4^k + \tilde V_1 k_2^m \tilde T^m_{2,3,4,5}
\eeq
is BRST-closed and satisfies \cite{Mafra:2010pn}
\beq
\label{Jvanish}
\langle \tilde J_{1|2|345}\rangle = 0\,.
\eeq
The pure spinor bracket $\langle \ldots\rangle$ here denotes the integration over the zero-modes of $\lambda^\alpha$ and $\theta^\alpha$ as
all non-zero-modes have been integrated out already (via operator product expansions). The prescription for zero mode integration has the schematic form $\langle (\lambda^3 \theta^5 ) \rangle =1$ and is reviewed in \cite{Mafra:2011nv}.
Therefore the vanishing of the left-moving BRST variation of \eqref{Cmdef} follows from the vanishing
of the pure spinor zero-mode integration of the superfield combination $\tilde J_{1|2|345}$ that builds up on the right-moving
sector.

Note that the vanishing of $\langle J_{1|2|345} + (2\leftrightarrow 3,4,5) \rangle$ yields
\begin{equation}\label{nowzero}
 \langle T_{12}T^i_3 T^i_4 T^i_5 + T_{13}T^i_2 T^i_4 T^i_5 + T_{14}T^i_2 T^i_3
T^i_5 + T_{15}T^i_2 T^i_3 T^i_4 +  k^1_m V_1 T^m_{2,3,4,5} \rangle = 0 \ .
\end{equation}
The left-hand side will be shown in appendix \ref{super1} to be BRST-exact.

It is interesting to observe that -- up to the left/right mixing terms -- \eqref{Jacdef}, \eqref{Jvanish} and \eqref{nowzero} can be viewed as kinematic
relations dual to Jacobi identities among certain colour factors\footnote{Pure spinor superspace expressions of the form
$\langle T_{\ldots}T_{\ldots}^iT_{\ldots}^jT_{\ldots}^k \rangle$ were argued in \cite{Mafra:2012kh} to be in correspondence with colour tensors built from structure
constants $f^{abc}$ contracted with one symmetrised four-trace $d^{a_1 a_2 a_3 a_4} := \frac{1}{6} \sum_{\si \in S_3} \te{Tr} \{ t^{a_1} t^{a_{\si(2)}}
t^{a_{\si(3)}} t^{a_{\si(4)}}\} $. The trace is taken in the fundamental representation of the gauge group generators $t^a$. These tensors obey four term
Jacobi relations such as $d^{abce} f^{edg}+d^{abde} f^{ecg}+d^{acde} f^{ebg}+d^{bcde} f^{eag}=0$ which formally resembles the first four terms in
(\ref{nowzero}) by identifying $\langle T_{12}T^i_3 T^i_4 T^i_5 \rangle \leftrightarrow f^{a_1a_2b} d^{ba_3 a_4 a_5}$.} (see the BCJ-like identity (7.33) of
\cite{Mafra:2012kh}). The fifth term $\sim k_m^i V_1 T^m_{2,3,4,5}$ of (\ref{Jacdef}) and \eqref{nowzero} ties in with the field theoretic identity (3.9) of
\cite{Bjerrum-Bohr:2013iza} connecting numerators of box diagrams with the vectorial part of a pentagon numerator contracting a loop momentum\footnote{Note
that a representation of the one-loop five-particle amplitude in ${\cal N}=4$ SYM theory was found in \cite{5ptBCJs} which satisfies all kinematic Jacobi
identities and where the loop momentum dependence in pentagon numerators vanishes.}. The derivation in \cite{Bjerrum-Bohr:2013iza} rests on demanding
kinematic Jacobi identities dual to Lie algebraic colour relations between cubic diagrams \cite{Bern:2008qj}. Hence, we are lead to interpret the superspace expression $\langle
T_{\ldots}T_{\ldots}^iT_{\ldots}^jT_{\ldots}^k \rangle$ as a box numerator (with a massive corner represented by the rank two building block $T_{pq}$ or
$T^i_{pq}$). Likewise, $\langle V_1 T^m_{2,3,4,5} \rangle$ qualifies to represent the loop momentum dependent part of a pentagon numerator. Given the vanishing of \eqref{nowzero}
in the BRST cohomology, we can regard it as a pure spinor superspace derivation of certain kinematic Jacobi relations at loop level.

\subsection{The five closed-string amplitude in terms of SYM trees}
\label{fivegrav}

By assembling the results from the previous subsections, we conclude that the supersymmetric closed-string five-particle
amplitude is given by
\begin{align}
&{\cal M}_{\te{1-loop}} =  2 \int \frac{ \dd^2 \tau}{\tau_2^5} \!\int  \dd^2 z_2 \, \ldots \, \dd^2 z_5 \ {\cal I}(s_{ij}) \notag \\
&\ \ \ \ \  \times \
\Bigl\{ \!\sum_{2\leq i<j \atop 2 \leq k<l}^5 (X_{ij}\tilde X_{kl} + \delta_{ik}\delta_{jl}\,
s_{ij}\Omega )\, \langle C_{1,ij} \tilde C_{1,kl}\rangle
- \Omega \langle C_{1,2,3,4,5}^m \, \tilde C_{1,2,3,4,5}^m \rangle \Bigr\}\,.
\label{XXX}
\end{align}
The superspace expressions $\langle s_{pq} C_{1,ij} \tilde C_{1,kl}\rangle$ and $\langle  C_{1,2,3,4,5}^m \tilde C_{1,2,3,4,5}^m\rangle$ have the same dimensions as $k^{10} A_{YM} \tilde A_{YM}$, which translates into dimensions of $R^5/k^2$ for their five-graviton components. Since $z_i$ integration over the $(X_{ij}\tilde
X_{kl} + \delta_{ik}\delta_{jl}\, s_{ij}\Omega)$ will be shown to have no leading contribution in $\ap$, the leading low
energy behaviour of (\ref{XXX}) is given entirely by $\langle C_{1,2,3,4,5}^m \tilde C_{1,2,3,4,5}^m \rangle$. In appendix \ref{super2} we will give a
superspace proof that this low-energy limit is totally symmetric in all the labels even though its
definition (\ref{Cmdef}) superficially treats the first external leg on a different footing.

So far, we have expressed the world-sheet integrand for the five-particle closed-string amplitude in terms of BRST
invariant kinematic factors $C_{1,ij}$ and $C_{1,2,3,4,5}^m \tilde C_{1,2,3,4,5}^m$ in pure spinor superspace. We
shall now translate these superfields into tree-level YM field-theory amplitudes \cite{Mafra:2010jq}.

The open-string BRST invariants $C_{1,ij}$ were found in \cite{Mafra:2012kh} to match with (permutations of) the
$(\ap)^2 \zeta_2$ corrections to disk amplitudes. According to (\ref{tree}), those in turn furnish linear combinations of the two independent YM
tree amplitudes $A_{YM}(1,2,3,4,5)$ and $A_{YM}(1,3,2,4,5)$,
\beq
\langle C_{1,23} \rangle =  s_{45} \ \bigl(s_{24} \,A_{YM}(1,3, 2, 4, 5)  -  s_{34}
   A_{YM}(1,2, 3, 4, 5) \bigr)
\label{CtoYM}
\eeq
with bilinear coefficients in Mandelstam variables \cite{Mafra:2011nv,Mafra:2011nw}. 
This guarantees that the $\langle C_{1,ij} \tilde C_{1,kl} \rangle$ terms in (\ref{XXX}) arising from $K^{(a)}$ can be expressed in terms of the four
independent kinematic factors formed by bilinears $\{ A_{YM}(1,2,3,4,5),A_{YM}(1,3,2,4,5) \} \times \{ \tilde
A_{YM}(1,2,3,4,5), \tilde A_{YM}(1,3,2,4,5) \} $.

The additional BRST invariant, $C_{1,2,3,4,5}^m \tilde C_{1,2,3,4,5}^m$ in (\ref{Cmdef}), involves contractions between
vectorial superfields from the left- and right-movers. Hence, the results from the open-string sector is not enough to determine the relation with $A_{YM}$ bilinears. Instead, we shall evaluate individual supercomponents of the last term in 
(\ref{XXX}) using the methods of \cite{Mafra:2010pn} and compare the result with the component representation of a $A_{YM}$ ansatz.

It turns out that the any type IIB component expansion encoded in the pure spinor superspace expression \eqref{Cmdef}
can be matched with a bilinear combination of $A_{YM}$ tree amplitudes. The expansion coefficients, however, depend on the overall $U(1)$ charge $q$ of the components of the maximal supergravity multiplet being scattered. Therefore one cannot hope to find a
single superspace expression like \eqref{CtoYM} which contains all supercomponents at the same time. In fact, the component evaluation based on \cite{Mafra:2010pn} shows that \eqref{Cmdef} can be expressed in terms of the tree amplitude (\ref{het1,30}) in the form
\beq
\langle C_{1,2,3,4,5}^m \, \tilde C_{1,2,3,4,5}^m \rangle = \left\{ \begin{array}{rl} +1 \,A_{YM}^t \, S_0 \, M_3 \, \tilde A_{YM} &: \ U(1) \ \te{conserved}, \ q=0\, . \\
-\tfrac{1}{3} \, A_{YM}^t \, S_0 \, M_3 \, \tilde A_{YM} &: \ U(1) \ \te{violated} , \ q=\pm 2\,. \end{array} \right.
\label{CmCmtoYM}
\eeq
The specific $2\times2$ matrices, $S_0$ and $M_3$, for the five-particle amplitude are given in \eqref{YMvec} and \eqref{Stree5pt}.
The five-graviton and four-graviton--one-dilaton component amplitudes were used to probe the $U(1)$-conserving and $U(1)$-violating sectors, respectively, including the relative prefactor $-1/3$ in (\ref{CmCmtoYM}). As we will see in section \ref{sec:u1vio}, the latter agrees with expectations based on type IIB S-duality introduced in section \ref{uonenonconserve}.

After substituting \eqref{CtoYM} and \eqref{CmCmtoYM} into \eqref{XXX}  we see that the   $U(1)$-conserving  and $U(1)$-violating components of the one-loop five-particle
amplitude both have the structure
\beq
{\cal M}_{\te{1-loop}}^{q} = A_{YM}^t \, {\cal S}_{\te{1-loop}}^{q} (s_{ij}) \, \tilde A_{YM}
\label{5g}
\eeq
with the  two-component vectors, $A_{YM}$ and $\tilde A_{YM}$, defined by (\ref{YMvec}) encoding all the polarisation dependence.
The $2\times 2$ matrix ${\cal S}_{\te{1-loop}}^{q} (s_{ij})$, on the other hand, only depends on the dimensionless Mandelstam
variables through world-sheet integrals and captures the $\ap$ dependence.  The structure of (\ref{5g}) resembles
that of the tree amplitude, but it should be stressed that the tree-level amplitude did not involve a contraction between left-moving and right-moving superfields. The BRST invariant (\ref{CmCmtoYM}) in the one-loop case with such a contraction can be expanded in a $A_{YM}^t  \tilde A_{YM}$ basis with coefficients depending on the $U(1)$ charge. Hence, the form of ${\cal
S}_{\te{1-loop}}^{q} (s_{ij})$ depends on the $U(1)$ charge-violation of the amplitude. In section \ref{lowtorus}  we will obtain the low energy expansions of the analytic part of ${\cal S}_{\te{1-loop}}^{q} (s_{ij})$ for both the  cases $q=0$  and $q=\pm 2$  and compare their coefficients with those in  the expansion of the five closed-string tree amplitude given in (\ref{het1,33})  (and derived in  \cite{Schlotterer:2012ny}).

\subsection{The distinction between type IIA and type IIB in ten dimensions}
\label{type2a}
The derivation of the five-particle closed-string amplitude \eqref{XXX}
in pure spinor superspace is equally valid for both the type IIA and type IIB theories.  For type IIA the left- and right-movers have opposite space-time chirality while for type IIB these chiralities are the same.  As a consequence, 
when extracting the supercomponent expansions of \eqref{XXX} the difference between type IIA and IIB comes entirely
from the different sign of the ten dimensional Levi-Civita tensors for the pure spinor correlators of the form $\langle (\lambda^3
\theta^5)\rangle$ given in the appendix of \cite{PSanomaly}. Since there are only 9 linearly independent vectors for either the left- or right-movers (5 polarisations and 4 momenta), the difference between these two theories is restricted to the
kinematic factor which involves at least one  contraction between the left- and right-moving superfields.  This arises in the first term on the right-hand side of (\ref{Cmdef}). More precisely,  one can show that\footnote{We are using the following shorthand notation for contractions of the $\ep$ tensor:
\[
\epsilon^{m}(k^1,k^2,k^3,k^4, \tilde e^1,\tilde e^2,\tilde e^3,\tilde e^4,\tilde e^5) := k^1_{n_1}k^2_{n_2}k^3_{n_3}k^4_{n_4} \tilde e_{n_5}^1\tilde e_{n_6}^2\tilde e_{n_7}^3\tilde e_{n_8}^4\tilde e_{n_9}^5 \epsilon^{n_1 n_2\ldots n_9 m}
\]The same kind of shorthand is used for the $t_8$ tensor in (\ref{BR4}).}
\cite{Mafra:2010pn}
\beq
\label{vtdef}
\langle \tilde V_1 \tilde T^m_{2,3,4,5} \rangle = \cdots +
\frac{1}{5760} \left\{ \begin{array}{rl}
{}+ \epsilon^{m}(k^1,k^2,k^3,k^4, \tilde e^1,\tilde e^2,\tilde e^3,\tilde e^4,\tilde e^5)   & \hbox{(type IIB)} \\
{}- \epsilon^{m}(k^1,k^2,k^3,k^4 ,\tilde e^1,\tilde e^2,\tilde e^3,\tilde e^4,\tilde e^5)  & \hbox{(type IIA)} \end{array} \right.
\eeq
where terms in the ellipsis are identical in type IIB and type IIA theory, and $e_m,\tilde e_n$ denote the bosonic polarisations of the left- and right-moving sector. So we see that  the  $\epsilon^{\ldots}$ term arising from the right-moving $\tilde V_1 \tilde T^m_{2,3,4,5}$ flips sign between the two theories. However, the $\epsilon^{\ldots}$ contribution to the left-moving factor $V_1  T^m_{2,3,4,5}$ in (\ref{Cmdef}) has the same sign in  type IIA as in type IIB.  

As a result,  the terms in $\langle V_1\, T_{2,3,4,5}^m \, \tilde V_1 \, \tilde T_{2,3,4,5}^m \rangle$ containing the product of two $\epsilon^{\ldots}$'s contracted on at least one index have opposite 
signs in the type IIA and type IIB cases.  Such  bilinears in $\epsilon^{\ldots}$ reduce to products of Kronecker $\delta$'s and are not parity-violating.  Analogous observations were demonstrated in the somewhat different sigma model calculations in \cite{Grisaru:1986px,Grisaru:1986vi}.  In the type IIB case the presence of these terms quadratic in $\epsilon^{\ldots}$ are important in ensuring that the amplitude can be expressed as a bilinear in Yang--Mills tree amplitudes, generalising the structure of the type IIB tree amplitude.  By contrast, in  the type IIA case, the sign of the $\epsilon^{\ldots}$ bilinears  is different and the amplitude cannot be expressed as a bilinear in Yang--Mills trees.

Furthermore, in  the type IIB case the parity-violating single $\epsilon^{\ldots}$ term cancels out of   (\ref{Cmdef})  whereas the type IIA amplitude does contain such a  term.  An example of such a parity-violating component amplitude is the amplitude with one  NSNS--two form and four gravitons,
\begin{align}
&{\cal M}^{BR^4}_{\te{1-loop}} \eq \frac{2}{(360)^2} \int \frac{ \dd^2 \tau }{\tau_2^5} \int \dd^2 z_2 \ldots \dd^2 z_5 \ {\cal I}(s_{ij}) \, \Om \notag \\
& \ \ \times \ 
\left\{ \!\!\!\begin{array}{rcl}
&0   & \hbox{(type IIB)} \\
&B^1_{mn} \epsilon^{mn} (k^2, e^2, k^3, e^3, k^4, e^4, k^5 ,e^5)t_8(k^2, \tilde e^2, k^3, \tilde e^3, k^4, \tilde e^4, k^5, \tilde e^5)   & \hbox{(type IIA)} \end{array} \right.
\label{BR4}
\end{align}
where $B^1_{mn} = e_{[m}^1 \otimes \tilde e_{n]}^1$ denotes the two--form polarisation and $t_8$ is defined in \cite{Green:1987mn}.
This reproduces the one-loop calculation in \cite{onelooptest}. It again cannot be expressed as a bilinear of ten-dimensional YM tree amplitudes. 

Upon compactification to dimensions $D<10$, however, maximally supersymmetric YM theory is not chiral and there is no distinction between the type IIA and type IIB theories. Therefore, the property of five-particle type IIB amplitudes that the polarisation dependence is contained in $A_{YM}^t \tilde A_{YM}$ also applies to the compactified type IIA theory.

Note further that the identity (\ref{nowzero}) motivates the interpretation  of the vector $\langle V_1 T_{2,3,4,5}^m \rangle$ as the loop momentum dependent piece of a pentagon numerator in field theory.

\section{Low energy expansion of type IIB one-loop amplitudes}
\label{lowtorus}

In this section we will consider the analytic terms in the low energy expansion of the  five-graviton one-loop amplitude (\ref{5g}) in powers of the Mandelstam
invariants.  We will concentrate on the type IIB case since  in section \ref{sec:sdual} we will be interested in studying how our results  match with considerations of $SL(2,\ZZ)$ duality. We will begin in section \eqref{world-sheetdiag} by reviewing the corresponding expansion of the four-graviton one-loop amplitude studied in
\cite{Green:2008uj}.  This involves the expansion of the loop integrand as a power series in the world-sheet Green functions.  At order $(\alpha')^{n+3}$ this
results in a sum of one-particle irreducible (1PI) vacuum diagrams with $n$ propagators joining the four points, corresponding to the external vertex
operators.  These diagrams are then integrated over the fundamental domain of the modular parameter $\tau$.  Care has to be taken to separate the non-analytic
parts of the amplitude that correspond to non-local terms in the effective action, which is an unambiguous procedure, at least at low orders in $\alpha'$.  
We will then, in section \eqref{fivepointdiag}, systematically expand the analytic part of the matrix ${\cal S}_{\te{1-loop}}^{q}(s_{ij})$ in the five-particle
amplitude (\ref{5g}) in terms of the same 1PI diagrams together with a few five-point generalisations. This extends partial results of \cite{Richards:2008jg}
and paves the way to identification of tree-level matrices $M_{2k+1}$ defined by (\ref{defM}) in the momentum expansion of the one-loop amplitude (\ref{5g}).

In order to include the possibility of describing amplitudes in compactifications on a $d$-torus to $D=10-d$ dimensions the loop integrand must be multiplied
by the standard lattice factor that accounts for Kaluza--Klein charges and winding modes in the loop measure.  This means that we should integrate over the
world-sheet torus with measure 
\beq
\int \dd\mu_d (\tau) := \int_{{\cal F}} \frac{\dd^2 \tau}{\tau_2^2}\, \Gamma_{d,d}(B,G;\tau)\,,
\label{torusmeasure}
\eeq
where the $\tau$ integral is over a fundamental domain ${\cal F}$ and 
\beq
\Gamma_{d,d}(B,G;\tau) := \sum_{m^i,n^i \in \Z^d\times \Z^d}  \exp\left(-\frac{\pi}{\tau_2}(G_{ij} + B_{ij})(m^i-\tau n^i)(m^j - \bar \tau n^j)    \right)\,.
\label{latticedef}
\eeq
and $G_{ij}$, $B_{ij}$ ($i,j = 1, \dots, d)$ are the metric and antisymmetric potential on the $d$-torus \cite{Green:1982sw}.
These scalar fields parameterise moduli space defined by the coset $SO(d,d, \Z)\backslash SO(d,d,\R)/(SO(d,\R)\times SO(d,\R))$ associated with T-duality.
Such compactifications on flat manifolds preserve all 32 supercharges.  

\subsection{The four-particle amplitude and its world-sheet diagrams}
\label{world-sheetdiag}

The four-particle genus-one amplitude involves only one type of world-sheet integral 
 \cite{Green:1982sw} (with Koba--Nielsen factor ${\cal I}(s_{ij})$ given by (\ref{KN})),
\begin{align}
I  := & \int  \dd\mu_d (\tau) \ \tau_2^{-3} \int \dd^2 z_2 \, \dd^2 z_3 \, \dd^2 z_4 \ {\cal I}(s_{ij}) \notag \\
=& \int  \dd\mu_d (\tau) \ \tau_2^{-3} \int  \dd^2 z_2 \, \dd^2 z_3 \, \dd^2 z_4 \ \prod_{i<j}^4 \left( \,  \sum_{n_{ij}=0}^{\infty} \, \frac{1}{n_{ij}!} \; s_{ij}^{n_{ij}} \, (\ln \chi_{ij})^{n_{ij}} \, \right) \,,
\label{int8}
\end{align}
where the integration measure is defined in (\ref{torusmeasure}).
The terms in the sum in the last factor in the integrand with a given value of $w= \sum_{i<j} n_{ij}$ involve $w$ powers of the propagator and contribute to
the terms in the expansion of order $(\alpha')^{w+3}$ relative to the classical supergravity terms.  In the following we will represent a propagator $\ln
\chi_{ij}$ by a line joining points $i$ and $j$,
\begin{center}
\vskip-7pt
\tikzpicture[scale=1.7]
\scope[xshift=-5cm,yshift=0.4cm]
\draw (0,0.5) node{$\bullet$} node[left]{$i$} ;
\draw (1,0.5) node{$\bullet$} node[right]{$j$} ;
\draw (0,0.5) -- (1,0.5) ;
\draw(2,0.5) node{$= \ \ln \chi_{ij} \ .$};
\endscope\endtikzpicture
\end{center}
\vskip-7pt
Terms in the sum in \eqref{int8} of order $(\alpha')^{w+3}$ are therefore represented by  diagrams with $w$ propagators, so we will also refer to them as having weight $w$.

The $z_i$ integrations within any diagram are conveniently performed by Fourier transforming  the propagator\footnote{Note that the normalization of $\ln \chi_{ij}$ chosen in this work differs by a factor of $4$ from \cite{Green:2008uj} and by a factor of $-2$ from \cite{Richards:2008jg}. As a consequence, any $w$ propagator diagram must be rescaled by factors of $4^w$ and $(-2)^w$ for comparison with \cite{Green:2008uj} and \cite{Richards:2008jg}, respectively.}, $\ln \chi_{ij}$ with its zero mode subtracted\footnote{Momentum conservation allows a simultaneous shift in all the two point functions in (\ref{KN}) by a $z$ independent function of $\tau,\tilde \tau$ without changing the Koba--Nielsen factor \cite{Green:1999pv}. This modification also drops out from derivatives $\pa \ln \chi_{ij}$.}, defined by (\ref{defeta}),
\begin{align}
\ln \chi_{ij} \eq -\frac{1}{\pi} \sum_{(m,n) \neq (0,0)} \frac{ \tau_2 }{|m\tau + n|^2} \, \exp \left( \frac{2\pi i}{\tau_2} \, (m \Re(z_{ij}) \tau_2 -(m\tau_1+n) \Im(z_{ij})) \right) \,. \label{chi1}  
\end{align}
The summation variables $m$ and $n$ are the integer components of the world-sheet momentum conjugate to $z_{ij}$, which are conserved at each vertex of the diagram. The $z_i$ integration within a diagram results in a
modular function of $\tau$ expressed as  a multiple sum over the internal world-sheet momenta.  This has to be integrated over a fundamental $\tau$ domain with measure $\mu_d(\tau)$ defined by (\ref{torusmeasure}).
Such diagrams were constructed in  \cite{Green:2008uj}  up to order $(\alpha')^9$ and  studied in ten space-time dimensions (as well as in the $S^1$ compactification to nine dimensions). 

{\it Note on conventions:} Within any diagram there may be multiple propagators joining a particular pair of points, $i$ and $j$.  This will be represented by the symbol
\begin{center}
\tikzpicture[scale=1.7]
\scope[xshift=-5cm,yshift=-0.4cm]
\draw (0,0.5) node{$\bullet$} node[left]{$i$} ;
\draw (1,0.5) node{$\bullet$} node[right]{$j$} ;
\draw (0,0.5) -- (1,0.5) ;
\draw (0.5,0.5) [fill=white] circle(0.15cm) ;
\draw (0.5,0.5) node{$n$};
\draw(2,0.5) node{$= \ (\ln \chi_{ij})^n \ .$};
\endscope
\endtikzpicture
\end{center}
\vskip-7pt
An immediate consequence of world-sheet momentum conservation at each vertex is that one-particle reducible diagrams integrate to zero.  Thus, any diagram
that contains a vertex connected by a single propagator gives a zero contribution\footnote{However, although this is true at fixed $\tau$, there is a
nonuniformity in the large-$\tau_2$ limit leading to important threshold singularities that contribute to the non-analytic part of the amplitude
\cite{Green:2008uj}, which we will not discuss further in this paper.}, such as $\ln \chi_{ij} \ln \chi_{jk}$ in the following figure,
\begin{center}
\vskip-7pt
\tikzpicture[scale=1.2]
\begin{scope}[xshift=-0.5cm]
\draw (0,0) node{$\bullet$} node[left]{$i$} ;
\draw (0,1) node{$\bullet$} node[left]{$j$} ;
\draw (0.7,0.5) node{$\bullet$} node[right]{$k$} ;
\draw (0,0) -- (0,1) ;
\draw (0,1) -- (0.7,0.5);
\draw(2.25,0.5) node{$= \ \ln \chi_{ij} \, \ln \chi_{jk} \ .$};
\end{scope}
\endtikzpicture
\end{center}

\vskip-7pt
The structure of the low order diagrams is summarised as follows. 
 The diagram with no propagators is simply the $n_{ij}=0$ term in the sum in \eqref{int8} and will be labelled $D_0:=1$.    The diagram consisting of a single propagator vanishes since the integral of \eqref{chi1} is zero at a fixed finite value of $\tau_2$,  so $D_1=0$.
There  is only one non-vanishing contribution at weight $w=2$, given by a diagram with a double line, 
\begin{center}
\tikzpicture[scale=1.6]
\draw (0,0) node{$\bullet$} ;
\draw (1,0) node{$\bullet$} ;
\draw (0,0) -- (1,0) ;
\draw (0.5,0) [fill=white] circle(0.15cm) ;
\draw (0.5,0) node{$2$};
\draw(1.6,0) node{$=: \ D_2\,.$};
\endtikzpicture
\end{center}

At weight $w=3$, there are two diagrams of different topologies,
\begin{center}
\vskip-7pt
\tikzpicture[scale=1.5]
\draw (0,0.5) node{$\bullet$} ;
\draw (1,0.5) node{$\bullet$} ;
\draw (0,0.5) -- (1,0.5) ;
\draw (0.5,0.5) [fill=white] circle(0.15cm) ;
\draw (0.5,0.5) node{$3$};
\draw(1.62,0.5) node{$=: \ D_3 \ ,$};
\begin{scope}[xshift=3cm]
\draw (0,0) node{$\bullet$} ;
\draw (0,1) node{$\bullet$} ;
\draw (0.7,0.5) node{$\bullet$} ;
\draw (0,0) -- (0,1) ;
\draw (0,0) -- (0.7,0.5);
\draw (0,1) -- (0.7,0.5);
\draw(1.4,0.5) node{$=: \ D_{111}\,,$};
\end{scope}
\endtikzpicture
\end{center}
\vskip-10pt
and weight $w=4$ gives diagrams of four distinct topologies\footnote{The  $D^2_2$ term can either be represented by a disconnected diagram or by an equivalent connected diagram,
\begin{center}
\vskip-10pt
\tikzpicture[scale=1.7]
\draw (-0.5,0.7) node{$\bullet$} ;
\draw (0.5,0.3) node{$\bullet$} ;
\draw (0.5,0.7) node{$\bullet$} ;
\draw (-0.5,0.3) node{$\bullet$} ;
\draw (-0.5,0.3) -- (0.5,0.3);
\draw (-0.5,0.7) -- (0.5,0.7) ;
\draw (0,0.3) [fill=white] circle(0.15cm) ;
\draw (0,0.3) node{$2$};
\draw (0,0.7) [fill=white] circle(0.15cm) ;
\draw (0,0.7) node{$2$};
\draw(1.25,0.5) node{$= \ D_2^2 \ = $};
\scope[xshift=2.5cm]
\draw (1.5,0.3) node{$\bullet$} ;
\draw (0.5,0.7) node{$\bullet$} ;
\draw (-0.5,0.3) node{$\bullet$} ;
\draw (-0.5,0.3) -- (0.5,0.7);
\draw (1.5,0.3) -- (0.5,0.7) ;
\draw (0,0.5) [fill=white] circle(0.15cm) ;
\draw (0,0.5) node{$2$};
\draw (1,0.5) [fill=white] circle(0.15cm) ;
\draw (1,0.5) node{$2$};
\endscope
\endtikzpicture
\end{center}
}:
\begin{center}
\tikzpicture[scale=1.6]
\begin{scope}[xshift=0cm, yshift=-1.5cm]
\draw (0,0.5) node{$\bullet$} ;
\draw (1,0.5) node{$\bullet$} ;
\draw (0,0.5) -- (1,0.5) ;
\draw (0.5,0.5) [fill=white] circle(0.15cm) ;
\draw (0.5,0.5) node{$4$};
\draw(1.57,0.5) node{$=: \ D_4 \ ,$};
\end{scope}
\begin{scope}[xshift=6.9cm, yshift=-1.5cm]
\draw (0.5,0) node{$\bullet$} ;
\draw (1,0) node{$\bullet$} ;
\draw (0.5,1) node{$\bullet$} ;
\draw (1,1) node{$\bullet$} ;
\draw (0.5,0) -- (0.5,1);
\draw (1,1) -- (1,0) ;
\draw (0.5,0.5) [fill=white] circle(0.15cm) ;
\draw (0.5,0.5) node{$2$};
\draw (1,0.5) [fill=white] circle(0.15cm) ;
\draw (1,0.5) node{$2$};
\draw(1.62,0.5) node{$=: \ D_2^2 $};
\end{scope}
\begin{scope}[xshift=2.45cm, yshift=-1.5cm]
\draw (0,0) node{$\bullet$} ;
\draw (0,1) node{$\bullet$} ;
\draw (0.7,0.5) node{$\bullet$} ;
\draw (0,0) -- (0,1) ;
\draw (0,0) -- (0.7,0.5);
\draw (0,1) -- (0.7,0.5);
\draw (0,0.5) [fill=white] circle(0.15cm) ;
\draw (0,0.5) node{$2$};
\draw(1.4,0.5) node{$=: \ D_{211} \ ,$};
\end{scope}
\begin{scope}[xshift=4.7cm, yshift=-1.5cm]
\draw (0,0) node{$\bullet$} ;
\draw (1,0) node{$\bullet$} ;
\draw (0,1) node{$\bullet$} ;
\draw (1,1) node{$\bullet$} ;
\draw (0,0) -- (1,0) ;
\draw (0,0) -- (0,1);
\draw (1,1) -- (1,0) ;
\draw (1,1) -- (0,1);
\draw(1.75,0.5) node{$=: \ D_{1111} \ ,$};
\end{scope}
\endtikzpicture
\end{center}

\vskip-5pt 
Using the Fourier expansion (\ref{chi1}) each propagator is associated with a factor of $-\tau_2/(\pi |m \tau + n|^2)$, where the components $m,n$
of the discrete world-sheet loop momentum are conserved at every vertex.  So the modular functions, $D_{ \ldots}$, are given by multiple sums over integer
loop momenta. For example, the lowest-weight cases are given by
\bea
D_2(\tau) &\eq &\frac{1}{\pi^2} \sum_{(m,n) \neq (0,0)} \frac{ \tau_2^2 }{|m\tau + n|^4},\qquad D_{111} (\tau) \eq - \frac{1}{\pi^3} \sum_{(m,n) \neq (0,0)} \frac{ \tau_2^3 }{|m\tau + n|^6}\nn\\
D_3(\tau) &\eq &-\frac{1}{\pi^3} \hskip-20pt   \sum_{(m_1,n_1), (m_2,n_2)  \neq (0,0)  \atop{ (m_1+m_2,n_1+n_2) \neq (0,0) } }
\hskip-10pt\frac{ \tau_2^3 }{|m_1 \tau + n_1|^2 \, |m_2 \tau + n_2|^2 \,|(m_1+m_2) \tau + ( n_1 + n_2) |^2}\,,
\label{diagramsums}
\eea 
with analogous expressions for the weight-four diagrams listed above and the higher-order diagrams that enter into the $\alpha'$ expansion of the four-particle amplitude.  In the $\ap$ expansion of the four-point one-loop amplitude, the coefficients of any product $M_{l_1} M_{l_2}\ldots$ are given by the integral of a particular linear combination of $D_{\ldots}$'s. We will use the notation
\beq
\lat^{(d)}_{l_1,l_2,\ldots} = \int \dd \mu_d(\tau) \ \sum_{} D_{\ldots}\,,
\label{xiint}
\eeq
where the $l_1,l_2,\ldots$ subscript refers to the accompanying $M_{l_1} M_{l_2}\ldots$. Explicit examples are given in \eqref{calDs} and other parts of
section \ref{sec:sdual}.  In the $D=10$ case (where $d=0$) the coefficients $ \lat^{(0)}_{\ldots}$ are constants that reduce to multi-zeta values in the cases
that we will consider, whereas the quantities $\lat^{(d)}_{\ldots}$ depend on the moduli of the $(10-d)$-dimensional theory.

The complete set of such sums that arise up to order $(\alpha')^9$ (as well as certain classes of higher weight diagrams) were analysed in \cite{Green:2008uj}
in sufficient detail to determine their contributions to the $D=10$ (or $d=0$) four-particle $\ap$ expansion.  Upon compactification on a circle of radius $r$
to $D=9$ (i.e., when $d=1$) the analysis in \cite{Green:2008uj} determined, for each power of $\ap$, the terms that are power-behaved in $r$ in the limit $r^2
\gg 1/\ap$.  For compactifications to lower dimensions (larger values of $d$) the coefficients of the terms in the $\ap$ expansion depend on the moduli that
parameterise the coset $SO(d,d)/(SO(d)\times SO(d))$ as discussed in \cite{Green:2010wi, Green:2010kv}

The evaluation of these multiple sums rapidly gets unwieldy and we do not have a general procedure for analysing the contribution of higher-weight diagrams.
However, it is possible to analyse completely certain infinite subsets of diagrams.  For example, the subset of weight-$k$ diagrams that form $k$-sided
polygons are non-holomorphic Eisenstein series',
\beq
 E_k(\tau) \eq (- \pi)^k  D_{ \underbrace{ 11\ldots 1}_{k}} \eq \sum_{(m,n) \neq (0,0)} \frac{ \tau_2^k }{|m\tau + n|^{2k}} \,.
\eeq

In the following we will extend the diagrammatic discussion to the five-particle amplitude up to order $(\ap)^9\, \cov^{10}R^5$.

\subsection{Five-particle world-sheet integrals and their diagrammatic expansion}
\label{fivepointdiag}

We now turn to consider the expansion of the five-particle amplitude.  The integrand of the amplitude (\ref{XXX}) comprises  terms proportional to $\Omega$, see (\ref{defomega}), and terms proportional to $X_{ij}\tilde X_{kl} $. The latter in turn is proportional to $\partial \log \chi_{ij}\bar\partial\log\chi_{kl}$ by (\ref{defeta}).  As a result, the diagrams that enter into the low energy expansion have the additional feature that they include
lines with holomorphic or anti-holomorphic derivatives acting on them.  In the following such a line will be represented by
\begin{center}
\vskip-7pt
\tikzpicture[scale=1.6]
\begin{scope}[xshift=-0.7cm, yshift=1cm]
\draw (0,0) node{$\bullet$} node[left]{1};
\draw (1,0) node{$\bullet$} node[right]{2};
\draw[dashed]  (0,0) -- (1,0) ;
\draw (0.5,0) node{$>$};
\draw (1.9,0) node{$= \ \partial \ln \chi_{12} $};
\end{scope}
\endtikzpicture
\end{center}
\vskip-7pt
where the arrow indicates a derivative (and it is not necessary to distinguish holomorphic or anti-holomorphic derivatives at the level of the five-particle amplitude).

\begin{figure}
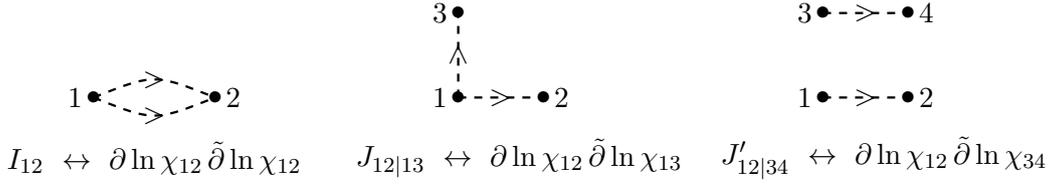

\centering
\tikzpicture [scale=1.6,line width=0.30mm]
\draw (0,0) node{$\bullet$} node[left]{1};
\draw (1,0) node{$\bullet$} node[right]{2};
\draw[dashed] (0,0) ..controls (0.5,0.2) .. (1,0) ;
\draw[dashed] (0,0) ..controls (0.5,-0.2) .. (1,0) ;
\draw (0.5,0.15) node{$>$};
\draw (0.5,-0.15) node{$>$};
\draw (0.5,-0.5) node{$I_{12} \ \leftrightarrow \ \partial \ln \chi_{12} \, \tilde \partial \ln \chi_{12}$};
\begin{scope}[xshift=3cm]
\draw (0,0) node{$\bullet$} node[left]{1};
\draw (0.7,0) node{$\bullet$} node[right]{2};
\draw (0,0.7) node{$\bullet$} node[left]{3};
\draw[dashed]  (0,0) -- (0.7,0) ;
\draw (0.35,0) node{$>$};
\draw[dashed]  (0,0) -- (0,0.7) ;
\draw(0,0.35) node[rotate=90] {$>$};
\draw (0.5,-0.5) node{$J_{12|13} \ \leftrightarrow \ \partial \ln \chi_{12} \, \tilde \partial \ln \chi_{13}$};
\end{scope}
\begin{scope}[xshift=6cm]
\draw (0,0) node{$\bullet$} node[left]{1};
\draw (0.7,0) node{$\bullet$} node[right]{2};
\draw (0,0.7) node{$\bullet$} node[left]{3};
\draw (0.7,0.7) node{$\bullet$} node[right]{4};
\draw[dashed]  (0,0) -- (0.7,0) ;
\draw[dashed]  (0,0.7) -- (0.7,0.7) ;
\draw (0.35,0) node{$>$};
\draw (0.35,0.7) node{$>$};
\draw (0.5,-0.5) node{$J'_{12|34} \ \leftrightarrow \ \partial \ln \chi_{12} \, \tilde \partial \ln \chi_{34}$};
\end{scope}
\endtikzpicture
\caption{Three different topologies of five-point integrals: Directed dashed lines represent both holomorphic and antiholomorphic derivatives $\pa \ln \chi_{ij}$ and $\tilde \pa \ln \chi_{ij}$.} 
\label{5pttopo}
\end{figure}

The complete five-particle amplitude involves the sum of a number of integrals over the moduli space of the torus: One of these is a totally symmetric integral \beq
K := -\frac{1}{\pi} \int \dd \mu_d(\tau) \ \tau_2^{-3} \int \dd^2 z_2 \ldots \dd^2 z_5
\ {\cal I}(s_{ij}) \, \Om 
\label{defk}
\eeq
and the others comprise 36 integrals involving a pair of holomorphic and antiholomorphic propagator derivatives $X_{ij} \tilde X_{pq} \sim \pa \ln \chi_{ij}  \tilde \pa \ln \chi_{pq}$.  
Depending on the relative labels $i,j \leftrightarrow p,q$, we have to distinguish three topologies \cite{Richards:2008jg} of integrands $\sim X_{ij} \tilde X_{pq}$ shown in figure \ref{5pttopo}. 
The integral expressions associated with these diagrams are given by
\begin{align}
I_{rs}& :=  \frac{1}{\pi} \int \dd \mu_d(\tau) \ \tau_2^{-3} \int  \dd^2 z_2 \ldots \dd^2 z_5 \ {\cal I}(s_{ij}) \ \pa \ln \chi_{rs} \, \tilde \pa \ln \chi_{rs}
\label{int26} \\ 
J_{rs|rt}&:= \frac{1}{\pi} \int \dd \mu_d(\tau) \ \tau_2^{-3} \int  \dd^2 z_2 \ldots \dd^2 z_5\ {\cal I}(s_{ij}) \ \pa \ln \chi_{rs} \, \tilde \pa \ln \chi_{rt}
\label{int27} \\
J'_{rs|tu}&:= \frac{1}{\pi} \int \dd \mu_d(\tau) \ \tau_2^{-3} \int  \dd^2 z_2 \ldots \dd^2 z_5 \ {\cal I}(s_{ij}) \ \pa  \ln \chi_{rs} \, \tilde \pa \ln \chi_{tu} \ .
\label{int28}
\end{align}
Expanding these integrals is a more substantial challenge. Integrals of $J$ and $J'$ type do not involve any poles in $s_{ij}$, and the power series expansion in $\ap$ can be performed at the level of the
Koba--Nielsen factor (\ref{KN}) in the integrand. In the case of the $I_{12}$ topology, however, we first have to subtract the $s_{12}^{-1}$ pole whose residue is given by the four-point integral (\ref{int8}) with momenta $k_1+k_2,k_3,k_4,k_5$ \cite{Richards:2008jg}:
\begin{align}
I_{12} &\eq  \frac{1}{s_{12}} \, I(k_1+k_2,k_3,k_4,k_5) \ + \ I_{12}^{\te{reg}}
\label{polea} \\
I_{12}^{\te{reg}} &\eq  \frac{1}{\pi} \int \dd \mu_d(\tau) \ \tau_2^{-3} \int \dd^2 z_2 \ldots \dd^2 z_5 \ \prod_{i<j}^5 \left( \,  \sum_{n_{ij}=0}^{\infty} \, \frac{1}{n_{ij}!} \; s_{ij}^{n_{ij}} \, (\ln \chi_{ij})^{n_{ij}} \, \right) \ \pa \ln \chi_{rs} \, \tilde \pa \ln \chi_{rs}
\label{poleb}
\end{align}
In other words, removal of the pole in (\ref{polea}) paves the way towards Taylor expanding the Koba--Nielsen factor. This generalises the treatment of
multiparticle disk integrals with kinematic poles in \cite{Broedel:2013tta}. The low energy behaviour of the regular part $I_{12}^{\te{reg}}$ can then be
obtained by the same methods as those applied to $J$ and $J'$. Hence, the residual task is the reduction of integrals of the form
\beq
\int \dd^2 z_2\ldots \dd^2 z_5 \ \pa \ln \chi_{rs} \, \tilde \pa \ln \chi_{tu} \ \prod_{i<j}^5 ( \ln \chi_{ij} )^{n_{ij}} \ , \ \ \ \ \ \ n_{ij} \in \NN
\label{proto}
\eeq
to an appropriate basis of multiple sums.

\begin{figure}
\centering
\begin{tikzpicture} [scale=0.8,line width=0.30mm]
\draw (-2.3,0)node{$0 \ =$};
\draw(0,0) node{$\bullet$} ;
\draw (1,1)node{$\bullet$}  -- (0,0);
\draw (1,-1)node{$\bullet$}  -- (0,0);
\draw[dashed] (-1,1)node{$\bullet$}  -- (0,0);
\draw (-1,-1)node{$\bullet$}  -- (0,0);
\draw (-0.5,0.5) node[rotate=-45]{$>$};
\draw (2,0)node{$+$};
\scope[xshift=4cm]
\draw(0,0) node{$\bullet$} ;
\draw[dashed] (1,1)node{$\bullet$}  -- (0,0);
\draw (1,-1)node{$\bullet$}  -- (0,0);
\draw (-1,1)node{$\bullet$}  -- (0,0);
\draw (-1,-1)node{$\bullet$}  -- (0,0);
\draw (0.5,+0.5) node[rotate=-135]{$>$};
\endscope
\draw (6,0)node{$+$};
\scope[xshift=8cm]
\draw(0,0) node{$\bullet$} ;
\draw (1,1)node{$\bullet$}  -- (0,0);
\draw[dashed] (1,-1)node{$\bullet$}  -- (0,0);
\draw (-1,1)node{$\bullet$}  -- (0,0);
\draw (-1,-1)node{$\bullet$}  -- (0,0);
\draw (0.5,-0.5) node[rotate=135]{$>$};
\endscope
\draw (10,0)node{$+$};
\scope[xshift=12cm]
\draw(0,0) node{$\bullet$} ;
\draw (1,1)node{$\bullet$}  -- (0,0);
\draw (1,-1)node{$\bullet$}  -- (0,0);
\draw (-1,1)node{$\bullet$}  -- (0,0);
\draw[dashed] (-1,-1)node{$\bullet$}  -- (0,0);
\draw (-0.5,-0.5) node[rotate=45]{$>$};
\endscope
\end{tikzpicture}
\caption{Momentum conservation at vertices of world-sheet diagrams following from $z_i$ integration.}
\label{momconserv}
\end{figure}
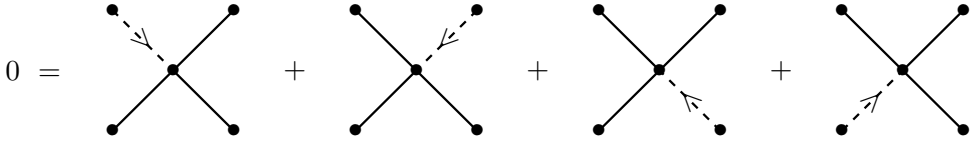

\begin{figure}
\centering
\begin{tikzpicture} [scale=1.5,line width=0.30mm]
\draw (0,0.2) node{$\bullet$} node[left]{1};
\draw (1,0.2) node{$\bullet$} node[right]{2};
\draw[dashed]  (0,0.2) -- (1,0.2) ;
\draw (0.4,0.2) node{$>$};
\draw (0.6,0.2) node{$>$};
\draw (0.5,-0.2) node{$= \ \partial \tilde \partial \ln \chi_{12}  \ ,$};
\draw (1.75,0) node{$\Longrightarrow$};
\scope[xshift=2.5cm]
\draw (0,-0.5) node{$\bullet$} -- (0,0.5) node{$\bullet$};
\draw (1,-0.5) node{$\bullet$} -- (1,0.5) node{$\bullet$};
\draw (0,-0.5) -- (1,-0.5); 
\draw (0,0.5) -- (1,0.5); 
\draw[dashed] (0,-0.5) -- (1,0.5);
\draw (0.4,-0.1)node[rotate = 45] {$>$};
\draw (0.6,0.1)node[rotate = 45] {$>$};
\draw (1.5,0) node{$=$};
\draw (2,-0.5)node{$\bullet$} ..controls (1.8,0) .. (2,0.5)node{$\bullet$} ;
\draw (2,-0.5) ..controls (2.2,0) .. (2,0.5) ;
\draw (2,-0.5) node{$\bullet$} ..controls (2.5,-0.7) ..  (3,-0.5) node{$\bullet$};
\draw (2,-0.5) node{$\bullet$} ..controls (2.5,-0.3) ..  (3,-0.5) node{$\bullet$};
%
\draw (3.5,0) node{$-$};
\draw (4,-0.5) node{$\bullet$} -- (4,0.5) node{$\bullet$};
\draw (5,-0.5) node{$\bullet$} -- (5,0.5) node{$\bullet$};
\draw (4,-0.5) -- (5,-0.5); 
\draw (4,0.5) -- (5,0.5); 
\endscope
\end{tikzpicture}
\caption{Graphical formulation of the Laplace equation: The first diagram on the right-hand side originates from the $\tau_2^{-1}$ factor
in $\pa \bar \pa \ln \chi_{ij}\sim (\de^2(z_{ij}) - \tau_2^{-1})$, the second diagram is due to the delta function admixture.}
\label{laplace}
\end{figure}
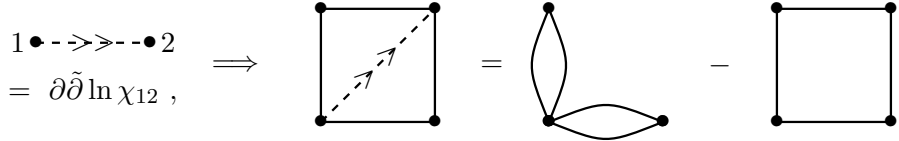

At low orders in the expansion expressions of the form \eqref{proto}, can be reduced to the expressions that define the $D_{\dots}$'s that arose in expanding the four-particle amplitude \eqref{int8}.  This reduction involves:
\begin{itemize}
\item  Integration by parts, or momentum conservation (as illustrated in figure \ref{momconserv}).  This can be used to express diagrams with non-coincident derivatives, as  in the integrals (\ref{int27}), (\ref{int28}) and (\ref{poleb}),  in terms of  those with a laplacian acting on one propagator.  However, this fails for some diagrams with more than 5 propagators, as will be discussed later.
\item The Laplace equation (\ref{defomega}) and its diagrammatic formulation shown in figure \ref{laplace} reduces any diagram with a double derivative on a propagator to a diagram with one less propagator.
 In diagrammatic language, the  action of the laplacian on a propagator, $\partial \bar
\partial \ln \chi_{ij} \sim (\de^2(z_{ij}) - \tau_2^{-1})$,  amounts to either shrinking or deleting it (with a relative sign), as in figure
\ref{laplace}. 

\end{itemize}
For diagrams with less than six propagators this procedure reduces the factor of $\pa \ln \chi_{rs} \, \tilde \pa \ln \chi_{tu} $ to a single propagator $\ln \chi_{ij}$ and the weight $w$ of the resulting expression is then related to the integer exponents $n_{ij}$ in (\ref{proto}) via $w=1+\sum_{i<j}n_{ij} $.  Certain diagrams with $w>4$ cannot be reduced to the diagrams   that arose in the case of the expansion of the four-particle amplitude and led to the expressions $D_{\ldots}$.

\subsection{Order by order expansion of five-point integrals}
\label{sec:expl}

In this section, we investigate low weight examples in detail in order to illustrate the procedure outlined above. The goal is to expand the five-point
integrals (\ref{defk}) to (\ref{int28}). In evaluating $K$, we may replace $\Om$ by $- \pi/\tau_2$ and proceed along the lines of section \ref{world-sheetdiag} where its four-particle relative (\ref{int8}) is analysed,
\begin{align}
& K \eq \int \dd \mu_d(\tau) \ \Big( \, D_0 \ + \ \frac{ D_2}{2}\, \sum_{i<j}s_{ij}^2  \ + \ D_{111} \, (s_{12}s_{13}s_{23} \, + \, 9 \ \te{perms.})   \notag \\
&\ \ + \ \frac{ D_3}{6}\, \sum_{i<j}s_{ij}^3 \ + \ \frac{ D_4 }{24}\, \sum_{i<j}s_{ij}^4 \ + \ \frac{ D_{2}^2 }{4} \, (s_{12}^2 s_{13}^2 \, + \, s_{12}^2 s_{34}^2 \, + \, 43 \ \te{perms.}) 
\label{Kint} \\
&\ \ + \ \frac{ D_{211} }{2} \, (s_{12}s_{13}s_{23}s_{123} \, + \, 9 \ \te{perms.}) \ + \ D_{1111} \, (s_{12}s_{23}s_{34} s_{41} \, + \, 14 \ \te{perms.}) \ + \ \ldots  \, \Big) \notag
\end{align}

Expanding the remaining integrals (\ref{int26}) to (\ref{int28}) is more complicated since they have an overall factor of  two propagators with a derivative acting on each of them.  The following enumerates the contributions that arise at each weight in the expansion.

\vskip0.2cm
\noindent
{\bf Weight 1}

The zeroth order term in the expansion has $n_{ij}=0$ in (\ref{proto}) and so involves the product of two propagators with a derivative acting on each of them. At fixed $\tau_2$ this integrates to zero as one can also see from the absence of a non-vanishing lattice sum $D_1$.  This not only ties in with the vanishing of the linearised $D^2\, R^4$ contribution to the graviton effective action,  which vanishes by integration by parts, but also implies the vanishing of its nonlinear contribution to the five-graviton amplitude, as well as the vanishing of the $R^5$ contribution to the five-graviton amplitude (as pointed out in \cite{Richards:2008jg}\footnote{The contributions  of $D^2\, R^4$ and $R^5$ to the five-particle amplitude in $D=4$ dimensions have also been argued to vanish by virtue of  supersymmetry in  theories with 32 supercharges \cite{Drummond:2003ex,Kallosh:2009jb, Elvang:2010jv}.}).
However,  as noted earlier, the fact that the propagator $\log \chi_{ij}$ (in (\ref{chi1})) is proportional to $\tau_2$ does lead to non-zero contributions to the non-analytic part of amplitudes from the large $\tau_2$ limit.  This is responsible for the threshold terms of the symbolic form $(s \ln s+ \ldots)\, R^4$ and $(\ln s+\ldots)\, R^5$ \cite{Green:2008uj}.

\vskip0.2cm
\noindent
{\bf Weight 2} 

In this case $\sum_{i<j} n_{ij}=1$ in (\ref{proto}) and the diagrams have three propagators. 
The following diagrams contribute to $I_{12}^{\te{reg}}$ and $J_{12|13}$  and are evaluated by integration by parts and use of the Laplace operator on one propagator. 
\begin{center}
\tikzpicture [scale=1.8,line width=0.30mm]
\draw (0,0)node{$\bullet$} ..controls (0.5,0.2) .. (1,0)node{$\bullet$} ;
\draw[dashed] (0,0)node{$\bullet$} ..controls (0.5,-0.2) .. (1,0)node{$\bullet$} ;
\draw[dashed] (0,0) -- (1,0);
\draw (0.5,0)node{$>$};
\draw (0.5,-0.15)node{$>$};
\draw (1.5,0) node{$=$};
\draw (2,0) node{$-\, \frac{1}{2}$};
\scope[xshift=2.5cm]
\draw (0,0)node{$\bullet$} ..controls (0.5,0.2) .. (1,0)node{$\bullet$} ;
\draw[dashed] (0,0)node{$\bullet$} ..controls (0.5,-0.2) .. (1,0)node{$\bullet$} ;
\draw (0,0) -- (1,0);
\draw (0.6,-0.15)node{$>$};
\draw (0.4,-0.15)node{$>$};
\endscope
\draw (4,0) node{$=$};
\draw (4.5,0) node{$\,\frac{1}{2} \, D_2$};
\scope[yshift=-1cm]
\draw (0,0)node{$\bullet$} -- (1,0)node{$\bullet$} ;
\draw (0.25,0.25)node[rotate=45]{$>$};
\draw (0.75,0.25)node[rotate=135]{$>$};
\draw (0.5,0.5)node{$\bullet$} ;
\draw[dashed] (0.5,0.5) -- (0,0);
\draw[dashed] (0.5,0.5) -- (1,0);
\draw (1.5,0.25) node{$=$};
\draw (2,0.25) node{$- $};
\scope[xshift=2.5cm]
\draw (0,0)node{$\bullet$} -- (1,0)node{$\bullet$} ;
\draw (0.2,0.2)node[rotate=45]{$>$};
\draw (0.3,0.3)node[rotate=45]{$>$};
\draw (0.5,0.5)node{$\bullet$} ;
\draw[dashed] (0.5,0.5) -- (0,0);
\draw  (0.5,0.5) -- (1,0);
\endscope
\draw (4,0.25) node{$=$};
\draw (4.5,0.25) node{$ - D_2$};
\endscope
\endtikzpicture
\end{center}
\vskip-8pt
For completeness, the following is a diagrammatic representation of the Laplace equation identities that are used in deriving the above expressions:
\begin{center}
\vskip-5pt
\tikzpicture [scale=1.8,line width=0.30mm]
\draw (0,0)node{$\bullet$} ..controls (0.5,0.2) .. (1,0)node{$\bullet$} ;
\draw[dashed] (0,0)node{$\bullet$} ..controls (0.5,-0.2) .. (1,0)node{$\bullet$} ;
\draw (0.4,-0.15)node{$>$};
\draw (0.6,-0.15)node{$>$};
\draw (0.5,0.2) [fill=white] circle(0.15cm) ;
\draw (0.5,0.2) node{$2$};
\draw (1.5,0) node{$=$};
\draw (2.5,0) node{$\times$};
\draw (3.5,0) node{$-$};
\draw (4,0)node{$\bullet$} ..controls (4.5,0.2) .. (5,0)node{$\bullet$} ;
%
\draw (4.5,0.2) [fill=white] circle(0.15cm) ;
\draw (4.5,0.2) node{$2$};
\draw (5.5,0) node{$=$};
\draw (6,0) node{$- \, D_2$};
\scope[yshift=-1cm]
\draw[dashed] (0,0)node{$\bullet$} -- (1,0)node{$\bullet$} ;
\draw (0.4,0)node{$>$};
\draw (0.6,0)node{$>$};
\draw (0.5,0.5)node{$\bullet$} ;
\draw (0.5,0.5) -- (0,0);
\draw (0.5,0.5) -- (1,0);
\draw (1.5,0.25) node{$=$};
\draw (2.5,0)node{$\bullet$} ..controls (2.6,0.25) .. (2.5,0.5)node{$\bullet$} ;
\draw (2.5,0)node{$\bullet$} ..controls (2.4,0.25) .. (2.5,0.5)node{$\bullet$} ;
\draw (3.5,0.25) node{$-$};
\draw (4,0)node{$\bullet$} -- (4.5,0.5)node{$\bullet$} ;
\draw (5,0)node{$\bullet$}-- (4.5,0.5) ;
%
\draw (5.5,0.25) node{$=$};
\draw (6,0.25) node{$+ \, D_2$};
\endscope
\endtikzpicture
\end{center}
\vskip-8pt
Propagators contracted to a point through the delta function in (\ref{defomega}) do not contribute in this case, which reflects the regularisation $\ln \chi_{ij}(0,\bar 0) \rightarrow 0$.  

Diagrams that arise in the expansion of $J'_{12|34}$ at this order vanish and hence $J'_{12|34} \big|_{w=2}=0$.  The above examples exhaust the
nonvanishing $w=2$ contributions of all the five-point integrals, resulting in 
\beq
I_{12}^{\te{reg}} \, \Big|_{w=2} \eq \frac{1}{2} \int \dd \mu_d(\tau) \ D_2 \, s_{12} \co J_{12|13}\, \Big|_{w=2} \eq - \int \dd \mu_d(\tau) \ D_2 \, s_{23}  \ .
\eeq
 
\vskip0.2cm
\noindent
{\bf Weight 3} 

The contributions of this order are terms  with  $\sum_{i<j} n_{ij}=2$  in (\ref{proto}) that correspond to diagrams with four propagators.   The following diagrams arise and can again be reduced  to standard form by integration by parts and use of the Laplace equation on one propagator. 
\begin{center}
\vskip-8pt
\tikzpicture [scale=1.7,line width=0.30mm]
\draw (0,0)node{$\bullet$} ..controls (0.5,0.2) .. (1,0)node{$\bullet$} ;
\draw[dashed] (0,0)node{$\bullet$} ..controls (0.5,-0.2) .. (1,0)node{$\bullet$} ;
\draw[dashed] (0,0)-- (1,0) ;
\draw (0.5,-0.15)node{$>$};
\draw (0.5,0)node{$>$};
\draw (0.5,0.2) [fill=white] circle(0.15cm) ;
\draw (0.5,0.2) node{$2$};
\draw (1.5,0) node{$=$};
\draw (1.9,0) node{$ \frac{1}{3}\, D_{3} $};
\scope[xshift=4cm]
\draw (0,0)--(0.5,0.5)node{$\bullet$}  ;
\draw (1,0)--(0.5,0.5) ;
\draw[dashed] (0,0)node{$\bullet$} ..controls (0.5,-0.2) .. (1,0)node{$\bullet$} ;
\draw[dashed] (0,0)-- (1,0) ;
\draw (0.5,-0.15)node{$>$};
\draw (0.5,0)node{$>$};
\draw (1.5,0) node{$=$};
\draw (2.4,0) node{$ D_{111} \ + \ \frac{1}{2}\, D_{3} $};
\endscope
\scope[yshift=-1cm]
\draw[dashed]  (0,0)--(0.5,0.5)node{$\bullet$}  ;
\draw(1,0)--(0.5,0.5) ;
\draw(0,0)node{$\bullet$} ..controls (0.5,-0.2) .. (1,0)node{$\bullet$} ;
\draw[dashed] (0,0)-- (1,0) ;
\draw (0.5,0)node{$>$};
\draw (0.25,0.25)node[rotate=45]{$>$};
\draw (1.5,0) node{$=$};
\draw (2,0) node{$ -\frac{1}{2}\, D_{3} $};
\endscope
\scope[xshift=4cm,yshift=-1cm]
\draw[dashed] (0,0)--(0.5,0.5)node{$\bullet$}  ;
\draw[dashed] (1,0)--(0.5,0.5) ;
\draw (0,0)node{$\bullet$} ..controls (0.5,-0.2) .. (1,0)node{$\bullet$} ;
\draw (0,0)-- (1,0) ;
\draw (0.25,0.25)node[rotate=45]{$>$};
\draw (0.75,0.25)node[rotate=135]{$>$};
\draw (1.5,0) node{$=$};
\draw (1.9,0) node{$ - D_{3} $};
\endscope
\scope[yshift=-1.8cm]
\draw[dashed] (0,-0.25)node{$\bullet$} -- (1,-0.25)node{$\bullet$};
\draw (0,0.25)node{$\bullet$} -- (1,0.25)node{$\bullet$};
\draw[dashed] (0,-0.25)  -- (0,0.25);
\draw (1,-0.25)  -- (1,0.25);
\draw (0.5,-0.25)node{$>$};
\draw (0,0)node[rotate=90]{$>$};
\draw (1.5,0) node{$=$};
\draw (2.0,0) node{$-D_{111}$};
\endscope
\scope[yshift=-1.8cm,xshift=4cm]
\draw[dashed] (0,-0.25)node{$\bullet$} -- (1,-0.25)node{$\bullet$};
\draw[dashed] (0,0.25)node{$\bullet$} -- (1,0.25)node{$\bullet$};
\draw (0,-0.25)  -- (0,0.25);
\draw (1,-0.25)  -- (1,0.25);
\draw (0.5,-0.25)node{$>$};
\draw (0.5,0.25)node{$>$};
\draw (1.5,0) node{$=$};
\draw (2.0,0) node{$-D_{111}$};
\endscope
\endtikzpicture
\end{center}
The Laplace equation identities that are used in determining these  expressions (following integration by parts) are 
the following.
\begin{center}
\tikzpicture [scale=1.8,line width=0.30mm]
\draw (0,0)node{$\bullet$} ..controls (0.5,0.2) .. (1,0)node{$\bullet$} ;
\draw[dashed] (0,0)node{$\bullet$} ..controls (0.5,-0.2) .. (1,0)node{$\bullet$} ;
\draw (0.4,-0.15)node{$>$};
\draw (0.6,-0.15)node{$>$};
\draw (0.5,0.2) [fill=white] circle(0.15cm) ;
\draw (0.5,0.2) node{$3$};
\draw (1.5,0) node{$=$};
\draw (2,0) node{$-D_3 \ ,$};
\scope[xshift=4cm]
\draw[dashed] (0,-0.25)node{$\bullet$} -- (1,-0.25)node{$\bullet$};
\draw (0,0.25)node{$\bullet$} -- (1,0.25)node{$\bullet$};
\draw (0,-0.25)  -- (0,0.25);
\draw (1,-0.25)  -- (1,0.25);
\draw (0.4,-0.25)node{$>$};
\draw (0.6,-0.25)node{$>$};
\draw (1.5,0) node{$=$};
\draw (2.1,0) node{$\,D_{111}$};
\endscope
\scope[xshift=0cm,yshift=-1cm]
\draw (0,0)node{$\bullet$} -- (1,0)node{$\bullet$} ;
\draw[dashed] (0,0)node{$\bullet$} ..controls (0.5,-0.2) .. (1,0)node{$\bullet$} ;
\draw (0.4,-0.15)node{$>$};
\draw (0.6,-0.15)node{$>$};
\draw (0.5,0.5)node{$\bullet$} ;
\draw  (0.5,0.5) -- (0,0);
\draw (0.5,0.5) -- (1,0);
\draw (1.5,0) node{$=$};
\draw (2,0) node{$- D_{111} \ ,$};
\endscope
\scope[xshift=4cm,yshift=-1cm]
\draw (0,0)node{$\bullet$}-- (1,0)node{$\bullet$} ;
\draw (0,0)node{$\bullet$} ..controls (0.5,-0.2) .. (1,0)node{$\bullet$} ;
\draw (0.5,0.5)node{$\bullet$} ;
\draw[dashed] (0.5,0.5) -- (0,0);
\draw (0.2,0.2)node[rotate=45]{$>$};
\draw (0.3,0.3)node[rotate=45]{$>$};
\draw (0.5,0.5) -- (1,0);
\draw (1.5,0) node{$=$};
\draw (2,0) node{$ D_{3} $};
\endscope
\endtikzpicture
\end{center}
\vskip-8pt

We conclude that the contributions of all diagrams at this order reduce to rational multiples of $D_3$ and $D_{111}$.  
Using the above results, the expansion of the integrals (\ref{int27}), (\ref{int28}) and (\ref{poleb})  to weight $w=3$ is
\begin{align}
J_{12|13} &\eq -\int \dd \mu_d(\tau) \ \Big(  \, D_2  s_{23} \, + \, D_{111} \, (s_{24}s_{34} +s_{25} s_{35}) \, + \, \frac{D_3}{2} \,  s_{23} \, (s_{12}
+ s_{13}+ s_{23})  \, + \, \cdots \Big)
\notag \\
J'_{12|34} &\eq \int \dd \mu_d(\tau) \ D_{111} \, ( s_{14}s_{23} - s_{13} s_{24} ) \, + \,  \cdots
\label{Jpexp}
 \\
I_{12}^{\te{reg}} &\eq  \int \dd \mu_d(\tau) \ \Big( \,\frac{D_2}{2} \, s_{12}  \, + \,  D_{111} \, (s_{13}s_{23}+ s_{14}s_{24} + s_{15}s_{25} ) \notag \\
& \ \ \ \ \ \ \ \ \ \ \ \ \ \ \ \ \ \  + \ \frac{D_3}{6} \,  \Bigl( s_{12}^2 \, + \, 3  (s_{13}s_{23} + s_{14}s_{24} +s_{15}s_{25} )  \Bigr) \ + \ \cdots \Big)\notag
\end{align} 
which  reproduce the expansion results in \cite{Richards:2008jg}.

A feature that is worth highlighting is that in reconstructing the expansion of the five-particle amplitude (\ref{XXX}), the left-right symmetric expression $X_{ij}\tilde X_{ij}$ is always accompanied by
$s_{ij}\Omega $. As a consequence, integrals of $I_{ij}$ type (\ref{int26}) only appear in combination with the $K$ integral (\ref{defk}) and important cancellations occur (e.g. the complete $D_0$ coefficient drops out). Therefore, the expansion of the piece of the amplitude that contains $I_{ij}$ is given, up to weight three, by
\begin{align}
 \frac{1}{\pi} \int \dd &\mu_d(\tau) \ \tau_2^{-3} \int \dd^2 z_2 \ldots \dd^2 z_5
\ {\cal I}(s_{ij}) \, \big[ \, X_{12} \, \tilde X_{12} \ + \ s_{12} \, \Om \, \big] \eq  s_{12} \, \int \dd \mu_d(\tau)\notag \\
& \Big( \, D_2 \,(s_{13} s_{23}+s_{14} s_{24}+s_{15} s_{25}) \ + \ \frac{D_3}{2} \, (s_{12} s_{13} s_{23}+s_{12} s_{14} s_{24}+s_{12} s_{15}
   s_{25}) \notag \\
   & \ \ + D_{111} \, (s_{13} s_{24} s_{34}+s_{13} s_{25} s_{35}+s_{14} s_{23} s_{34}+s_{14} s_{25}
   s_{45}  +  s_{15} s_{23} s_{35}+s_{15} s_{24} s_{45})  \notag \\
 & \ \ + \ \frac{D_3}{2} \, (s_{13}^2 s_{23}+s_{13} s_{23}^2 + \,s_{14} s_{24}^2+s_{14}^2 s_{24}+s_{15}
   s_{25}^2+s_{15}^2 s_{25}) \ + \ \cdots \, \Big) \,.
\end{align}

\vskip0.2cm
\noindent
{\bf Weight 4} 

The diagrammatic framework is easily applied to higher weight contributions. Terms with $\sum_{i<j} n_{ij}=3$  in (\ref{proto}) give rise to diagrams with five propagators, two of which have derivatives acting on them. If we ignore the derivatives, there are eight distinct  five-propagator diagrams, which  are listed  in appendix \ref{sec:A1}.   These diagrams (without arrows)  will be directly relevant for the $w=5$ case as described below.  However,  attaching  derivatives to two lines, which can be done in many ways, and using the integration by parts procedure and  Laplace equation, reduces the diagrams to the standard weight-four $D_{\ldots}$ that were encountered in the case of the four-particle amplitude.  As an example, consider attaching two arrows to the diagram that is equal to $D'_{1111}$ in appendix \ref{sec:A1}, which has a  particularly convoluted topology.  The  weight-four contributions that result are the following

\begin{center}
\tikzpicture [scale=1.3,line width=0.30mm]
\draw (1,-0.5) node{$\bullet$} -- (1,0.5) node{$\bullet$};
\draw[dashed] (0,-0.5) node{$\bullet$} -- (0,0.5) node{$\bullet$};
\draw (0,-0.5) -- (1,-0.5); 
\draw[dashed] (0,0.5) -- (1,0.5); 
\draw (0,-0.5) -- (1,0.5);
\draw (0,0)node[rotate = 90]{$>$};
\draw (0.5,0.5)node {$<$};
\draw (1.5,0) node{$\ = \ -$};
\scope[xshift=2.1cm]
\draw (1,-0.5) node{$\bullet$} -- (1,0.5) node{$\bullet$};
\draw[dashed] (0,-0.5) node{$\bullet$} -- (0,0.5) node{$\bullet$};
\draw (0,-0.5) -- (1,-0.5); 
\draw (0,0.5) -- (1,0.5); 
\draw (0,-0.5) -- (1,0.5);
\draw (0,-0.1)node[rotate = 90]{$>$};
\draw (0,0.1)node[rotate = 90]{$>$};
\endscope
\draw (3.7,0) node{$\ \ =  - D_{211}$};
\scope[yshift=-1.5cm]
\draw (1,-0.5) node{$\bullet$} -- (1,0.5) node{$\bullet$};
\draw[dashed] (0,-0.5) node{$\bullet$} -- (0,0.5) node{$\bullet$};
\draw (0,-0.5) -- (1,-0.5); 
\draw (0,0.5) -- (1,0.5); 
\draw[dashed] (0,-0.5) -- (1,0.5);
\draw (0,0)node[rotate = 90]{$>$};
\draw (0.5,0)node[rotate = 45] {$>$};
\draw (1.5,0) node{$\ = \ -$};
\scope[xshift=2.1cm]
\draw (1,-0.5) node{$\bullet$} -- (1,0.5) node{$\bullet$};
\draw (0,-0.5) node{$\bullet$} -- (0,0.5) node{$\bullet$};
\draw (0,-0.5) -- (1,-0.5); 
\draw (0,0.5) -- (1,0.5); 
\draw[dashed] (0,-0.5) -- (1,0.5);
\draw (0.6,0.1)node[rotate = 45] {$>$};
\draw (0.4,-0.1)node[rotate = 45] {$>$};
\draw (1.25,0) node{$ -$};
\endscope
\scope[xshift=3.7cm]
\draw (1,-0.5) node{$\bullet$} -- (1,0.5) node{$\bullet$};
\draw (0,-0.5) node{$\bullet$} -- (0,0.5) node{$\bullet$};
\draw[dashed] (0,-0.5) -- (1,-0.5); 
\draw (0,0.5) -- (1,0.5); 
\draw[dashed] (0,-0.5) -- (1,0.5);
\draw (0.5,0)node[rotate = 45] {$>$};
\draw (0.5,-0.5)node {$>$};
\endscope
\draw (5.9,0) node{$\ \ = \ \frac{1}{2} \, (D_{1111} - D_2^2)$};
\endscope
\scope[yshift=-3cm]
\draw (1,-0.5) node{$\bullet$} -- (1,0.5) node{$\bullet$};
\draw[dashed] (0,-0.5) node{$\bullet$} -- (0,0.5) node{$\bullet$};
\draw[dashed] (0,-0.5) -- (1,-0.5); 
\draw (0,0.5) -- (1,0.5); 
\draw (0,-0.5) -- (1,0.5);
\draw (0,0)node[rotate = 90]{$>$};
\draw (0.5,-0.5)node[rotate = 0] {$>$};
\draw (1.5,0) node{$\ \ = \ -$};
\scope[xshift=2.1cm]
\draw (1,-0.5) node{$\bullet$} -- (1,0.5) node{$\bullet$};
\draw[dashed] (0,-0.5) node{$\bullet$} -- (0,0.5) node{$\bullet$};
\draw (0,-0.5) -- (1,-0.5); 
\draw (0,0.5) -- (1,0.5); 
\draw[dashed] (0,-0.5) -- (1,0.5);
\draw (0.5,0)node[rotate = 45] {$>$};
\draw (0,0)node[rotate = 90] {$>$};
\draw (1.25,0) node{$ -$};
\endscope
\scope[xshift=3.7cm]
\draw (1,-0.5) node{$\bullet$} -- (1,0.5) node{$\bullet$};
\draw[dashed] (0,-0.5) node{$\bullet$} -- (0,0.5) node{$\bullet$};
\draw (0,-0.5) -- (1,-0.5); 
\draw (0,0.5) -- (1,0.5); 
\draw (0,-0.5) -- (1,0.5);
\draw (0,0.1)node[rotate = 90] {$>$};
\draw (0,-0.1)node[rotate = 90] {$>$};
\endscope
\draw (6.6,0) node{$ = - D_{211} \, + \, \frac{1}{2} \, (D_2^2 - D_{1111})$};
\endscope
\endtikzpicture
\end{center}
These expressions make use of the following  identities involving  the action of the  Laplace operator on a propagator 
\begin{center}
\tikzpicture [scale=1.3,line width=0.30mm]
\scope[xshift=2.5cm]
\draw (0,-0.5) node{$\bullet$} -- (0,0.5) node{$\bullet$};
\draw (1,-0.5) node{$\bullet$} -- (1,0.5) node{$\bullet$};
\draw (0,-0.5) -- (1,-0.5); 
\draw (0,0.5) -- (1,0.5); 
\draw[dashed] (0,-0.5) -- (1,0.5);
\draw (0.4,-0.1)node[rotate = 45] {$>$};
\draw (0.6,0.1)node[rotate = 45] {$>$};
\draw (1.5,0) node{$=$};
\draw (2,-0.5)node{$\bullet$} ..controls (1.8,0) .. (2,0.5)node{$\bullet$} ;
\draw (2,-0.5) ..controls (2.2,0) .. (2,0.5) ;
\draw (2,-0.5) node{$\bullet$} ..controls (2.5,-0.7) ..  (3,-0.5) node{$\bullet$};
\draw (2,-0.5) node{$\bullet$} ..controls (2.5,-0.3) ..  (3,-0.5) node{$\bullet$};
%
\draw (3.5,0) node{$-$};
\draw (4,-0.5) node{$\bullet$} -- (4,0.5) node{$\bullet$};
\draw (5,-0.5) node{$\bullet$} -- (5,0.5) node{$\bullet$};
\draw (4,-0.5) -- (5,-0.5); 
\draw (4,0.5) -- (5,0.5); 
\draw (6.3,0)node{$= \ D_2^2 \ - \ D_{1111}$};
\endscope
\scope[xshift=2.5cm,yshift=-1.5cm]
\draw (0,-0.5) node{$\bullet$} -- (0,0.5) node{$\bullet$};
\draw (1,-0.5) node{$\bullet$} -- (1,0.5) node{$\bullet$};
\draw (0,-0.5) -- (1,-0.5); 
\draw[dashed] (0,0.5) -- (1,0.5); 
\draw (0,-0.5) -- (1,0.5);
\draw (0.4,0.5)node{$>$};
\draw (0.6,0.5)node {$>$};
\draw (1.5,0) node{$=$};
\draw (2,-0.5)node{$\bullet$} ..controls (2.6,-0.1) .. (3,0.5)node{$\bullet$} ;
\draw (2,-0.5) ..controls (2.4,0.1) .. (3,0.5) ;
\draw (2,-0.5) -- (3,-0.5) node{$\bullet$};
\draw (3,-0.5) -- (3,0.5) node{$\bullet$};
%
\draw (3.5,0) node{$-$};
\draw (4,-0.5) node{$\bullet$} -- (4,0.5) node{$\bullet$};
\draw (5,-0.5) node{$\bullet$} -- (5,0.5) node{$\bullet$};
\draw (4,-0.5) -- (5,-0.5); 
\draw (4,-0.5) -- (5,0.5); 
\draw (5.9,0)node{$= \ D_{211}$};
\endscope
\endtikzpicture
\end{center}
We see that  any diagram of this topology appearing in the expansions of $I_{12}^{\te{reg}}$, $J_{12|13}$ and $J'_{12|34}$ can be reduced to rational combinations  of $(D_{1111}-D_2^2)$ and $D_{211}$. 
The analysis of the remaining topologies involving five propagators (such as $D_5, D_2D_3,\ldots$ in the terminology of appendix \ref{sec:A1}) is simpler and we will not display it here.
The complete expansion of the  five-point integrals up to $w=4$ is fully displayed in appendix \ref{app:w4}.

\vskip0.2cm
\noindent
{\bf Weights 5 and 6} 

Starting at  weight five, not all the derivatives in the $\partial \ln \chi_{ij} \tilde \partial \ln \chi_{pq}$ integrand can
be eliminated by this method. 
For $w=5$ (terms in \eqref{proto} with $\sum_{i<j}n_{ij} = 4$) the diagrams have six propagators, two of which have derivatives.  The derivatives can be eliminated from most types of diagrams as shown in appendix \ref{sec:A2}, 
although some of these were not encountered in the expansion of the four-particle amplitude because they involve more than four vertices.  This leave two types of diagrams from which the derivatives cannot be eliminated by the above procedure. The precise definition of these diagrams is ambiguous since they are  only defined up to terms arising from integration by parts, but they can be chosen to be the ones shown below
\begin{center}
\tikzpicture[scale=1.4]
\draw (0,0) node{$\bullet$} ;
\draw (0,1) node{$\bullet$} ;
\draw (0.5,0.5) node{$\bullet$} ;
\draw[dashed] (0.1,0) -- (0.5,0.4);
\draw[dashed] (0.1,1) -- (0.5,0.6);
\draw (0,0) -- (0.5,0.5);
\draw (0,1) -- (0.5,0.5);
\draw(0.3,0.8) node[rotate=315]{$>$};
\draw(0.3,0.2) node[rotate=225]{$>$};
\draw (0,0) -- (0,1);
\draw (0,0.5) [fill=white] circle(0.15cm) ;
\draw (0,0.5)  node{$2$};
\draw(1.2,0.5) node{$=: \ D^{\partial}_{222}$};
\begin{scope}[xshift=4cm, yshift=0cm]
\draw (0,0) node{$\bullet$} ;
\draw (1,0) node{$\bullet$} ;
\draw (0,1) node{$\bullet$} ;
\draw (1,1) node{$\bullet$} ;
\draw (0,0) -- (1,0) ;
\draw[dashed] (0,0) -- (0,1);
\draw(0,0.5) node[rotate=270]{$>$};
\draw (1,1) -- (1,0) ;
\draw (1,1) -- (0,1);
\draw (1,0.9) -- (0.1,0);
\draw[dashed] (0.9,1) -- (0,0.1);
\draw(0.45,0.55) node[rotate=45]{$>$};
\draw(1.7,0.5) node{$=: \ D''^{\partial}_{1111}$};
\end{scope}
\endtikzpicture
\end{center}
Together with the eight diagrams shown in appendix \ref{sec:A1}, the diagrams of this weight span a ten-dimensional vector space of lattice sums.

A similar analysis of the $w=6$ contributions (diagrams with seven propagators, two of which have derivatives) leads to the 20 diagrams from which derivatives have been eliminated that are listed in appendix \ref{sec:A2}, and leaves the following seven undetermined diagrams from which the derivatives cannot be eliminated by integration by parts. 
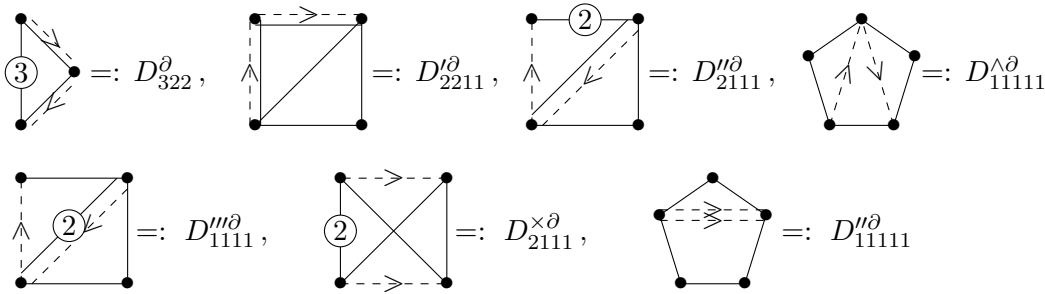
\begin{center}
\tikzpicture[scale=1.4]
\draw (0,0) node{$\bullet$} ;
\draw (0,1) node{$\bullet$} ;
\draw (0.5,0.5) node{$\bullet$} ;
\draw[dashed] (0.1,0) -- (0.5,0.4);
\draw[dashed] (0.1,1) -- (0.5,0.6);
\draw (0,0) -- (0.5,0.5);
\draw (0,1) -- (0.5,0.5);
\draw(0.3,0.8) node[rotate=315]{$>$};
\draw(0.3,0.2) node[rotate=225]{$>$};
\draw (0,0) -- (0,1);
\draw (0,0.5) [fill=white] circle(0.15cm) ;
\draw (0,0.5)  node{$3$};
\draw(1.1,0.5) node{$\ \ =: \, D^{\pa}_{322}  \, ,$};
\begin{scope}[xshift=2.2cm, yshift=0cm]
\draw (0,0) node{$\bullet$} ;
\draw (1,0) node{$\bullet$} ;
\draw (0,1) node{$\bullet$} ;
\draw (1,1) node{$\bullet$} ;
\draw (0,0) -- (1,0) ;
\draw (0.05,0) -- (0.05,1);
\draw[dashed] (-0.05,0) -- (-0.05,1);
\draw(-0.05,0.5) node[rotate=90]{$>$};
\draw (1,1) -- (1,0) ;
\draw (1,0.95) -- (0,0.95);
\draw[dashed] (1,1.05) -- (0,1.05);
\draw(0.5,1.05) node[rotate=0]{$>$};
\draw (1,1) -- (0,0);
\draw(1.6,0.5) node{$\ \ =: \, D'^{\pa}_{2211}\,,$};
\end{scope}
\begin{scope}[xshift=4.8cm, yshift=0cm]
\draw (0,0) node{$\bullet$} ;
\draw (1,0) node{$\bullet$} ;
\draw (0,1) node{$\bullet$} ;
\draw (1,1) node{$\bullet$} ;
\draw (0,0) -- (1,0) ;
\draw (1,1) -- (1,0) ;
\draw (1,1) -- (0,1);
\draw (0.9,1) -- (0,0.1);
\draw[dashed] (1,0.9) -- (0.1,0);
\draw(0.55,0.45) node[rotate=225]{$>$};
\draw[dashed] (0,0) -- (0,1);
\draw(0,0.5) node[rotate=90]{$>$};
\draw (0.5,1) [fill=white] circle(0.15cm) ;
\draw (0.5,1)  node{$2$};
\draw(1.6,0.5) node{$\ \ =: \, D''^{\pa}_{2111}\,,$};
\end{scope}
\begin{scope}[xshift=0cm, yshift=-1.5cm]
\draw (0,0) node{$\bullet$} ;
\draw (1,0) node{$\bullet$} ;
\draw (0,1) node{$\bullet$} ;
\draw (1,1) node{$\bullet$} ;
\draw (0,0) -- (1,0) ;
\draw[dashed] (0,0) -- (0,1);
\draw(0,0.5) node[rotate=90]{$>$};
\draw (1,1) -- (1,0) ;
\draw (1,1) -- (0,1);
\draw(0.65,0.55) node[rotate=225]{$>$};
\draw[dashed] (1,0.9) -- (0.1,0);
\draw (0.9,1) -- (0,0.1);
\draw (0.45,0.55) [fill=white] circle(0.15cm) ;
\draw (0.45,0.55)  node{$2$};
\draw(1.7,0.5) node{$=: \ D'''^{\pa}_{1111}\,,$};
\end{scope}
\begin{scope}[xshift=3cm, yshift=-1.5cm]
\draw (0,0) node{$\bullet$} ;
\draw (1,0) node{$\bullet$} ;
\draw (0,1) node{$\bullet$} ;
\draw (1,1) node{$\bullet$} ;
\draw[dashed] (0,0) -- (1,0) ;
\draw(0.5,0) node[rotate=0]{$>$};
\draw(0.5,1) node[rotate=0]{$>$};
\draw (0,0) -- (0,1);
\draw (1,1) -- (1,0) ;
\draw[dashed] (1,1) -- (0,1);
\draw (1,1) -- (0,0);
\draw (1,0) -- (0,1);
\draw (0,0.5) [fill=white] circle(0.15cm) ;
\draw (0,0.5)  node{$2$};
\draw(1.7,0.5) node{$=: \ D_{2111}^{\times \pa}\,,$};
\end{scope}
\begin{scope}[xshift=6cm, yshift=-1.5cm]
\draw (0,0.65) node{$\bullet$} ;
\draw (1,0.65) node{$\bullet$} ;
\draw (0.2,0) node{$\bullet$} ;
\draw (0.8,0) node{$\bullet$} ;
\draw (0.5,1) node{$\bullet$} ;
\draw(0.5,0.6) node[rotate=0]{$>$};
\draw(0.5,0.7) node[rotate=0]{$>$};
\draw(0.5,1)--(1,0.65);
\draw(1,0.65) --(0.8,0);
\draw(0.8,0)--(0.2,0);
\draw(0.2,0)--(0,0.65);
\draw(0,0.65)--(0.5,1);
\draw[dashed](0,0.6) -- (1,0.6);
\draw[dashed](0,0.7) -- (1,0.7);
\draw(1.76,0.5) node{$=: \ D''^{\partial}_{11111}$};
\end{scope}
\begin{scope}[xshift=7.4cm]
\draw (0,0.65) node{$\bullet$} ;
\draw (1,0.65) node{$\bullet$} ;
\draw (0.2,0) node{$\bullet$} ;
\draw (0.8,0) node{$\bullet$} ;
\draw (0.5,1) node{$\bullet$} ;
\draw(0.5,1)--(1,0.65);
\draw(1,0.65) --(0.8,0);
\draw(0.8,0)--(0.2,0);
\draw(0.2,0)--(0,0.65);
\draw(0,0.65)--(0.5,1);
\draw[dashed] (0.2,0) --(0.5,1); 
\draw[dashed] (0.8,0) --(0.5,1); 
\draw(0.35,0.5) node[rotate=70]{$>$};
\draw(0.65,0.5) node[rotate=290]{$>$};
\draw(1.65,0.5) node{$=: \, D^{\wedge \partial}_{11111}$};
\end{scope}
\endtikzpicture
\end{center}
So we see that the diagrams of weight 6 span a 27-dimensional vector space of lattice sums.
Again there are ambiguities in the definition of these diagrams since they  are only defined modulo integration by parts, corresponding to redefinitions of the basis.  

\section{Low energy type IIB amplitudes and S-duality}
\label{sec:sdual}

In this section, we will compare our results for the low energy expansion of four- and five-particle typeIIB amplitudes up to order $(\alpha')^9$. This
is of particular interest in the type IIB theory, in which the constraints of $SL(2,\Z)$ S-duality are expected to relate different orders in perturbation
theory in an interesting manner.  Since the amplitudes at tree level and one loop both take the form $A_{YM}^t {\cal S}_{\te{tree}} \tilde A_{YM}$ and
$A_{YM}^t {\cal S}_{\te{1-loop}} \tilde A_{YM}$, we can restrict our discussion to the quantities ${\cal S}_{\ldots}$. It turns out that several combinations
of multiple sums $D_{\ldots}$ arise in the $\ap$ expansions of ${\cal S}_{\te{1-loop}}$ in both the $N=4$ and $N=5$ cases.  It is also striking that the set
of tree-level kinematic invariants at any order in $\alpha'$ (encoded in the $M$ matrices) is reproduced in the one-loop amplitudes, although extra invariants
also appear at one loop in the $N=5$ case starting at order $(\alpha')^7\, \cov^6R^5$.  In the case of $U(1)$-conserving amplitudes we will see that the ratio
of the coefficients of the tree-level invariants and the coefficients of the same invariants at one-loop is identical in the $N=4$ and $N=5$ cases.  This has
suggestive implications for the pattern of non-perturbative $SL(2,\Z)$ S-duality of the type IIB amplitudes that extends the well-established $N=4$ pattern.

We will also determine the $\alpha'$ expansion of the amplitudes that violate the $U(1)$ charge by $q=\pm 2$ units (which arises when $N\ge 5$) up to order
$(\alpha')^9$.  We will see that the pattern of rational coefficients of such terms extends the systematics of of S-duality at order $(\alpha')^3$ described
in section \ref{dualityreview} in a compelling manner.

\subsection{The four-particle one-loop amplitude}

In order to make contact between four-particle and five-particle closed-string amplitudes at tree-level and one-loop, we summarise the four-particle one-loop
results of \cite{Green:2008uj}. In our present convention, the four-point amplitude
\beq
{\cal M}_{\te{1-loop}} \eq 2\pi I \cdot A_{YM} S_0 M_3 \tilde A_{YM}
\eeq
proportional to the integral $I$ defined by (\ref{int8}) allows to read off
\begin{align}
{\cal S}_{\te{1-loop}} =   S_0  \, \Bigl( \,
  \lat^{(d)}_3 \, M_3
&+ \lat^{(d)}_5 \, M_5
+ \lat^{(d)}_{3,3} \, M_3^2
+ \lat^{(d)}_7 \, M_7
+  \lat^{(d)}_{\{3, 5 \} } \, \{M_3,M_5 \} \notag \\
&+ \lat^{(d)}_9 \, M_9 + \lat^{(d)}_{3,3,3} \, M_3^3 +  \cdots  \, \Bigr)  \ ,
 \label{fourgravcons}
\end{align}
where $S_0$ and $M_w$ are scalar functions of Mandelstam invariants defined in (\ref{Stree4pt}). The accompanying coefficients,  $\lat^{(d)}_{\ldots}$,
are combinations of the integrated  multiple sums, $\int \dd\mu_d(\tau)\, D_{\ldots}(\tau)$, and their dependence on the moduli of the $(10-d)$-dimensional
theory (which has has been suppressed) enters through the measure factor $\dd\mu_d(\tau)$.
The precise combinations in \eqref{fourgravcons} are
\begin{align}
\lat^{(d)}_3 &:= 2\pi \int \dd \mu_d(\tau) \, D_0 \notag \\
\lat^{(d)}_5 &:= 4\pi \int \dd \mu_d(\tau) \, D_2 \notag \\
\lat^{(d)}_{3,3} &:= -2\pi \int \dd \mu_d(\tau) \, (4 D_{111} + D_3)\notag \\
\lat^{(d)}_7 &:=  \tfrac{\pi}{3} \, \int \dd \mu_d(\tau) \, (D_4 + 9 D_2^2 + 6 D_{1111}) \label{calDs} \\
\lat^{(d)}_{\{3,5 \} } &:=- \tfrac{\pi}{12} \, \int \dd \mu_d(\tau) \,( D_5 +16 D_{311}-12 D_{221} + 12 D_{2111}-24 D'_{1111}+14 D_3 D_2 + 48 D_{111} D_2 ) \notag \\
\lat^{(d)}_9 & :=  \tfrac{\pi}{90} \int \dd \mu_d(\tau) \, ( D_6-90 D_{2211} + 90 D_2^3 + 120 D_{3111} - 10 D_3^2 + 45 D_4D_2 )\notag \\
\lat^{(d)}_{3,3,3} & :=  \tfrac{\pi}{540} \int \dd \mu_d(\tau) \,(7 D_6+ 540 (D_{411}-2 D_{321}+D_{222})+2610 D_{2211} - 450 D_2^3   - 1320 D_{3111} \notag\\
&\quad{} + 2160 D_3D_{111} + 470 D_{3}^2  - 225 D_4D_2  - 6480 D'_{2111} +  3240 D''_{1111} + 1080 D^{\times}_{1111}) \,. \notag
\end{align}
The diagrams associated with multiple sums with up to four propagators were displayed in section \ref{world-sheetdiag},
 whereas the more complicated weight-five and six $D_{\ldots}$ entering $\lat^{(d)}_{\{3,5 \} }$, $\lat^{(d)}_9$ and $
\lat^{(d)}_{3,3,3}$ are reviewed in appendix \ref{morediag}.

The expressions for $\lat^{(d)}_w$ were analysed explicitly in \cite{Green:2008uj} in $D=10$ and $D=9$ dimensions, after compactification on a circle of
radius $r$.  For example, the $\lat^{(0)}_w$'s that arise up to weight four in $D=10$, where the measure $\dd\mu_0(\tau)$ trivialises, are given in terms of
\begin{align}
&\int \dd\mu_0(\tau) \, D_0 \approx \frac{\pi}{3} \co &\int \dd\mu_0(\tau) \,  D _2 \ \approx 0 \notag \\
&\int \dd\mu_0(\tau) \,  D_3  \approx - \frac{ \pi \zeta_3}{3} \co &\int \dd\mu_0(\tau) \,  D_{111} \ \approx  0  \label{vanish10} \\
&\int \dd\mu_0(\tau) \, D_4  \approx  0 \co & \int \dd\mu_0(\tau) \,  D_{211}  \approx 0 \notag \\
&\int \dd\mu_0(\tau) \,  D_{1111}  \approx 0 \co &\int \dd\mu_0(\tau) \,  D^2_{2}  \approx 0 \, . \notag
\end{align}
The $\approx$ sign indicates that the integrations actually diverge at the large $\tau_2$ boundary of the fundamental domain, but they are regulated by taking
into account the non-analytic thresholds that arise from the large-$\tau_2$ limit.  This was discussed in detail in \cite{Green:1999pv,Green:2008uj} and is an
indication of the important interplay between the analytic and non-analytic parts of the amplitude.  We refer the reader to \cite{Green:2008uj} for details of
the integrals that arise at weights five and six.

Table \ref{coll4pt} summarises the $M_r$ matrices that correspond to the $\cov^{2k}R^4$ interactions in the expansion of the $N=4$ tree-level and one-loop
amplitudes up to order $(\ap)^9$ that are given in (\ref{het1,33}) and (\ref{fourgravcons}), together with the corresponding tree-level and one-loop $D=10$
coefficients.

\begin{table}[t]
\[
\begin{array}{|c|c|c|c|c|} \hline
\te{4pt} \ M_w\te{'s}  &\te{Interaction} &\te{Tree} &\te{ 1-loop generic $d$ } &$ $d=0$\, $ 
 \\ \hline \hline
1 &R &1 &0 &0\\ \hline
M_3 &R^4 &2 \zeta_3 &\lat^{(d)}_3& 4\zeta_2\\ \hline
M_5 &\cov^4 R^4 &2 \zeta_5 &\lat^{(d)}_5  & 0\\ \hline
M_3^2 &\cov^6 R^4 &2 \zeta_3^2 &\lat^{(d)}_{3,3}&4\zeta_2\zeta_3 \\ \hline
M_7 &\cov^8 R^4 &2 \zeta_7 &\lat^{(d)}_7 &0\\ \hline
\{ M_3,M_5\} &\cov^{10}R^4 &2 \zeta_3 \zeta_5 &\lat^{(d)}_{\{3,5\}}& \frac{97}{90} \zeta_2\zeta_5\\ \hline
M_9 &\cov^{12} R^4 &2 \zeta_9 &\lat^{(d)}_{9}& \frac{16}{15}\zeta_2 \zeta_3^2 \\ \hline
M_3^3 &\cov^{12} R^4 &\tfrac{4}{3} \zeta_3^3 &\lat^{(d)}_{3,3,3}  &\frac{151}{90} \zeta_2\zeta_3^2 \\ \hline
\end{array}   
\]
\vskip-5pt
\caption{\small Summary of  tree-level and one-loop contributions to terms in the effective action of the form $\cov^{2k}R^4$,
which can be extracted from the four-graviton components of the $\alpha'$ expansion of the superamplitudes shown in (\ref{het1,33})
and (\ref{fourgravcons}). The $M_9,M_3^3$ degeneracy at order $(\ap)^9$ gives rise to two distinct $\cov^{12}R^4$ operators which differ
in the tensor contractions of the derivatives and curvature.}
\label{coll4pt}
\end{table}

Notably, the values of the coefficients in table \ref{coll4pt} up to order $M_3^2$ match with the tree-level and one-loop contributions contained in the
modular invariant coefficient functions, $\calE_3$, $\calE_5$ and $\calE_{3,3}$. As reviewed in section \ref{dualityreview} \cite{Green:1997tv, Green:1998by,
Sinha:2002zr}, these accompany the interactions $R^4, \cov^4 R^4$ and $\cov^6 R^4$ in the $D=10$ type IIB effective action. In particular, it follows from 
(\ref{vanish10}) that the $D=10$ coefficients, $\lat_5^{(0)}$ and $\lat_7^{(0)}$, vanish. As a consequence, the interactions of order $\cov^4 R^4$ and $\cov^8
R^4$, as well as their supersymmetric partners, do not receive one-loop contributions to their analytic parts in ten dimensions\footnote{However, as discussed
in \cite{Green:2008uj} there is a non-analytic threshold term of the schematic form $(\alpha')^7\, s^4 \ln(\alpha' s)\,R^4$ that is related to a logarithmic
divergence in $\int \dd \mu_0(\tau) D_2^2$.}. Nevertheless, generic compactifications to $D<10$ lead to non-vanishing $\lat^{(d)}_5$ and $\lat^{(d)}_7$. This
is denoted in tables \ref{operators} and \ref{uoneviolating} by the wavy underlining of those terms that are zero only in $D=10$ dimensions.

\subsection{The $U(1)$-conserving five-particle  amplitudes}
\label{sec:here}

We have generalised the procedure used for the four-particle amplitude to expand the five-particle integrals (\ref{int26})-(\ref{int28}) and (\ref{Kint}) to
all weights $w\leq6$, using integration by parts identities (\ref{exampTwo})
as a cross-check. In $U(1)$-conserving five-curvature components, this leads to the following $\ap$ expansion
for the matrix ${\cal S}_{\te{1-loop}}$ in the five point amplitude (\ref{5g}),
\begin{align}
{\cal S}_{\te{1-loop}}^{q =0}
 =  &\, S_0\, \Bigl(
 \,\lat^{(d)}_3 \, M_3
+ \lat^{(d)}_5 \, M_5
+ \lat^{(d)}_{3,3} \, M_3^2
+ \lat^{(d)}_7 \, M_7
+ \lat^{(d)}_{7'} \, M_7' 
+ \lat^{(d)}_{ \{3,5 \}} \, \{M_3,M_5 \} 
 \notag \\
&{} \ \ \ \  +  \lat^{(d)}_{8'}\, M_8' + \lat^{(d)}_9 \, M_9 + \lat^{(d)}_{3,3,3} \, M_3^3  + \lat^{(d)}_{9'} \, M_9'  + \lat^{(d)}_{9''} \, M_9'' \ +  \cdots  \Bigr)\,,
 \label{fivegravcons}
\end{align}
In the present $N=5$ case, the quantities $S_0$, $M_k$ and $M'_k$ are $2\times 2$ matrices. The unprimed matrices $M_k$ are
defined by (\ref{defM}) in a tree-level context whereas $M_k'$ do not appear at genus zero. Their entries are degree $k$ polynomials
in Mandelstam invariants (and, by (\ref{manddef}), in $\ap$) whose explicit form can be obtained from an extra Mathematica file accompanying
the arXiv submission of this paper. Strikingly, the  combinations of terms that make up the coefficients $\lat^{(d)}_k$ of the
matrices $M_k$  in \eqref{fivegravcons} are exactly the same as those in the four-particle amplitude, (\ref{fourgravcons}) and \eqref{calDs}.
However, the  following additional combinations of multiple sums arise
\begin{align}
\lat^{(d)}_{7'} &:=  -10\pi \int \dd \mu_d(\tau)\, ( \tfrac{1}{96} D_4+\tfrac{19}{32} D_2^2 - D_{211}-\tfrac{99}{80} D_{1111}) \notag  \\
\lat^{(d)}_{8'} &:=  \tfrac{24\pi}{5} \int \dd \mu_d(\tau)\, \Big( \, \tfrac{ 7}{96} D_5+\tfrac{1}{6} D_{311}-\tfrac{1}{8}
D_{221}+\tfrac{1}{8} D_{2111}+\tfrac{1}{4} D'_{1111} \notag\\
& \quad{} +\tfrac{13}{48} D_3 D_2 + D_{11111}- \tfrac{1}{4}D^{\partial}_{222}-D''^{\partial}_{1111} \, \Big) \notag \\
\lat^{(d)}_{9'} &:= -2\pi \int \dd \mu_d(\tau)\, \Big( \,- \tfrac{667}{12960}D_6 -4 D_{1111}D_2 - 3 D_{111}^2 + \tfrac{2}{3} D_{11,11,11}+ D_{21111} - 5 D_{211}D_2  \notag \\
&\quad{} -\tfrac{1}{2} D_{222}+ \tfrac{ 55}{144}D_{2211}  + \tfrac{191}{144} D_2^3 - \tfrac{235}{108} D_{3111} - \tfrac{7}{6}  D_{321} + \tfrac{7}{6} D_3D_{111} 
- \tfrac{1}{6}D_{411} + \tfrac{257}{288} D_4D_2 \notag \\
&\quad{}+ \tfrac{667}{1296} D_3^2  -  2 D'_{11111} + D'_{2111} + D''_{1111} - D^{\times}_{1111} + 2 D''^{\partial}_{11111} + \tfrac{1}{6}D^{\pa}_{322} + 4 D'^{\pa}_{2211}  \label{smoke2}  \\
&\quad {} - 6 D''^{\pa}_{2111}+ 4 D'''^{\pa}_{1111} - 2 D_{2111}^{\times \pa} - 2 D^{\wedge \partial}_{11111} \, \Big)\notag \\
\lat^{(d)}_{9''}&:=-2\pi \int \dd \mu_d(\tau)\,  \Big( \,\tfrac{1}{10368}D_6-D_{1111}D_2 -\tfrac{1}{2} D_{111}^2 + \tfrac{1}{3}D_{11,11,11} + D_{21111} -\tfrac{1}{4} D_{211}D_2 
 \notag \\
 &\quad{}  + \tfrac{355}{576} D_{2211} -\tfrac{1}{8} D_{222} + \tfrac{101}{576} D_2^3  - \tfrac{ 211}{432} D_{3111} - \tfrac{1}{2}D_{321} + \tfrac{1}{2}D_3D_{111} + \tfrac{1291}{5184} D_3^2
 + \tfrac{3}{8} D_{411}  \notag \\
 &\quad + \tfrac{341}{1152} D_4D_2  - 2 D'_{11111} - 2 D'_{2111} + D''_{1111} - 2 D''^{\pa}_{2111} + D'''^{\pa}_{1111} - D_{2111}^{\times \pa} - 2 D^{\wedge \partial}_{11111} \, \Big) \notag
\end{align}
The terms in (\ref{fivegravcons}) are associated with the contributions to the low energy expansion of the five-particle amplitude due to (supersymmetrised)
combinations of $\cov^{2w}R^4$ and $\cov^{2w-2}R^5$ interactions (with $w$ denoting the weight of the accompanying multiple sum at order $(\ap)^{w+3}$).  It
is worth noting the following points:
\begin{itemize}
\item the five-particle matrices $M_{k}$ arise in both the tree-level and one-loop amplitudes, (\ref{het1,33}) and (\ref{fivegravcons}), respectively;
\item the specific linear combinations $\lat^{(d)}_{\ldots}$  of one-loop multiple sums that arose in the four-particle case (\ref{calDs}) also contribute to the  five-particle one-loop amplitude.
\end{itemize}
We therefore conclude that the $\cov^{2w}R^4$ and $\cov^{2w-2}R^5$ interactions associated with the $w \leq 3$ results in (\ref{fivegravcons}) involve exactly
the same tree-level and one-loop coefficients.  For $4\le w\le6$ one of the kinematic invariants in $\cov^{2w-2}R^5$ (the one involving only unprimed $M_k$
matrices) has the same coefficient as $\cov^{2w}R^4$. Such perturbative results provide useful input for constraining the exact form of the modular invariant
coefficient functions ${\cal E}_{\ldots}$ in the type IIB theory, generalising the arguments in section \ref{dualityreview}. After discussing analogous
results concerning $U(1)$-violating five-particle amplitudes in the following section, we will see, in section \ref{sec:s}, how these results may be
interpreted in terms of $SL(2,\Z)$ duality of the type IIB amplitude.

A class of $\cov^{2w-2}R^5$ interactions which does not occur at tree level is signalled by novel matrices $M_7',M_8',M_9'$ and $M_9''$ in
(\ref{fivegravcons}), which share a lot of algebraic properties with the $M_k$.  For example, they again preserve the BCJ and KK relations \cite{Bern:2008qj}
among the $A_{YM}$ they act on.  But they do not appear in the tree amplitude (\ref{het1,33}). In fact, these matrices were identified in the research leading
to \cite{Broedel:2013tta} as the unique deformations of the constituents $M_7,P_8,M_9$ of disk amplitudes that are consistent with their factorisation
properties, cyclicity and monodromy relations \cite{BjerrumBohr:2009rd,Stieberger:2009hq}.

A small comment: There is an apparent mismatch between the new five-particle kinematic invariants associated with the matrices $M_7', M_8',M_9'$ and $M_9''$
and the classification of candidate counterterms in four dimensional ${\cal N}=8$ supergravity given in \cite{Elvang:2010jv}.  In particular, this reference
rules out an independent $\cov^6 R^5$ operator based on four dimensional spinor helicity methods.  The fact that we have found a term of the form $A^t_{YM}
S_0 M_7' \tilde A_{YM}$ is, however, compatible with the analysis of \cite{Elvang:2010jv} since this term vanishes in the phase space of four dimensional
on-shell kinematics whereas in dimensions $D\geq 5$ the interaction is non-vanishing. We are grateful to the authors of \cite{Elvang:2010jv} for helpful email
correspondence which clarified this issue.

\subsection{The $U(1)$-violating five-particle  amplitudes}
\label{sec:u1vio}

Also $U(1)$-violating  five-particle closed-string amplitudes, such as the amplitude with two $G$'s and three gravitons,  can be expressed in terms of
$A_{YM}$ bilinears. According to (\ref{CmCmtoYM}), their coefficients encoded in the matrix ${\cal S}_{\te{1-loop}}^{q}$ are different for R-symmetry
charges $q=0$ and $q=\pm2$. In the latter case, we arrive at the following low energy expansion
\begin{align}
{\cal S}_{\te{1-loop}}^{q =  \pm 2}  &=  S_0\, \Bigl(
 -\frac{1}{3} \lat^{(d)}_3 \, M_3
+ \frac{1}{5}\lat^{(d)}_5 \, M_5
+ \frac{1}{3}\lat^{(d)}_{3,3} \, M_3^2
+ \frac{3}{7}\lat^{(d)}_7 \, M_7
+ \hat \lat^{(d)}_{7'} \, M_7' 
 + \frac{1}{2}\lat^{(d)}_{\{3,5\}} \, \{M_3,M_5 \} 
\notag \\
& +  \hat \lat^{(d)}_{8'} \, M_8' 
+ \frac{5}{9} \lat^{(d)}_9\, M_9 + \frac{5}{9} \lat^{(d)}_{3,3,3} \, M_3^3 +  \hat \lat^{(d)}_{9'} \, M_9' + \hat \lat^{(d)}_{9''}  M_9'' + \hat \lat^{(d)}_{9'''}  M_9''' \ +  \cdots  \Bigr)
 \label{5ptviolating}
\end{align}
Apart from the terms with coefficients $\lat^{(d)}_{\dots}$,  which arose in the $U(1)$-conserving amplitudes (\ref{fourgravcons}) and (\ref{fivegravcons}),
novel linear combinations, ${\hat{\lat}}^{(d)}_{\ldots}$, of multiple sums appear (which are different from
$\lat^{(d)}_{7'}$, $\ldots$, $\lat^{(d)}_{9''}$ of the $q=0$ case in \eqref{smoke2}), which accompany  the $M'_k$, $M''_9$ and $M'''_9$  matrices,
\begin{align}
\hat \lat^{(d)}_{7'} &:= \pi \int \dd \mu_d(\tau) \, \Big( \, \tfrac{7}{48} D_4-\tfrac{251}{16} D_2^2 +18 D_{211}+\tfrac{207}{8} D_{1111} \, \Big) \notag \\
\hat \lat^{(d)}_{8'} &:= \pi \int \dd \mu_d(\tau) \, \Big( \, \tfrac{107}{576} D_5 - \tfrac{479}{180} D_{311} + \tfrac{767}{240} D_{221} - \tfrac{767}{240} D_{2111}+ \tfrac{817}{120} D'_{1111} + \tfrac{313}{30} D_{11111} \notag \\
& \ \ + \tfrac{6}{5} D_{111}D_2  - \tfrac{4067}{1440} D_3D_2 - \tfrac{5}{6} D''^{\partial}_{1111} - \tfrac{5}{24} D^{\partial}_{222} \, \Big)\notag \\
\hat \lat^{(d)}_{9'}  & :=-2\pi \int \dd \mu_d(\tau) \, \Big( \, - \tfrac{4847}{116640} D_6 -3 D_{111}^2 + \tfrac{2}{3} D_{11,11,11} + 3 D_{21111} - 10 D_{211}D_2   \notag \\
& \ \ + \tfrac{2579}{1296} D_{2211} - \tfrac{1}{6} D_{222} + \tfrac{7195}{1296} D_2^3   - \tfrac{6143}{972} D_{3111} - \tfrac{3}{2} D_{321} + \tfrac{7}{6} D_3D_{111} + \tfrac{2903}{11664} D_3^2  \notag \\
 & \ \ - \tfrac{1}{2}D_{411}  + \tfrac{2605}{2592} D_4D_2  - 6 D'_{11111} + 
 7 D'_{2111} - D''_{1111} - \tfrac{7}{3} D^{\times}_{1111} + 2 D''^{\partial}_{11111}\notag \\
 & \ \   + \tfrac{1}{6}D^{\pa}_{322} + 4 D'^{\pa}_{2211} - 
 6 D''^{\pa}_{2111} + 4 D'''^{\pa}_{1111} - 2 D_{2111}^{\times \pa} - 2 D^{\wedge \partial}_{11111} \, \Big) \label{smoke3} \\
\hat \lat^{(d)}_{9''} & :=-2\pi\int \dd \mu_d(\tau) \, \Big( \, \tfrac{6073}{466560} D_6+3 D_{1111}D_2 + \tfrac{3}{2} D_{111}^2 + \tfrac{1}{3}D_{11,11,11} + 3 D_{21111} - \tfrac{9}{4} D_{211}D_2    \notag \\
& \ \  - \tfrac{7}{24} D_{222}
+ \tfrac{ 2351}{5184} D_{2211} + \tfrac{8665}{5184} D_2^3 - \tfrac{8831}{3888} D_{3111} - \tfrac{1}{6}D_3D_{111} - \tfrac{2185}{46656} D_3^2- 6 D'_{11111}   \notag \\
& \ \ + \tfrac{457}{10368} D_4D_2  -\tfrac{1}{6} D_{321}  + \tfrac{5}{24} D_{411} +
 4 D'_{2111} - D''_{1111} - \tfrac{4}{3} D^{\times}_{1111} - 2 D''^{\pa}_{2111} + D'''^{\pa}_{1111}\notag \\
 &\ \  - D_{2111}^{\times \pa} - 2 D^{\wedge \partial}_{11111} \, \Big)\notag \\
 \hat \lat^{(d)}_{9'''}  &:= -2\pi \int \dd \mu_d(\tau) \, \Big( \, - \tfrac{7}{270} D_6-12 D_{1111}D_2 + 4 D_{11,11,11} - \tfrac{11}{3} D_{2211} - \tfrac{1}{3}D_2^3 + \tfrac{44}{9} D_{3111}  \notag \\ 
 &\ \  + \tfrac{7}{27} D_3^2 + \tfrac{5}{6} D_4D_2 \, \Big) \ . \notag
\end{align}
The coefficients of terms in the expression \eqref{5ptviolating} have a striking pattern when compared with the coefficients of
terms in the $U(1)$-conserving sector. In particular, the coefficient along with (products of) unprimed matrices $M_{k}$ that
arise at order $(\ap)^{w+3}$ has a factor
\beq
\frac{w-1}{w+3}
\label{extrafac}
\eeq
in \eqref{5ptviolating}  relative to the corresponding coefficient in (\ref{fivegravcons}). This simple pattern
appears to fit in well with considerations based on $SL(2,\Z)$, the $D=10$ type IIB S-duality group, as we will
argue at the end of the following section.

\subsection{S-duality of $U(1)$-conserving amplitudes}
\label{sec:s}

Section \ref{dualityreview} reviewed the manner in which the moduli dependence of perturbative contributions to the terms in the low energy expansion of the ten-dimensional
type IIB four-particle amplitude fit into duality-invariant functions, ${\cal E}_{\ldots}(\Omega)$.  In particular, we presented explicit expressions for
the low dimension terms, ${\cal E}_3\, R^4$, ${\cal E}_5\, \cov^4 R^4$ and ${\cal E}_{3,3}\, \cov^6R^4$.  Here we would like to see to
what extent combining information from the $N=5$ loop amplitudes with previous results on tree amplitudes and $N=4$ loop amplitudes may extend our
understanding of the non-perturbative structure of amplitudes in the type IIB theory. 

In order to characterise the $N$-particle effective action, we need some notation that distinguishes the distinct invariants that are contained in
interactions such as $\cov^{2w} R^4$ and $\cov^{2w-2} R^5$ when $w>3$.  We will therefore introduce the following shorthand notation:
\smallskip
\begin{itemize}
\item Terms in the  order $(\ap)^{w+3}$ $N$-graviton amplitude are generated by  effective interactions of the form $\cov^{2w-2\ell}R^{4+\ell}$ with $\ell=0,1,\ldots,w$. The family of interactions which gives rise to a product of matrices $M_{p_1}M_{p_2}\dots$ of order $(\alpha')^{w+3}$  in the amplitude  (with $w=\sum_j p_j -3$) is denoted by $( \bigoplus_{\ell=0}^w \cov^{2w-2\ell}R^{4+\ell})_{p_1,p_2,\dots}$ with subscripts $p_1,p_2,\dots$ referring to the matrices.
\item More generally, $(\cov^{2w}R^4 \oplus \ldots)_{[p_1,p_2]\dots}$  or $(\cov^{2w}R^4 \oplus \ldots)_{\{p_1,p_2\}\dots}$ indicate kinematic structures associated with (anti-)commutators of matrices $[M_{p_1},M_{p_2}],\dots$ or $\{M_{p_1},M_{p_2}\},\dots$, respectively.
\item{A quantity such as $(\cov^{2w-2}R^5 \oplus \ldots)_{p'_1,p'_2,\ldots}$ is associated with primed matrices, $M'_{p_1}M'_{p_2}\ldots$.}
\end{itemize}
The results for $U(1)$-conserving five-particle amplitudes described above suggest that the same ${\cal E}_{\ldots}$ defined at the four-point level by
(\ref{symmamp}) accompany the combinations $(\cov^{2w}R^4 \oplus \cov^{2w-2}R^5)_{p_1, \dots}$ of appropriate dimensions, at least to the orders $2p+3q \leq
6$ investigated here. Our incomplete knowledge of one-loop $N \geq 6$ point amplitudes makes the analogous $\cov^{2w-4}R^6,\cov^{2w-6}R^7,\ldots$ inaccessible
to the present analysis.

The absence of $M_7',M_8',M_9'$ and $M_9''$ in tree-level amplitudes implies that they arise with new $SL(2,\Z)$-invariant coefficients to be denoted by
${\cal E}_{w'}$ and ${\cal E}_{9''}$ in the following.  These must contain perturbative terms that begin at one loop, and their one-loop contribution must
have a coefficient that is given by the integrated multiple sums, $\lat^{(d)}_{w'}$ and $\lat^{(d)}_{9''}$, in (\ref{smoke2}).

Since the matrices $M_7',M_8',M_9'$ and $M_9''$ do not contribute to collinear limits\footnote{The absence of poles in open string-like expressions $M'_w
\tilde A_{YM}$ implies locality of $A_{YM}^t S_0 M'_w \tilde A_{YM}$ for the following reason: The latter object is totally symmetric (thanks to the field
theory monodromy relations preserved by $M_w'$) but it can only have the poles of $A_{YM}^t$. Since some of the pole channels (such as $s_{15}^{-1},
s_{24}^{-1}$ and $s_{34}^{-1}$ in the five-particle case) are absent in $A_{YM}^t S_0$, the other ones which superficially occur in $A_{YM}^t$ or $S_0$ must
also be cancelled due to the total symmetry of $A_{YM}^t S_0 M'_w \tilde A_{YM}$.}, $\lat^{(d)}_{7'},\lat^{(d)}_{8'},\lat^{(d)}_{9'}$ and $\lat^{(d)}_{9''}$
are associated with local five-field operators of dimension $(\cov^6 R^5)_{7'}$, $(\cov^8 R^5)_{8'}$, $(\cov^{10} R^5)_{9'}$ and $(\cov^{10} R^5)_{9''}$,
respectively. 

In the following we will restrict our discussion to the ten-dimensional case with S-duality group $SL(2,\ZZ)$ although this should extend to the higher-rank groups associated with compactification on a $d$-torus.
The low-energy expansion of the exact $SL(2,\Z)$-invariant amplitudes for $N\le 5$ must include the perturbative terms corresponding to the tree-level and analytic parts of the one-loop amplitudes described above.  This suggests that  these terms should be incorporated into an  $SL(2,\Z)$-invariant  effective action in Einstein frame of the  schematic form,
\begin{align}
S^{q=0}_{\te{eff}} \, \Big|_{\te{local}} &\eq \int \dd^{10} x \ \sqrt{-g} \, \Big(  R +  {\cal E}_{3} \, (R^4)_{3}  +  {\cal E}_{5}\, (\cov^4 R^4 + \cov^2 R^5)_5+  {\cal E}_{3,3} \, (\cov^6 R^4+\cov^4 R^5)_{3,3}  \notag \\
&  + \ {\cal E}_{7} \, (\cov^8 R^4+\cov^6 R^5)_{7} + {\cal E}_{7'} \,( \cov^6 R^5)_{7'} +  {\cal E}_{\{3,5\}} \, (\cov^{10} R^4+\cov^8 R^5)_{ \{ 3,5 \} }+  {\cal E}_{8'} \, (\cov^8 R^5)_{8'}\notag \\
&  
+  {\cal E}_{9} \, (\cov^{12} R^4+\cov^{10} R^5)_{9}  + {\cal E}_{3,3,3} \, (\cov^{12} R^4+\cov^{10} R^5) _{3,3,3}   \notag \\
& +  {\cal E}_{9'} \, (\cov^{10} R^5)_{9'} +  {\cal E}_{9''} \, (\cov^{10} R^5)_{9''}   + {\cal O}(\ap^{10}) \, \Big) \,,
\label{effq0}
\end{align}
with $g$ denoting the determinant of the space-time metric. The coefficients ${\cal E}_{\ldots}(\Omega)$ and ${\cal E}_{w'}(\Omega)$ and  ${\cal E}_{9''}(\Omega)$ are $SL(2,\Z)$-invariant functions that have perturbative expansions that contain the required tree-level $MZV$'s and one-loop $\lat^{(0)}_{\dots}$ contributions.    There is strong evidence that the first three of these functions,  ${\cal E}_{3}(\Omega)$, ${\cal E}_{5}(\Omega)$ and ${\cal E}_{3,3}(\Omega)$ have the form  reviewed in section \ref{dualityreview},   which was based on the $N=4$ case.  
The purpose of \eqref{effq0} is to indicate the general pattern of modular functions that are expected to arise,
based on the tree-level and one-loop results\footnote{Of course, the full amplitude also includes the host of terms that are
non-analytic in the Mandelstam invariants that we have been ignoring, which should appear as complicated non-local terms in the effective action.}.
Although we do not know the  form of the modular functions ${\cal E}_{\ldots}$ (that arose in the $N=4$ case) and the new ones, ${\cal E}_{w'}(\Omega)$ and  ${\cal E}_{9''}(\Omega)$,   we now know that they must contain  the MZVs coefficients of tree-level powers of $\Omega_2$  (\ref{het1,33}) and the lattice sums in the one-loop powers of $\Omega_2$  (\ref{fivegravcons}).

 \begin{table}[t]
\[
\begin{array}{|c|c|c|c|c|} \hline
\te\ M_w\te{'s}   &\te{Interactions} &\te{Tree} &\te{1-loop} &\te{Coefficient}  \\ \hline \hline
1 &R &1 &0 &1\\ \hline
M_3 &(R^4)_3 &2 \zeta_3 &\lat^{(0)}_3 &{\cal E}_3 \\ \hline
M_5 &(\cov^4 R^4 \oplus \cov^2 R^5)_5  &2 \zeta_5 &\lat^{(0)}_5 &{\cal E}_5 \\ \hline
M_3^2 &(\cov^6 R^4 \oplus \cov^4 R^5)_{3,3} &2 \zeta_3^2 &\lat^{(0)}_{3,3} &{\cal E}_{3,3} \\ \hline
M_7 &(\cov^8 R^4  \oplus \cov^6 R^5)_{7}&2 \zeta_7 &\lat^{(0)}_7 &{\cal E}_7 \\ \hline
M'_7 & (\cov^6 R^5)_{7'} &0 &\lat^{(0)}_{7'} &{\cal E}_{7'} \\ \hline 
\{ M_3,M_5\} &(\cov^{10}R^4  \oplus \cov^8 R^5)_{\{3,5\}} &2 \zeta_3 \zeta_5 &\lat^{(0)}_{\{3,5\}} &{\cal E}_{\{3,5\}} \\ \hline 
 M_8' &(\cov^8 R^5)_{8'}&0 &\lat^{(0)}_{8'} &{\cal E}_{8'} \\ \hline 
 M_9 &(\cov^{12} R^4 \oplus \cov^{10} R^5)_{9} &2 \zeta_9 &\lat^{(0)}_9  &{\cal E}_9 \\ \hline
 M_3^3 & (\cov^{12} R^4 \oplus \cov^{10} R^5)_{3,3,3} &\tfrac{4}{3} \zeta_3^3 &\lat^{(0)}_{3,3,3} & {\cal E}_{3.3.3}  \\ \hline
M_9' &(\cov^{10} R^5)_{9'} &0 &\lat^{(0)}_{9'} &{\cal E}_{9'}  \\ \hline
M_9'' & (\cov^{10} R^5)_{9''} &0 &\lat^{(0)}_{9''} &{\cal E}_{9''}  \\ \hline
\end{array}
\]
\vskip-5pt
\caption{\small  Tree-level and one-loop contributions to $U(1)$-conserving terms in the $D=10$ low energy effective action of the form $\cov^{2k}R^5$, which can be extracted from the five-graviton components of the superamplitudes expanded in (\ref{het1,33}) and (\ref{fivegravcons}).}
\label{coll5pt}
\end{table}
For clarity, the coefficients of the $\ap$ expansion of the $U(1)$-conserving $4$-particle and $5$-particle tree-level and one-loop amplitudes in
\eqref{effq0} are summarised in table \ref{coll5pt}.  The second column of the table indicates the combinations of $\cov^{2w} R^4$ and $\cov^{2w-2} R^5$
interactions that have the same coefficients\footnote{These are particular components of the complete set of interactions that result from the pure spinor
construction, which naturally produces a supersymmetric expression that contains all interactions that are related by supersymmetry.}.  The precise structure
of the kinematic invariants associated with the interactions is defined by the polynomials of $M_k$ associated with them, which are listed in the first
column.
The tree and one-loop coefficients are listed in the third and fourth columns, respectively.  The last column of the table indicates how the
coefficients of these perturbative terms fit into $SL(2,\Z)$-invariant functions, ${\cal E}_{\dots}$.  
 
It is notable that the ratio of one-loop to tree-level contributions to the $\cov^4 R^4$ and $\cov^2 R^5$ interactions are identical, which fits in with the
fact that they are in the same supermultiplet.  They should therefore both be associated with the same modular function, ${\cal E}_{5}$.  Similarly, the ratio
of tree-level and one-loop contributions to $\cov^6 R^4$ and $\cov^4 R^5$ are identical and they should be contained in the same modular invariant function,
${\cal E}_{3,3}$.  More generally, interactions of $5$-particle interactions of the form $\cov^{2w-2}R^5$ with $w>3$ have more kinematic invariants than the
corresponding $4$-particle interaction, $\cov^{2w}R^4$.  However, as shown in the second column of the table, for each $w$ there is at least one five-point
invariant that pairs with a corresponding $4$-particle invariant and these particular invariants are presumably related by supersymmetry.  In that case they
should also be associated with a single modular function ${\cal E}_{\ldots}$.  The five-particle invariants listed in column $2$ that are not paired with
corresponding $4$-particle interactions are ones that do not have tree-level contributions.  They must therefore be contained in distinct modular functions
${\cal E}_{w'}$ and ${\cal E}_{9''}$ in which the one-loop term is the leading perturbative contribution.  

\subsection{S-duality of $U(1)$-violating type IIB amplitudes}

Analogous considerations apply to the terms in the $\ap$ expansion of the $U(1)$-violating amplitude discussed in section \ref{sec:u1vio} where the one-loop
coefficients were presented in \eqref{5ptviolating}.  Properties of the one-loop coefficients of $U(1)$-violating five-particle interactions of the form
$\cov^{2w}\, G^{2}\, R^3$ together with their tree-level counterparts are listed in table \ref{coll5ptpt}\footnote{Similar to the $U(1)$ preserving cases (see
early section \ref{sec:s}), we use the notation $(\cov^{2w}G^2 R^3)_{p_1,p_2,\ldots}$ and $(\cov^{2w}G^2 R^3)_{p'_1,p'_2,\ldots}$ to specify interactions
related to $M_{p_1}M_{p_2}\ldots$ and $M'_{p_1}M'_{p_2}\ldots$ terms in the amplitude (\ref{5ptviolating}).} Again these are particular examples of the
complete set of interactions that are related by supersymmetry.  In this case the degeneracy of kinematic invariants of a given weight grows faster than in
the $U(1)$-conserving case.  For example, there are $5$ invariants ar order $\cov^{12} G^2 R^3$, which accounts for the coefficient $5$ in the entry in the
second row and last column of table \ref{uoneviolating}.  In the last column of table \ref{coll5ptpt} we have speculated as to how these perturbative terms
might be incorporated into $(1,-1)$ modular forms, analogous to $E_{3/2}^{(1,-1)}(\Omega)$ defined in \eqref{uonecharge} in the context of the terms of order
$G^2 R^3$.  In particular, for those interactions that are partners of corresponding $U(1)$-conserving four-particle interactions (i.e., that have the same
structure of the unprimed $M_k$ matrices) it is tempting to make the ansatz that the $(1,-1)$ modular form is the one obtained by applying a covariant
derivative \eqref{covderiv} to the corresponding modular function, ${\cal E}_{\ldots}$, so it has the form $\calD {\cal E}_{\ldots}$.

This ansatz relates the terms that are power-behaved in $\Omega_2$ that contribute to the constant terms (the zero modes with respect to $\Omega_1$) of the
coefficient functions in the expansion of the $U(1)$-violating ($q=-2$) amplitude to those of the $U(1)$-conserving ($q=0$) amplitude, as follows.  The zero
Fourier mode of a coefficient ${\cal E}_{\ldots}$ of a $U(1)$-conserving interaction $\cov^{2w}R^4$, with $w =2p+3q$, has power behaved terms given by 
\beq
{\cal F}^w_0  \ \ =: \ \ a_w\, \Omega_2^{(w+3)/2} + b_w\, \Omega_2^{(w-1)/2} + O(\Omega_2^{(w-5)/2})\,.
\label{genpert}
\eeq
The zero mode of the corresponding $(-1,1)$ form, $\calD {\cal E}_{\ldots}$, is given by 
\bea
{\cal D}\, {\cal F}^w_0 &=&  a_w \, \tfrac{3+w}{2}\, \Om_2^{(3+w)/2} \ + \ b_w\, \tfrac{w-1}{2} \,\Om_2^{(w-1)/2}\nn\\
&=:& \ \ A_w \,  \Om_2^{(3+w)/2} \ + \ B_w\, \Om_2^{(w-1)/2} \ + \ O( \Om_2^{(w-5)/2} ) \ .
\label{chargetwo}
\eea 
We see that the ratio of  tree-level and one-loop amplitudes for the $q=-2$ process is related to that of the $q =0$  processes by  
\beq
\frac{B_w/A_w}{b_w/a_w} \eq \frac{w-1}{w+3} \,.
\label{ratios}
\eeq
This provides a natural explanation for the result \eqref{extrafac} that we observed in the five-particle amplitudes (\ref{fivegravcons}) and
(\ref{5ptviolating}). In fact, this also accounts for the factor of $-\frac{1}{3}$ in the kinematic identity (\ref{CmCmtoYM}). 

 \begin{table}[t]
\[
\begin{array}{|c|c|c|c|c|} \hline
\te\ M_w\te{'s}   &\te{Interactions} &\te{Tree} &\te{1-loop} &\te{$(1,-1)$ modular form} 
\\ \hline \hline
1 &\times &0 &0 &0 \\ \hline
M_3 &(G^2 R^3)_3 &2 \zeta_3 &-\tfrac{1}{3} \lat^{(0)}_3 &{\cal D}  {\cal E}_3  \\ \hline
M_5 &(\cov^4 G^2 R^3)_5  &2 \zeta_5 &\tfrac{1}{5}  \lat^{(0)}_5 &{\cal D}  {\cal E}_5 \\ \hline
M_3^2 &(\cov^6 G^2 R^3)_{3,3} &2 \zeta_3^2 &\tfrac{1}{3}  \lat^{(0)}_{3,3} &{\cal D}  {\cal E}_{3,3}\\ \hline
M_7 & (\cov^8 G^2 R^3)_7&2 \zeta_7 &\tfrac{3}{7}  \lat^{(0)}_7 &{\cal D}  {\cal E}_{7} \\ \hline
M'_7 & (\cov^8 G^2 R^3)_{7'} &0 &\hat  \lat^{(0)}_{7'} &{\cal E}^+_{7'} \\ \hline
\{ M_3,M_5\} &(\cov^{10} G^2 R^3)_{\{3,5\}} &2 \zeta_3 \zeta_5 &\tfrac{1}{2} \lat^{(0)}_{\{3,5\}} &{\cal D}  {\cal E}_{\{3,5\}}
 \\ \hline 
M_8' &(\cov^{10} G^2 R^3)_{8'}&0 &\hat  \lat^{(0)}_{8'}  &{\cal E}^+_{8'} \\ \hline 
M_9 &  (\cov^{12}G^2 R^3)_9 &2 \zeta_9 &\tfrac{5}{9}  \lat^{(0)}_{9} &{\cal D}  {\cal E}_{9}  \\ \hline
 M_3^3 & (\cov^{12} G^2 R^3)_{3,3,3} &\tfrac{4}{3} \zeta_3^3 &\tfrac{5}{9}  \lat^{(0)}_{3,3,3} & {\cal D}  {\cal E}_{3,3,3}  \\ \hline
M_9' &(\cov^{12} G^2 R^3)_{9'} &0 &\hat  \lat^{(0)}_{9'}  &{\cal E}^+_{9'}\\ \hline
M_9'' & (\cov^{12} G^2 R^3)_{9''} &0 &\hat  \lat^{(0)}_{9''} &{\cal E}^+_{9''} \\ \hline
M_9''' & (\cov^{12} G^2 R^3)_{9'''} &0 &\hat  \lat^{(0)}_{9'''}&{\cal E}^+_{9'''}  \\ \hline
\end{array}
\]
\vskip-5pt
\caption{\small Tree-level and one-loop contributions to the coefficients of  interactions  of the form $\cov^{2k}G^2 R^3$,  which can be extracted from the $\alpha'$ expansion of the $U(1)$-violating  five-particle amplitudes given in (\ref{het1,33}) and (\ref{5ptviolating}).}
\label{coll5ptpt}
\end{table}

The kinematic structures that arise at one loop but have no tree-level partners are associated with matrices $M'_{7}$, $M'_{8}$, $M'_{9}$, $M''_{9}$ and
$M'''_{9}$ listed in column 1 of table \ref{coll5ptpt}.  Their coefficients are $(1,-1)$ forms, denoted by ${\cal E}^+_{w'}$, ${\cal E}^+_{w''}$ and ${\cal
E}^+_{w'''}$ must have zero modes that contain the multiple sums (\ref{smoke3}), but they do not have any obvious connection with the ${\cal E}_{w'}$ and
${\cal E}_{9''}$ coefficients of the expansion of the $U(1)$-conserving amplitude that arise in (\ref{effq0}).

Finally, we note that with the above ansatz, we can write a schematic form for the local part of the $U(1)$-violating effective action for the low energy
expansion of the five-particle amplitude up to order $(\ap)^9$ in the form 
\begin{align}
S^{q=2}_{\te{eff}}& \, \Big|_{\te{local}} \eq \int \dd^{10} x \ \sqrt{-g} \, \Big( \, {\cal D}{\cal E}_3 \, (G^2 R^3)_3 \ + \ {\cal D}{\cal E}_5 \, (\cov^4 G^2 R^3)_5  \ + \ {\cal D}{\cal E}_{3,3} \, (\cov^6 G^2 R^3)_{3,3}  \notag \\
& + \ {\cal D} {\cal E}_{7} \, (\cov^8 G^2 R^3)_{7}  \ + \ {\cal E}^+_{7'} \, (\cov^8 G^2 R^3)_{7'} \ + \ {\cal D}  {\cal E}_{\{3,5\}} \, (\cov^{10} G^2 R^3)_{\{3,5\}}\ + \ {\cal E}^+_{8'} \, (\cov^{10} G^2 R^3)_{8'}
\notag \\
& + \ {\cal D} {\cal E}_{9} \, (\cov^{12} G^2 R^3)_9    \ + \ {\cal D}  {\cal E}_{3,3,3} \, (\cov^{12} G^2 R^3)_{3,3,3} \ + \ {\cal E}^+_{9'} \, (\cov^{12} G^2 R^3)_{9'} \notag \\
&  + \ {\cal E}^+_{9''} \, (\cov^{12} G^2 R^3)_{9''} \ + \ {\cal E}^+_{9'''} \, (\cov^{12} G^2 R^3)_{9'''}   \ + \ {\cal O}(\ap^{10}) \, \Big) \,.
\label{effq2}
\end{align}


\section{A glimpse of the six-particle amplitude}
\label{sixpoint}

The CFT calculations of section \ref{purespinor} and the set of integrals appearing in the five-particle amplitude (\ref{XXX}) provide a convenient starting
point for developing the structure of the integrand at higher multiplicity. According to \cite{Mafra:2012kh}, the $N$ point open-string correlator at one loop
is built from $N-4$ powers of propagator derivatives $X_{ij}$ as defined in (\ref{defeta}), and the closed-string integrand augments its holomorphic square
by interactions between left- and right-movers. Sections \ref{intbypart} and \ref{Kcsection} give two examples in which such interactions stem either from
integration by parts or from the OPE involving $\Pi^m \bar \Pi^n$ fields from opposite sectors as in (\ref{PiPibar}). Both cases introduce a factor of $\Om$
given in (\ref{defomega}) into the integrand instead of a product $X_{ij} \tilde X_{kl}$ due to separate OPEs within the left- and right moving sectors.

In the prescription (\ref{prescription}) for the six closed-string correlator on the torus, zero mode saturation is compatible with up to two contractions
between left- and right-movers. As a consequence, the six-point integrand involves three classes of $z$ dependencies:
\begin{align}
{\cal M}^6_{\te{1-loop}} \sim \int &\dd \mu_d(\tau) \ \tau_2^{-3} \int \dd^2 z_2 \ldots \dd^2 z_6 \ {\cal I}(s_{ij}) \notag \\
&\times \ \bigg\{ \, X_{ij} \, X_{kl} \, \tilde X_{pq} \, \tilde X_{rs} \, K_{ij,kl,pq,rs} \ + \ \Om \,X_{ij} \, \tilde X_{pq} \, K^\Om_{ij,pq} \ + \ \Om^2 \, K^{\Om^2} \, \bigg\} \,,
\label{6ptcorr}
\end{align}
where the kinematic factor  $K_{ij,kl,pq,rs}$ has the dimension of $R^6/k^8$ (where $k$ is a momentum), $ \, K^\Om_{ij,pq} $ has the dimension of $R^6/k^6$
and $ K^{\Om^2}$ has the dimension of $R^6/k^4$.
A careful evaluation of these kinematic factors in superspace is left for future work \cite{6ptwip}. 
 In this case the presence of  anomalies (associated with BRST non-invariant kinematic factors) in the open-string six-particle loop amplitude  leads to delicate issues in building the non-anomalous closed-string   amplitude. However,
 as we will argue in the following, we can still extract
general statements on the six-particle low energy effective action from the classification (\ref{6ptcorr}) of contributing world-sheet integrals, even without
precise knowledge of the kinematic factors. The analytic $\ap$ dependence of (\ref{6ptcorr}) is sufficient to derive the selection rules summarised in table
\ref{operators}.
  
First of all, the momentum expansion of the $\Om^2$ and $\Om X_{ij}  \tilde X_{pq}$ integrals can be almost literally\footnote{Starting from weight $w=5$, the possibility of having six vertex diagrams introduces extra terms into the six-point version of $\Om X_{ij}  \tilde X_{pq}$ integrals which cannot appear in the five-particle setting.} inferred from the results in (\ref{Kint}) and appendix \ref{app:w4} on five-particle integrals over $\Om$ and $X_{ij}  \tilde X_{pq}$, respectively. Up to weight $w=4$, they are shown to introduce no extra multiple sums beyond $D_0,D_2,D_{111},D_3,D_{1111},D_{211},D_2^2$ and $D_4$ that are known from the four-particle amplitude (see section \ref{world-sheetdiag}). The only potential source of new multiple sums are the integrands of schematic form $X_{ij}  X_{kl} \tilde X_{pq}  \tilde X_{rs}$ due to the product of left- and right moving correlators.

\begin{figure}[ht]
\begin{tikzpicture}
\draw(0,0) node{$\phantom{a}$};
\draw (-0.2,0) -- (14.5,0);
\draw (-0.2,-2) -- (14.5,-2);
\draw (-0.2,-4) -- (14.5,-4);
\begin{scope}[xshift=1.2cm]
\draw (0,1) node{$\! \! \! \! \! \! \! \! \! \! \! \!  \! \! \! \! \! \! \! \! \! \! \! \! {\cal I}^1_{ij,pqrs}$}; 
\draw (7,1) node{$  {\cal I}^2_{ijkl,pq} $}; 
\draw (0,-1) node{$\! \! \! \! \! \! \! \! \! \! \! \!  \! \! \! \! \! \! \! \! \! \! \! \!  {\cal I}^3_{ijk,pqr}$}; 
\draw (7,-1) node{$  {\cal I}^4_{ij,pqr}$}; 
\draw (0,-3) node{$\! \! \! \! \! \! \! \! \! \! \! \!  \! \! \! \! \! \! \! \! \! \! \! \!  {\cal I}^5_{ijklm}$}; 
\draw (7,-3) node{$  {\cal I}^6_{ij,kl}$}; 
\draw (0,-5) node{$\! \! \! \! \! \! \! \! \! \! \! \!  \! \! \! \! \! \! \! \! \! \! \! \!  {\cal I}^7_{ijkl}$}; 
\draw (6.2,1.7) -- (6.2,-5.7);
\draw (1.5,1.5) node{$\pa \ln \chi_{ij} \, \bar \pa \ln \chi_{ij}$}; 
\draw (1.5,0.5) node{$\pa \ln \chi_{pq}\, \bar \pa \ln \chi_{rs}$}; 
\draw (9,1.5) node{$\pa \ln \chi_{ij} \, \bar \pa \ln \chi_{jk}$}; 
\draw (9,0.5) node{$\pa \ln \chi_{kl}\, \bar \pa \ln \chi_{pq}$}; 
\draw (1.5,-0.5) node{$ \pa \ln \chi_{ij} \, \bar \pa \ln \chi_{jk}$}; 
\draw (1.5,-1.5) node{$\pa \ln \chi_{pq}\, \bar \pa \ln \chi_{qr}$}; 
\draw (9,-0.5) node{$ \pa \ln \chi_{ij} \, \bar \pa \ln \chi_{ij} $}; 
\draw (9,-1.5) node{$\pa \ln \chi_{pq}\, \bar \pa \ln \chi_{qr} $}; 
\draw (1.5,-2.5) node{$\pa \ln \chi_{ij} \, \bar \pa \ln \chi_{jk}$}; 
\draw (1.5,-3.5) node{$\pa \ln \chi_{kl}\, \bar \pa \ln \chi_{lm}$}; 
\draw (9,-2.5) node{$ \pa \ln \chi_{ij} \, \bar \pa \ln \chi_{ij}$}; 
\draw (9,-3.5) node{$\pa \ln \chi_{kl}\, \bar \pa \ln \chi_{kl}$}; 
\draw (1.5,-4.5) node{$\pa \ln \chi_{ij} \, \bar \pa \ln \chi_{jk}$}; 
\draw (1.5,-5.5) node{$\pa \ln \chi_{kl}\, \bar \pa \ln \chi_{li}$}; 
\begin{scope}[xshift=3.5cm,yshift=1cm,scale=1]
\draw (0,-0.5) node{$\bullet$} node[left]{$i$};
\draw (2,-0.5) node{$\bullet$} node[right]{$j$} ;
\draw (0,0) node{$\bullet$} node[left]{$p$};
\draw (2,0) node{$\bullet$} node[right]{$q$} ;
\draw (0,0.5) node{$\bullet$} node[left]{$r$};
\draw (2,0.5) node{$\bullet$} node[right]{$s$} ;
\draw[dashed] (0,0) -- (2,0) ;
\draw[dashed] (0,0.5) -- (2,0.5) ;
\draw[dashed] (0,-0.5) ..controls (1,-0.3) .. (2,-0.5) ;
\draw[dashed] (0,-0.5) ..controls (1,-0.7) .. (2,-0.5) ;
\draw (1,-0.35) node{$>$};
\draw (1,-0.65) node{$>$};
\draw (1,0) node{$>$};
\draw (1,0.5) node{$>$};
\end{scope}
\begin{scope}[xshift=11cm,yshift=1cm,scale=1]
\draw (0,0.5) node{$\bullet$} node[left]{$j$};
\draw (0,-0.5) node{$\bullet$} node[left]{$k$} ;
\draw (1,0.5) node{$\bullet$} node[right]{$i$};
\draw (1,-0.5) node{$\bullet$} node[right]{$l$} ;
\draw (2,0.5) node{$\bullet$} node[right]{$q$};
\draw (2,-0.5) node{$\bullet$} node[right]{$p$} ;
\draw (0.5,-0.5) node{$>$};
\draw (0.5,0.5) node{$<$};
\draw (0,0) node[rotate=-90]{$>$};
\draw (2,0) node[rotate=90]{$>$};
\draw[dashed] (0,0.5) -- (1,0.5) ;
\draw[dashed] (0,0.5) -- (0,-0.5) ;
\draw[dashed] (0,-0.5) -- (1,-0.5) ;
\draw[dashed] (2,-0.5) -- (2,0.5) ;
\end{scope}
\begin{scope}[xshift=3.5cm,yshift=-1cm,scale=1]
\draw (0,0.5) node{$\bullet$} node[left]{$i$};
\draw (0,-0.5) node{$\bullet$} node[left]{$j$};
\draw (1,-0.5) node{$\bullet$} node[right]{$k$};
\draw (1,0.5) node{$\bullet$} node[left]{$p$};
\draw (2,0.5) node{$\bullet$} node[right]{$q$};
\draw (2,-0.5) node{$\bullet$} node[right]{$r$};
\draw (0.5,-0.5) node{$>$};
\draw (1.5,0.5) node{$>$};
\draw (0,0) node[rotate=-90]{$>$};
\draw (2,0) node[rotate=90]{$<$};
\draw[dashed] (0,0.5) -- (0,-0.5) ;
\draw[dashed] (1,-0.5) -- (0,-0.5) ;
\draw[dashed] (2,0.5) -- (2,-0.5) ;
\draw[dashed] (1,0.5) -- (2,0.5) ;
\end{scope}
\begin{scope}[xshift=11cm,yshift=-1cm,scale=1]
\draw (0,-0.5) node{$\bullet$} node[left]{$i$};
\draw (2,-0.5) node{$\bullet$} node[right]{$j$};
\draw (1,0) node{$\bullet$} node[above]{$q$};
\draw (0,0.5) node{$\bullet$} node[left]{$p$};
\draw (2,0.5) node{$\bullet$} node[right]{$r$};
\draw[dashed] (0,0.5) -- (1,0);
\draw[dashed] (2,0.5) -- (1,0);
\draw (0.5,0.25) node[rotate=-30]{$>$};
\draw (1.5,0.25) node[rotate=30]{$>$};
\draw[dashed] (0,-0.5) ..controls (1,-0.3) .. (2,-0.5) ;
\draw[dashed] (0,-0.5) ..controls (1,-0.7) .. (2,-0.5) ;
\draw (1,-0.35) node{$>$};
\draw (1,-0.65) node{$>$};
\end{scope}
\begin{scope}[xshift=3.5cm,yshift=-3cm,scale=1]
\draw (0,0) node{$\bullet$} node[left]{$k$};
\draw (1,0.5) node{$\bullet$} node[left]{$j$};
\draw (2,0.5) node{$\bullet$} node[right]{$i$};
\draw (1,-0.5) node{$\bullet$} node[left]{$l$};
\draw (2,-0.5) node{$\bullet$} node[right]{$m$};
\draw[dashed] (0,0) -- (1,0.5);
\draw[dashed] (2,0.5) -- (1,0.5);
\draw[dashed] (0,0) -- (1,-0.5);
\draw[dashed] (2,-0.5) -- (1,-0.5);
\draw (0.5,-0.25) node[rotate=-30]{$>$};
\draw (0.5,0.25) node[rotate=-150]{$>$};
\draw (1.5,0.5) node{$<$};
\draw (1.5,-0.5) node{$>$};
\end{scope}
\begin{scope}[xshift=11cm,yshift=-3cm,scale=1]
\draw (0,-0.5) node{$\bullet$} node[left]{$i$};
\draw (2,-0.5) node{$\bullet$} node[right]{$j$};
\draw (0,0.5) node{$\bullet$} node[left]{$k$};
\draw (2,0.5) node{$\bullet$} node[right]{$l$};
\draw[dashed] (0,-0.5) ..controls (1,-0.3) .. (2,-0.5) ;
\draw[dashed] (0,-0.5) ..controls (1,-0.7) .. (2,-0.5) ;
\draw (1,-0.35) node{$>$};
\draw (1,-0.65) node{$>$};
\draw[dashed] (0,0.5) ..controls (1,0.3) .. (2,0.5) ;
\draw[dashed] (0,0.5) ..controls (1,0.7) .. (2,0.5) ;
\draw (1,0.35) node{$>$};
\draw (1,0.65) node{$>$};
\end{scope}
\begin{scope}[xshift=3.5cm,yshift=-5cm,scale=1]
\draw (0,-0.5) node{$\bullet$} node[left]{$i$};
\draw (2,-0.5) node{$\bullet$} node[right]{$j$};
\draw (0,0.5) node{$\bullet$} node[left]{$l$};
\draw (2,0.5) node{$\bullet$} node[right]{$k$};
\draw[dashed] (0,-0.5) -- (2,-0.5) ;
\draw[dashed] (0,0.5) -- (0,-0.5) ;
\draw[dashed] (0,0.5) -- (2,0.5) ;
\draw[dashed] (2,-0.5) -- (2,0.5) ;
\draw (1,-0.5) node{$>$};
\draw (1,0.5) node{$<$};
\draw (0,0) node[rotate=-90]{$>$};
\draw (2,0) node[rotate=90]{$>$};
\end{scope}
\end{scope}
\end{tikzpicture}
\caption{Possible topologies of six-particle integrals with four $\ln \chi$ derivatives where both the $ \partial \ln \chi_{ij}$'s and the $\bar \partial \ln \chi_{ij}$'s carry four different labels. The integration measure for the above expressions can be found in the first line of (\ref{6ptcorr}). A distinction between $\partial$ and $\tilde \partial$ is not needed since dashed lines with alike derivatives never end on the same vertex}
\label{sixpttop}
\end{figure}
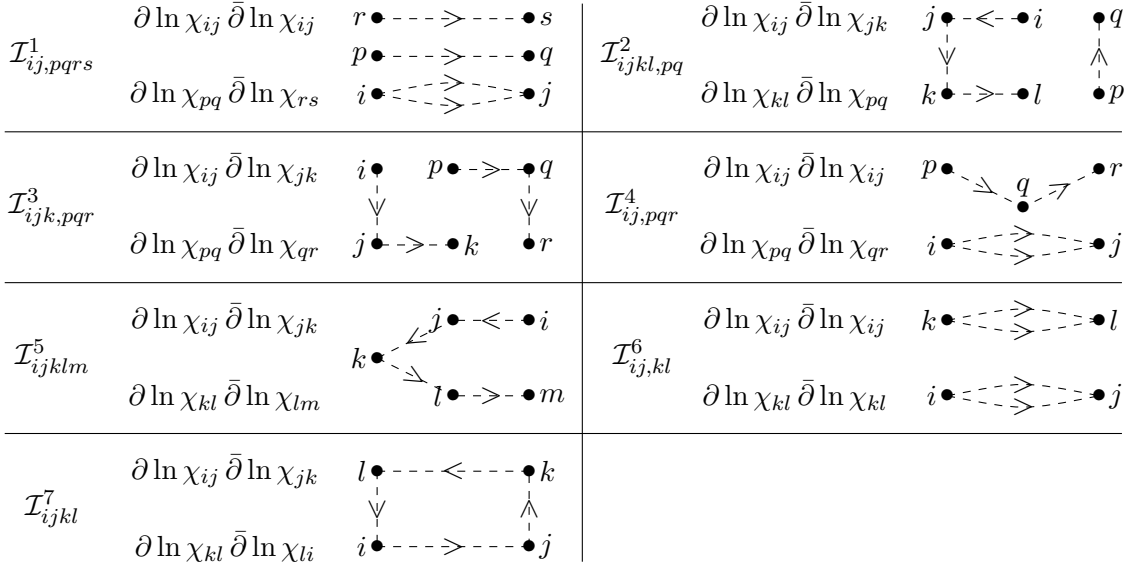

The six-point open-string correlator computed in \cite{Mafra:2012kh} is characterised by two topologies of BRST invariants and accompanying $X_{ij}$
bilinears: The final expression for its integrand comprises 20 permutations of $X_{23} (X_{24}+X_{34})$ and 15 permutations of $X_{23} X_{45}$ with respect to integrated labels
$2,3,4,5,6$. Integration by parts can be used to eliminate the former topology via $X_{23} (X_{24}+X_{34})= X_{23}(X_{41}+X_{45}+X_{46})$, possibly at the expense
of introducing $\Om$ admixtures from the right-movers, see (\ref{ibpcompact}). Hence, it is sufficient to expand elementary integrals $X_{23}X_{45} \tilde X_{pq} \tilde X_{rs}$ (where all the labels  $\{p,q,r,s\}$ are pairwise distinct) to understand the low energy behaviour of the six closed-string integrals. Within this
topology, there are seven inequivalent ways of arranging the right moving labels $\{p,q,r,s\}$ relative to the left moving ones $\{2,3,4,5\}$, which can be
conveniently visualised through the diagrams in figure~\ref{sixpttop} (which generalise the five-particle specific figure \ref{5pttopo}).

The integrals ${\cal I}^{1}_{ij,pqrs}, {\cal I}^{4}_{ij,pqr}$ and ${\cal I}^{7}_{ij,pq}$ defined in figure \ref{sixpttop} have kinematic poles with residues given by lower point amplitudes along the lines of (\ref{polea}) and (\ref{poleb}) (see appendix \ref{sixptw4a} for details). The intrinsic six-particle information stems from regular parts ${\cal I}^{\te{reg}}$ where the Koba--Nielsen factor is Taylor expanded before performing the $z_i$ integration. Using the diagrammatic methods introduced in sections \ref{fivepointdiag} and \ref{sec:expl}, we have expanded the seven inequivalent ${\cal I}^{\te{reg}}$ functions up to weight four, the results are displayed in appendix \ref{sixptw4}.

As shown in appendix \ref{sixptw4}, the 1PI diagrams $D_0,D_2,D_{111},D_3,D_{1111},D_{211},D_2^2$ and $D_4$ are sufficient to
express all the regular parts of the integrals in figure \ref{sixpttop} up to weight $w\leq4$. The fact that there are no further diagrams, apart from those that contributed to the  four- and five-particle amplitude,s that  contribute to the six-particle function up to this weight   imposes upper bounds $1,2$ and $4$ on the number of independent $R^6,\cov^2 R^6$ and $\cov^4 R^6$ interactions, respectively, see table \ref{operators}. In particular, in view of the vanishing integrated multiple sums at weight $w=2,4$ in $D=10$ -- see (\ref{vanish10}) -- the  $R^6$ and $\cov^4 R^6$ interactions must be absent in ten space-time dimensions. If the kinematic factors along with $D_{111},D_3$ as well as $D_{1111},D_{211},D_2^2$ and $D_4$ satisfy linear relations as they do for the five-particle closed-string amplitudes
(\ref{fivegravcons}) and (\ref{5ptviolating}), then there could be less such operators, and the numbers could possibly match with the five-field operators in
table \ref{operators} with the same mass dimension. Settling these questions is one motivation for detailed evaluation of $N$-point amplitudes in superspace with $N>5$
\cite{6ptwip}.

\section{Summary and comments}
\label{summary}

In this paper we have investigated one-loop amplitudes for the scattering of five (and to some extent, six) massless superparticles  in type IIA and type IIB  closed superstring theories.  The
world-sheet integrand (\ref{XXX}) was computed in pure spinor superspace and expressed in terms of a minimal set of functions.  The structure of the type IIB
amplitude shown in \eqref{closedfiveamp},
\beq
{\cal M}_{\te{1-loop}} = A^{t}_{YM} \, {\cal S}_{\te{1-loop}}(s_{ij}) \, \tilde A_{YM},
\eeq
 has the form of a bilinear in YM tree amplitudes contracted with a matrix function of the Mandelstam invariants,
just as for the closed-string tree amplitudes sketched in \eqref{closedamp}.  The type IIA five-particle amplitude has extra pieces that do not have this form, as discussed in section \ref{type2a}.  One of these is a parity-conserving piece that arises from the product of two $\epsilon$ tensors contracted on at least one index.  The second is a parity-violating component that was also constructed in section \ref{type2a} and includes the familiar $BR^4$ component interaction.

The one-loop expression, $ {\cal S}_{\te{1-loop}}(s_{ij})$, is defined in terms of integrals over a world-sheet torus. Their contributions to local terms in the low energy
effective action involving four or five powers of the (supersymmetrised) curvatures were obtained up to order $(\ap)^9$ and compared with analogous tree-level
expressions. In fact, all of the kinematic invariants that appear in the $\ap$ expansion of the tree amplitude up to order $(\ap)^{9}$ were found to
reappear in the one-loop $N$-particle amplitude with $N=4$ and $N=5$.  

Whereas at tree-level the coefficients of the terms in the $\alpha'$ expansion are (rationals multiplying) MZVs, the coefficients, $\lat^{(d)}_w$, of the
$\ap$ expansion of the one-loop amplitudes are rational combinations of integrated multiple sums $D_{\ldots}(\tau)$. We systematically reduced the weight $w\leq 6$ multiple sums in the five-particle amplitude to a tentative basis. This has the flavour of a higher-genus generalisation of the $\QQ$ basis reduction
of MZVs.

Some implications for S-duality ($SL(2,\Z)$ covariance) of the $D=10$ type IIB theory were considered in section \ref{sec:sdual}.  One important fact that emerged from the
five-particle amplitude is that for every tree-level interaction of the form $\cov^{2w-2}R^5$ associated with a particular combination of $M_i$ matrices, there
is a corresponding one-loop counterpart.  Moreover, the ratio of the coefficients of these five-curvature terms is precisely the same as the ratio of
tree-level to one-loop coefficients of the $\cov^{2w} R^4$ interaction that was extracted from the four-particle amplitude.  This strongly suggests that these
terms arise in the combination ${\cal E}_{l_1,l_2,\dots}(\Omega) (\cov^{2w} R^4\oplus 
\cov^{2w-2}R^5)$, with $\sum_i l_i = w+3$ in the effective action, where  ${\cal E}_{l_1,l_2,\dots}(\Omega) $ is a common modular invariant coefficient. Furthermore, the presence of new kinematic invariants at one-loop that are absent at tree-level requires modular invariant
coefficients that have no genus-zero contributions. This concerns order $(\alpha')^7 \cov^6R^5$ and higher.

We also considered tree-level and one-loop features of the modular forms associated with $U(1)$-violating interactions, such as $\cov^{2w}G^2 R^3$.  In
particular, the coefficients in the $U(1)$-violating and $U(1)$-conserving processes shown in \eqref{5ptviolating} and \eqref{fivegravcons} fit perfectly
with the expected pattern, as described in \eqref{genpert}-\eqref{ratios}. 

In order to get a better understanding of the S-duality systematics of type IIB amplitudes it would be interesting to extend our knowledge of kinematic
factors and one-loop world-sheet integrals to higher multiplicity and higher orders in $\ap$.  For example, we have not yet checked whether the
one-loop coefficients of the tree-level matrices $M_{2k+1}$ (and products thereof) are truly independent of the number $N$ of external legs for $N>5$ i.e. whether the
full families $\{ \cov^{2k-2l} R^{N+l}, \ l=0,1,\ldots,k\}$ of interactions are really accompanied by the same modular function in the exact type IIB
effective action. Multiparticle amplitudes also allow for a broader range of $U(1)$ charges and therefore provide further information for pinning down the modular forms of
non-zero weight in the effective action. Higher powers in $\ap$ are expected to enter via new matrices $M'_{w>9}$ but also via products of various lower order
matrices present in (\ref{fivegravcons}). It would be interesting to explore the matrix multiplicative patterns at orders beyond $(\ap)^9$.

On the other hand, even up to weight $w \leq 6$ the coefficients (\ref{calDs}), (\ref{smoke2}) and (\ref{smoke3}) in our expressions point to hidden
systematics underlying the multiple sums $D_{\ldots}$. Their unwieldy rational prefactors are reminiscent of the $\ap$ expansion of the tree amplitude
\cite{Schlotterer:2012ny} where the MZVs have not yet been mapped to the more transparent alphabet of noncommutative generators $f_i$ \cite{BrownDecomp}. Our
results  suggest that there is a more natural language to describe and arrange the multiple sums, and one might speculate whether further hidden
structures become visible after performing the modular $\tau$ integrals. Finally, in view of the all orders result in $\ap$ obtained for open-string trees
from the Drinfeld associator \cite{Broedel:2013aza} (extending the work of \cite{Drummond:2013vz}), it would be desirable to develop a unified description of
$\ap$ corrections to higher genus string amplitudes. In the same way as the Drinfeld associator encodes the universal monodromy of the Knizhnik-Zamolodchikov
equation governing correlators on the disk and sphere, one might envision a reduction of loop amplitudes to universal monodromies associated with the Riemann
surface in question.

Finally, it is important to stress that we have concentrated entirely on analytic contributions to the low energy expansion of the amplitude, but it is
important to develop a better understanding of the interplay between these terms and the non-analytic contributions (discussed to a limited extent in
\cite{Green:2010wi}), which are crucial in understanding the nonlocal structure of the quantum effective action.

\section*{Acknowledgement}

We are very grateful to Francis Brown and Don Zagier, as well as to Herbert Gangl and the other organisers of the Newton Institute workshop on
``Grothendieck-Teichmueller Groups, Deformation and Operads'', for many mathematical insights. OS is grateful to Johannes Br\"odel and Stephan Stieberger for
collaboration on related topics. The research leading to these results has received funding from the European Research Council under the European Community's
Seventh Framework Programme (FP7/2007-2013) / ERC grant agreement no. [247252].

\appendix

\section*{Appendix}

\section{Comparison of conventions}
\label{compconv}

The notation for the $N=4$ case in section \ref{dualityreview} differs somewhat from earlier conventions in the literature.  This change in convention is necessary in order to uniformly  describe higher multiplicities with $N\ge 4$.    We will here review the correspondence between these different conventions.

In \cite{Green:2008uj} the expansion of the analytic part of the amplitude was written in the form (in Einstein frame)
\begin{align}
T(s_{ij};\Omega)  \, \Big|_{\te{analytic}} &=\sum_{p=0,q=-1}^\infty  \, \calE_{(p,q)}(\Omega) \, \sigma_2^p \sigma_3^q \notag \\
&= \frac{3}{\si_3} + {\cal E}_{(0,0)}+ {\cal E}_{(1,0)} \si_2 + {\cal E}_{(0,1)} \si_3 + {\cal O}(\ap^7) \,,
\label{oldsymmamp}
\end{align}
where the dependence on the Mandelstam variables $s:= s_{12}=s_{34}$, $t:= s_{14}=s_{23}$ and $u :=s_{13}=s_{24}$ (with $s+t+u=0$), is contained in powers of $\sigma_2 =  (s^2 + t^2 + u^2)$ and $\sigma_3 =  (s^3 + t^3 + u^3)$.
In order to make contact with the notation in \eqref{symmamp} note that \cite{Green:1999pv}
\beq
M_w  =  -\frac{s^w+t^w+u^w}{w} = -  \sum_{2p+3q=w}\frac{(p+q-1)!} {p!q!}\,
\left(\frac{\sigma_2}{2}\right)^p\, \left(\frac{\sigma_3}{3}\right)^q\,.
\label{mdefs}
\eeq
The duality-invariant coefficients $\calE_{(p,q)}(\Omega)$ in \eqref{oldsymmamp} are linear combinations of those $\calE_{l_1,l_2, \ldots}$ in \eqref{symmamp} with $\sum_i l_i  = 2p+3q  +3$. The tree-level supergravity amplitude is given by the $p=0$, $q=-1$ term with coefficient $\calE_{(0,-1)}(\Omega) = 3$.

The representation \eqref{oldsymmamp} of the four-particle amplitude makes use of the fact that any symmetric polynomial of the four-point Mandelstam invariants can be written as a product of the form $\sigma_2^p\sigma_3^q$. The kinematic combinations involving four fields are unique below order $(\ap)^{9}\, \cov^{12} \, R^4$.  At that order the expansions of  both the tree-level and one-loop four-point amplitudes, the superpositions of a $(\sigma_2)^3 $ term and $(\sigma_3)^2$ term \cite{Green:2008uj} ties in with the two-fold degeneracy in the MZV content $\zeta_3^3$ and $\zeta_9$ of the tree amplitude at order $(\ap^9)$, as indicated in the last column of table \ref{operators}. At higher orders, however, the $\sigma_2^p\sigma_3^q$ monomials might not capture the full variety of $M_l$ products. For example, at weight  $(\ap)^{10}\, \cov^{14} \, R^4$ there are two elements, $M_5^2$ and $\{M_3, M_7\}$, whereas only the element $\sigma_2^2\sigma_3^2$ exists at that weight in the $\sigma_2^p\sigma_3^q$ basis\footnote{In the $N=4$ case where $M_l\in \R$ the element $\{M_3,M_7\}$ is equivalent to $2M_3M_7$, but for higher $N$ it is important that it is a symmetrised product of $(N-3)!\times (N-3)!$-dimensional matrices.}. This shortcoming of the $\sigma_2^p\sigma_3^q$ basis as well as the straightforward $N\geq 5$ point generalizability of the $M_l$'s motivate us to only refer to the basis formed by products of $M_l$'s in this paper.

\section{Pure spinor superspace calculations}
\label{superspacecalc}

This appendix gives a more detailed discussion of certain pure spinor superspace identities which were not proved in the main text.

\subsection{The pentagon numerator equation in superspace}
\label{super1}

In this subsection we provide a pure spinor superspace derivation for the field theory relation between box numerators and loop momentum dependent parts of
pentagon numerators, see (3.9) of \cite{Bjerrum-Bohr:2013iza}. For this purpose, we demonstrate that the left-hand side of (\ref{nowzero}) is BRST-exact. As a
starting point, consider $0=\langle Q\left(A^1_m T^m_{2,3,4,5}\right)\rangle$. Using \eqref{QTm} it is easy to show that
\begin{equation}\label{idproof}
0 =\langle Q\left(A^1_m T^m_{2,3,4,5}\right)\rangle = \langle (\lambda\g^m W^1)T^m_{2,3,4,5} + k^1_m V^1 T^m_{2,3,4,5} + (k^2\cdot A^1)V^2 T_3^i T_4^j T_5^k
\end{equation}
$$
+ (k^3\cdot A^1)V^3 T_2^i T_4^j T_5^k + (k^4\cdot A^1)V^4 T_2^i T_3^j T_5^k
+ (k^5\cdot A^1)V^5 T_2^i T_3^j T_4^k\rangle.
$$
Using the tree-level building blocks $L_{1j}$ \cite{Mafra:2010ir,Mafra:2010jq} one rewrites
\begin{equation}\label{Lonej}
(k^j\cdot A^1)V^j = - L_{1j} - A^j_m (\lambda\g^m W^1), \qquad j=2,3,4,5
\end{equation}
to obtain, after a few trivial cancellations, that
\begin{align}
0 =\langle Q\left(A^1_m T^m_{2,3,4,5}\right)\rangle &= \langle (\lambda\g^m W^1)W^m_{2,3,4,5} + k^1_m V^1 T^m_{2,3,4,5} \cr 
&\quad - L_{12} T_3^i T_4^j T_5^k -  L_{13} T_2^i T_4^j T_5^k -  L_{14} T_2^i T_3^j T_5^k -
L_{15} T_2^i T_3^j T_4^k\rangle.
\end{align}
The bosonic component of the term $\langle(\lambda\g^m W^1)W^m_{2,3,4,5}\rangle$ was shown in \cite{PSanomaly} to be proportional to $\epsilon_{10}F^5$ so it vanishes identically
using the momentum phase space of five particles.
Therefore,
\begin{align}
&\langle  T_{12}T^i_3 T^i_4 T^i_5 + T_{13}T^i_2 T^i_4 T^i_5 + T_{14}T^i_2 T^i_3
T^i_5 + T_{15}T^i_2 T^i_3 T^i_4 + k^1_m V_1 T^m_{2,3,4,5}\rangle \cr
&=\langle Q\Big[A^1_m T^m_{2,3,4,5} - \half \Big[(A^1\cdot A^2) T_3^i T_4^j T_5^k + (2\leftrightarrow 3,4,5)\Big]\rangle
\end{align}
where we used $L_{1j} = - T_{1j} - \half Q(A^1\cdot A^j)$, finishing the proof that \eqref{nowzero} is BRST-exact.

\subsection{Permutation symmetry of the five-graviton amplitude}
\label{super2}

This subsection is devoted to a superspace proof that the expression (\ref{Cmdef}) for the leading low energy contribution of the five-graviton amplitude (\ref{XXX}) is totally symmetric even though label 1 associated with the unintegrated vertex appears to enter on special footing. For this purpose, we rewrite
\begin{align}
C_{1,2,3,4,5}^m& \tilde C_{1,2,3,4,5}^m - C_{2,1,3,4,5}^m \tilde C_{2,1,3,4,5}^m = (V_1 T_2^i \tilde V_1 \tilde T_2^i - V_2 T_1^i \tilde V_2 \tilde T_1^i) \Big( \, \frac{ T^j_{34} T_5^k \tilde T^j_{34} \tilde T_5^k }{s_{34}} + (3 \leftrightarrow 4,5)  \, \Big)
\notag \\
&+ (V_1 T^m_{2,3,4,5} - V_2 T^m_{1,3,4,5}) \tilde V_1 \tilde T_{2,3,4,5}^m + V_2 T_{1,3,4,5}^m ( \tilde V_1 \tilde T^m_{2,3,4,5} - \tilde V_2 \tilde T^m_{1,3,4,5}) \label{someeq} \\
&+ \Big( \, \frac{ |T_{13} T_2^i T_4^j T_5^k |^2 -  |V_2 T^i_{13} T_4^j T_5^k |^2 }{s_{13}} - \frac{ |T_{23} T_1^i T_4^j T_5^k |^2 -  |V_1 T^i_{23} T_4^j T_5^k |^2 }{s_{23}} + (3 \leftrightarrow 4,5) \, \Big) \notag 
\end{align}
(where e.g. $|T_{13} T_2^i T_4^j T_5^k |^2 := T_{13} T_2^i T_4^j T_5^k \tilde T_{13} \tilde T_2^i \tilde T_4^j \tilde T_5^k$) and insert identities
\begin{align}
V_1 T_2^i \tilde V_1 \tilde T_2^i - V_2 T_1^i \tilde V_2 \tilde T_1^i &= \frac{ Q T_{12}^i \tilde V_1 \tilde T_2^i + V_2 T_1^i Q \tilde T_{12}^i}{s_{12}}  \\
\langle V_1 T^m_{2,3,4,5} - V_2 T^m_{1,3,4,5} \rangle &= \frac{k_{12}^m \langle T_{21} T_3^i T_4^j T_5^k \rangle + (k_3^m \langle V_3 T_{21}^i T_4^j T_5^k \rangle + (3\leftrightarrow 4,5))  }{s_{12}} \\
\langle k_1^m V_1 T_{2,3,4,5}^m \rangle &= -\langle T_{12} T_3^i T_4^j T_5^k + (2\leftrightarrow 3,4,5) \rangle \\
 \langle k_2^m V_1 T_{2,3,4,5}^m \rangle &= \langle T_{12} T_3^i T_4^j T_5^k \rangle - \langle V_1 T_{23}^i T_4^j T_5^k + (3\leftrightarrow 4,5) \rangle
\end{align}
as well as
\begin{align}
\frac{1 }{s_{13}}\langle  |T_{13} T_2^i T_4^j T_5^k |^2 -  |V_2 T^i_{13} T_4^j T_5^k |^2 \rangle =  &\frac{1}{s_{12}}   \langle (V_3 T_{12}^i - T_{12} T_3^i)T_4^j T_5^k \tilde T_{13} \tilde T_2^i  \tilde T_4^j  \tilde T_5^k \rangle \notag \\
  + &\frac{1}{s_{12}}  \langle V_2 T_{13}^i T_4^j T_5^k (\tilde V_3 \tilde T_{12}^i -\tilde T_{12} \tilde T_3^i) \tilde T_4^j \tilde T_5^k\rangle \ .
\end{align}
After making repeated use of these manipulations, all the terms on the right-hand side of (\ref{someeq}) are proportional to $s_{12}^{-1}$ and cancel pairwise.

\section{Further 1PI world-sheet diagrams}
\label{morediag}

Starting at weight $w=5$, it is essential to have a systematic classification scheme for the large number of world-sheet 1PI diagrams. A first criterion is
the distinction between indecomposable and possibly disconnected diagrams (such as $D_2^2$, see section \ref{world-sheetdiag}). Secondly, we can characterise 
diagrams according to the number of vertices that have propagators ending on them.  Note that the weight-$w$ diagram involving the greatest number of 
vertices is the $w$-gon.  This includes the pentagon at $w=5$, which does not contribute to the four-particle amplitude, and the hexagon at $w=6$, which does
not contribute to the four- or five-particle ampitude.

\medskip
The following table summarises the number of inequivalent diagrams at $w \leq 8$ (both the overall number and the number of disconnected representatives). Their graphical representation for $w=5$ and $w=6$ will be given in the subsequent subsections.
\[
\begin{array}{|c|c|c|c|c|c|c|c|c|c|} \hline
\te{weight} \ w &0 &1&2&3&4 &5&6&7&8 \\ \hline
\te{overall} \ \# \ \te{diagrams} & \ 1 \ & \ 0 \ & \ 1 \ & \ 2 \ & \ 4 \ & \ 8 \ & \ 20 \ & \ 42 \ &109 \\ \hline
\ \# \ \te{indecomposables} \ &1 &0 &1 &2 &3 &6 &13 &28 &71 \\\hline
\end{array} 
\]

\subsection{Weight $w=5$}
\label{sec:A1}

There are the following 8 diagrams (6 of them indecomposable):
\begin{itemize}
\item 2 disconnected diagrams

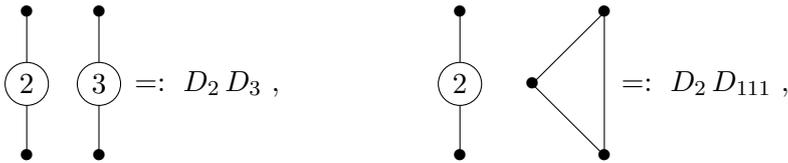
\tikzpicture[scale=1.9]
\begin{scope}
\draw (0.5,0) node{$\bullet$} ;
\draw (1,0) node{$\bullet$} ;
\draw (0.5,1) node{$\bullet$} ;
\draw (1,1) node{$\bullet$} ;
\draw (0.5,0) -- (0.5,1);
\draw (1,1) -- (1,0) ;
\draw (0.5,0.5) [fill=white] circle(0.15cm) ;
\draw (0.5,0.5) node{$2$};
\draw (1,0.5) [fill=white] circle(0.15cm) ;
\draw (1,0.5) node{$3$};
\draw(1.75,0.5) node{$=: \ D_2 \, D_3 \ ,$};
\end{scope}
\begin{scope}[xshift=3.5cm]
\draw (0,0) node{$\bullet$} ;
\draw (1,0) node{$\bullet$} ;
\draw (0,1) node{$\bullet$} ;
\draw (1,1) node{$\bullet$} ;
\draw (0.5,0.5) node{$\bullet$} ;
\draw (0,0) -- (0,1);
\draw (1,1) -- (1,0) ;
\draw (1,1) -- (0.5,0.5) ;
\draw (1,0) -- (0.5,0.5) ;
\draw (0,0.5) [fill=white] circle(0.15cm) ;
\draw (0,0.5) node{$2$};
\draw(1.7,0.5) node{$=: \ D_2\,D_{111} \ ,$};
\end{scope}
\endtikzpicture

\item 3 diagrams with two or three vertices

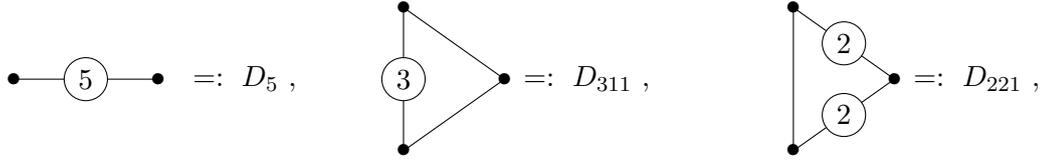
\tikzpicture[scale=1.9]
\draw (0,0.5) node{$\bullet$} ;
\draw (1,0.5) node{$\bullet$} ;
\draw (0,0.5) -- (1,0.5) ;
\draw (0.5,0.5) [fill=white] circle(0.15cm) ;
\draw (0.5,0.5) node{$5$};
\draw(1.6,0.5) node{$=: \ D_5 \ ,$};
\begin{scope}[xshift=2.7cm]
\draw (0,0) node{$\bullet$} ;
\draw (0,1) node{$\bullet$} ;
\draw (0.7,0.5) node{$\bullet$} ;
\draw (0,0) -- (0,1) ;
\draw (0,0) -- (0.7,0.5);
\draw (0,1) -- (0.7,0.5);
\draw (0,0.5) [fill=white] circle(0.15cm) ;
\draw (0,0.5) node{$3$};
\draw(1.27,0.5) node{$=: \ D_{311} \ ,$};
\end{scope}
\begin{scope}[xshift=5.4cm]
\draw (0,0) node{$\bullet$} ;
\draw (0,1) node{$\bullet$} ;
\draw (0.7,0.5) node{$\bullet$} ;
\draw (0,0) -- (0,1) ;
\draw (0,0) -- (0.7,0.5);
\draw (0,1) -- (0.7,0.5);
\draw (0.35,0.25) [fill=white] circle(0.15cm) ;
\draw (0.35,0.25)  node{$2$};
\draw (0.35,0.75) [fill=white] circle(0.15cm) ;
\draw (0.35,0.75)  node{$2$};
\draw(1.27,0.5) node{$=: \ D_{221} \ ,$};
\end{scope}
\endtikzpicture

\item 3 diagrams with four or five vertices

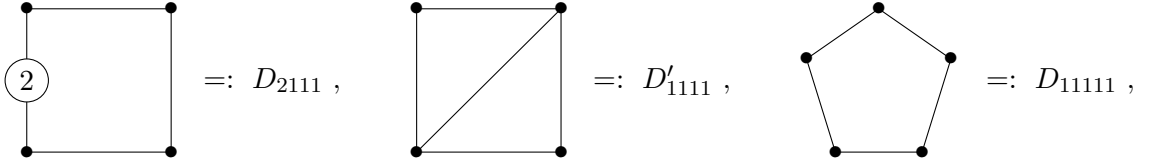
\tikzpicture[scale=1.9]
\begin{scope}
\draw (0,0) node{$\bullet$} ;
\draw (1,0) node{$\bullet$} ;
\draw (0,1) node{$\bullet$} ;
\draw (1,1) node{$\bullet$} ;
\draw (0,0) -- (1,0) ;
\draw (0,0) -- (0,1);
\draw (1,1) -- (1,0) ;
\draw (1,1) -- (0,1);
\draw (0,0.5) [fill=white] circle(0.15cm) ;
\draw (0,0.5)  node{$2$};
\draw(1.7,0.5) node{$=: \ D_{2111} \ ,$};
\end{scope}
\begin{scope}[xshift=2.7cm]
\draw (0,0) node{$\bullet$} ;
\draw (1,0) node{$\bullet$} ;
\draw (0,1) node{$\bullet$} ;
\draw (1,1) node{$\bullet$} ;
\draw (0,0) -- (1,0) ;
\draw (0,0) -- (0,1);
\draw (1,1) -- (1,0) ;
\draw (1,1) -- (0,1);
\draw (0,0)--(1,1);
\draw(1.7,0.5) node{$=: \ D'_{1111} \ ,$};
\end{scope}
\begin{scope}[xshift=5.4cm]
\draw (0,0.65) node{$\bullet$} ;
\draw (1,0.65) node{$\bullet$} ;
\draw (0.2,0) node{$\bullet$} ;
\draw (0.8,0) node{$\bullet$} ;
\draw (0.5,1) node{$\bullet$} ;
\draw(0.5,1)--(1,0.65);
\draw(1,0.65) --(0.8,0);
\draw(0.8,0)--(0.2,0);
\draw(0.2,0)--(0,0.65);
\draw(0,0.65)--(0.5,1);
\draw(1.76,0.5) node{$=: \ D_{11111} \ ,$};
\end{scope}
\endtikzpicture

\end{itemize}

\subsection{Weight $w=6$}
\label{sec:A2}

There are  the following 20 diagrams (13 of them  indecomposable):
\begin{itemize}
\item 7 disconnected diagrams 
   \begin{itemize} 
   \item 1 with 2+2+2 partition of the six propagators: $D_2^3$
   \item 3 with 4+2 partition of the six propagators: $D_4 D_2, D_{211} D_2,  D_{1111}D_2$
   \item 3 with 3+3 partition of the six propagators: $D_3^2,D_3D_{111},D_{111}^2$
   \end{itemize}
   %

\item 4 diagrams with two or three vertices

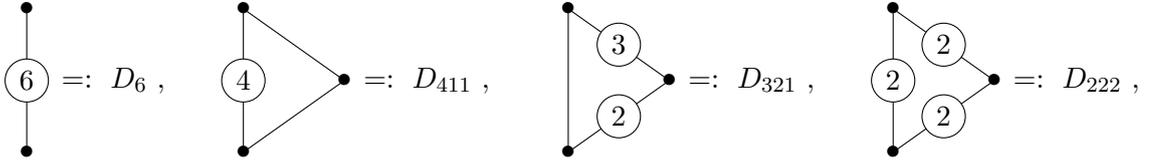
\tikzpicture[scale=1.9]
\draw (0,0) node{$\bullet$} ;
\draw (0,1) node{$\bullet$} ;
\draw (0,0)--(0,1);
\draw (0,0.5) [fill=white] circle(0.15cm) ;
\draw (0,0.5) node{$6$};
\draw(0.6,0.5) node{$=: \ D_6 \ ,$};
\begin{scope}[xshift=1.5cm]
\draw (0,0) node{$\bullet$} ;
\draw (0,1) node{$\bullet$} ;
\draw (0.7,0.5) node{$\bullet$} ;
\draw (0,0) -- (0,1) ;
\draw (0,0) -- (0.7,0.5);
\draw (0,1) -- (0.7,0.5);
\draw (0,0.5) [fill=white] circle(0.15cm) ;
\draw (0,0.5)  node{$4$};
\draw(1.27,0.5) node{$=: \ D_{411} \ ,$};
\end{scope}
\begin{scope}[xshift=3.75cm]
\draw (0,0) node{$\bullet$} ;
\draw (0,1) node{$\bullet$} ;
\draw (0.7,0.5) node{$\bullet$} ;
\draw (0,0) -- (0,1) ;
\draw (0,0) -- (0.7,0.5);
\draw (0,1) -- (0.7,0.5);
\draw (0.35,0.25) [fill=white] circle(0.15cm) ;
\draw (0.35,0.25)  node{$2$};
\draw (0.35,0.75) [fill=white] circle(0.15cm) ;
\draw (0.35,0.75)  node{$3$};
\draw(1.27,0.5) node{$=: \ D_{321} \ ,$};
\end{scope}
\begin{scope}[xshift=6cm]
\draw (0,0) node{$\bullet$} ;
\draw (0,1) node{$\bullet$} ;
\draw (0.7,0.5) node{$\bullet$} ;
\draw (0,0) -- (0,1) ;
\draw (0,0) -- (0.7,0.5);
\draw (0,1) -- (0.7,0.5);
\draw (0.35,0.25) [fill=white] circle(0.15cm) ;
\draw (0.35,0.25)  node{$2$};
\draw (0.35,0.75) [fill=white] circle(0.15cm) ;
\draw (0.35,0.75)  node{$2$};
\draw (0,0.5) [fill=white] circle(0.15cm) ;
\draw (0,0.5)  node{$2$};
\draw(1.27,0.5) node{$=: \ D_{222} \ ,$};
\end{scope}
\endtikzpicture

\item 5 diagrams with four vertices 

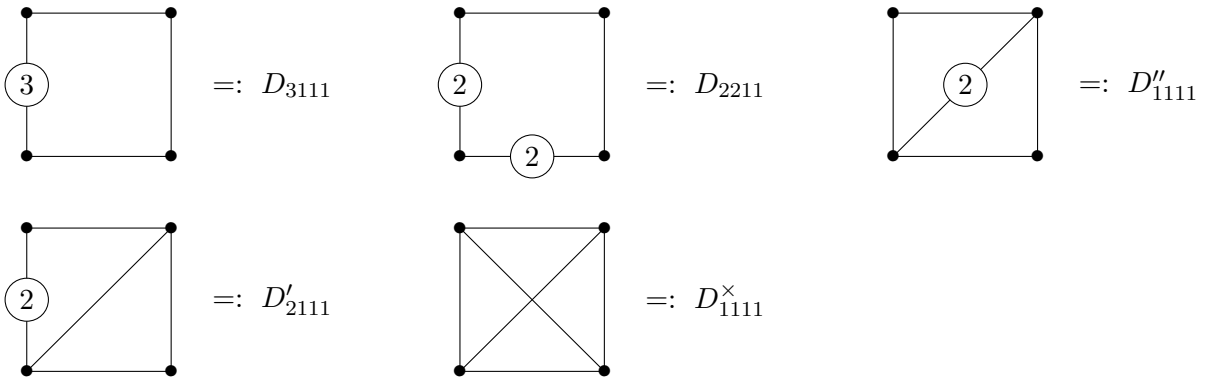
\tikzpicture[scale=1.9]
\begin{scope}[xshift=0cm]
\draw (0,0) node{$\bullet$} ;
\draw (1,0) node{$\bullet$} ;
\draw (0,1) node{$\bullet$} ;
\draw (1,1) node{$\bullet$} ;
\draw (0,0) -- (1,0) ;
\draw (0,0) -- (0,1);
\draw (1,1) -- (1,0) ;
\draw (1,1) -- (0,1);
\draw (0,0.5) [fill=white] circle(0.15cm) ;
\draw (0,0.5)  node{$3$};
\draw(1.7,0.5) node{$=: \ D_{3111}$};
\end{scope}
\begin{scope}[xshift=3cm]
\draw (0,0) node{$\bullet$} ;
\draw (1,0) node{$\bullet$} ;
\draw (0,1) node{$\bullet$} ;
\draw (1,1) node{$\bullet$} ;
\draw (0,0) -- (1,0) ;
\draw (0,0) -- (0,1);
\draw (1,1) -- (1,0) ;
\draw (1,1) -- (0,1);
\draw (0,0.5) [fill=white] circle(0.15cm) ;
\draw (0,0.5)  node{$2$};
\draw (0.5,0) [fill=white] circle(0.15cm) ;
\draw (0.5,0)  node{$2$};
\draw(1.7,0.5) node{$=: \ D_{2211}$};
\end{scope}
%
\begin{scope}[xshift=0cm, yshift=-1.5cm]
\draw (0,0) node{$\bullet$} ;
\draw (1,0) node{$\bullet$} ;
\draw (0,1) node{$\bullet$} ;
\draw (1,1) node{$\bullet$} ;
\draw (0,0) -- (1,0) ;
\draw (0,0) -- (0,1);
\draw (1,1) -- (1,0) ;
\draw (1,1) -- (0,1);
\draw (0,0.5) [fill=white] circle(0.15cm) ;
\draw (0,0.5)  node{$2$};
\draw(0,0)--(1,1);
\draw(1.7,0.5) node{$=: \ D'_{2111}$};
\end{scope}
\begin{scope}[xshift=6cm, yshift=0cm]
\draw (0,0) node{$\bullet$} ;
\draw (1,0) node{$\bullet$} ;
\draw (0,1) node{$\bullet$} ;
\draw (1,1) node{$\bullet$} ;
\draw (0,0) -- (1,0) ;
\draw (0,0) -- (0,1);
\draw (1,1) -- (1,0) ;
\draw (1,1) -- (0,1);
\draw (1,1) -- (0,0);
\draw (0.5,0.5) [fill=white] circle(0.15cm) ;
\draw (0.5,0.5)  node{$2$};
\draw(1.7,0.5) node{$=: \ D''_{1111}$};
\end{scope}
\begin{scope}[xshift=3cm, yshift=-1.5cm]
\draw (0,0) node{$\bullet$} ;
\draw (1,0) node{$\bullet$} ;
\draw (0,1) node{$\bullet$} ;
\draw (1,1) node{$\bullet$} ;
\draw (0,0) -- (1,0) ;
\draw (0,0) -- (0,1);
\draw (1,1) -- (1,0) ;
\draw (1,1) -- (0,1);
\draw (1,1) -- (0,0);
\draw (1,0) -- (0,1);
\draw(1.7,0.5) node{$=: \ D^{\times}_{1111}$};
\end{scope}
\endtikzpicture

\item 4 diagrams with five or six vertices 

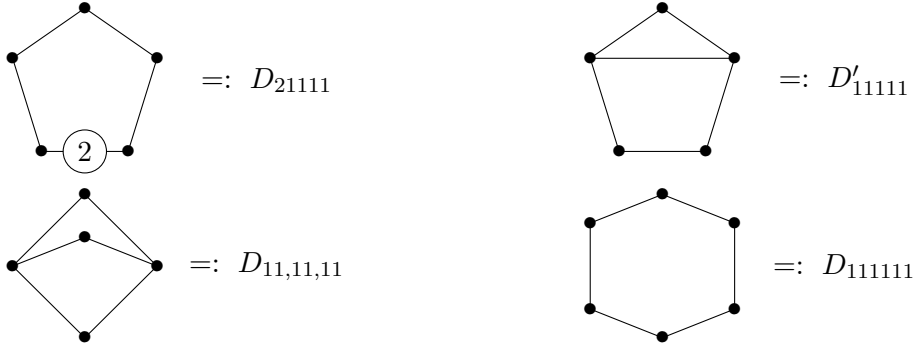
\tikzpicture[scale=1.9]
\begin{scope}[xshift=0cm]
\draw (0,0.65) node{$\bullet$} ;
\draw (1,0.65) node{$\bullet$} ;
\draw (0.2,0) node{$\bullet$} ;
\draw (0.8,0) node{$\bullet$} ;
\draw (0.5,1) node{$\bullet$} ;
\draw(0.5,1)--(1,0.65);
\draw(1,0.65) --(0.8,0);
\draw(0.8,0)--(0.2,0);
\draw(0.2,0)--(0,0.65);
\draw(0,0.65)--(0.5,1);
\draw (0.5,0) [fill=white] circle(0.15cm) ;
\draw (0.5,0)  node{$2$};
\draw(1.76,0.5) node{$=: \ D_{21111}$};
\end{scope}
\begin{scope}[xshift=4cm]
\draw (0,0.65) node{$\bullet$} ;
\draw (1,0.65) node{$\bullet$} ;
\draw (0.2,0) node{$\bullet$} ;
\draw (0.8,0) node{$\bullet$} ;
\draw (0.5,1) node{$\bullet$} ;
\draw(0.5,1)--(1,0.65);
\draw(1,0.65) --(0.8,0);
\draw(0.8,0)--(0.2,0);
\draw(0.2,0)--(0,0.65);
\draw(0,0.65)--(0.5,1);
\draw (0,0.65)--(1,0.65);
\draw(1.76,0.5) node{$=: \ D'_{11111}$};
\end{scope}
\begin{scope}[xshift=0cm, yshift=-1.3cm]
\draw (0,0.5) node{$\bullet$} ;
\draw (1,0.5) node{$\bullet$} ;
\draw (0.5,0) node{$\bullet$} ;
\draw (0.5,0.7) node{$\bullet$} ;
\draw (0.5,01) node{$\bullet$} ;
\draw(0,0.5)--(0.5,0);
\draw(0,0.5)--(0.5,1);
\draw(0,0.5)--(0.5,0.7);
\draw(1,0.5)--(0.5,0);
\draw(1,0.5)--(0.5,1);
\draw(1,0.5)--(0.5,0.7);
\draw(1.76,0.5) node{$=: \ D_{11,11,11}$};
\end{scope}
\begin{scope}[xshift=4cm, yshift=-1.3cm]
\draw (0,0.2) node{$\bullet $} ;
\draw (1,0.2) node{$\bullet$} ;
\draw (0,0.8) node{$\bullet$} ;
\draw (1,0.8) node{$\bullet$} ;
\draw (0.5,1) node{$\bullet$} ;
\draw (0.5,0) node{$\bullet$} ;
\draw (0,0.2) -- (0.5,-0);
\draw (1,0.2) -- (0.5,-0);
\draw (1,0.8) -- (1,0.2);
\draw (0,0.2) -- (0,0.8);
\draw (0,0.8) -- (0.5,1);
\draw (1,0.8) -- (0.5,1);
\draw(1.76,0.5) node{$=: \ D_{111111}$};
\end{scope}
\endtikzpicture

\end{itemize}

Note that polygons with neighbouring propagators interchanged are indistinguishable under the translation invariant torus integration measure:

\tikzpicture[scale=1.9]
\begin{scope}[xshift=0cm]
\draw (0,0) node{$\bullet$} ;
\draw (1,0) node{$\bullet$} ;
\draw (0,1) node{$\bullet$} ;
\draw (1,1) node{$\bullet$} ;
\draw (0,0) -- (1,0) ;
\draw (0,0) -- (0,1);
\draw (1,1) -- (1,0) ;
\draw (1,1) -- (0,1);
\draw (0,0.5) [fill=white] circle(0.15cm) ;
\draw (0,0.5)  node{$2$};
\draw (0.5,0) [fill=white] circle(0.15cm) ;
\draw (0.5,0)  node{$2$};
\end{scope}
\draw(1.5,0.5) node{$=$};
\begin{scope}[xshift=2cm]
\draw (0,0) node{$\bullet$} ;
\draw (1,0) node{$\bullet$} ;
\draw (0,1) node{$\bullet$} ;
\draw (1,1) node{$\bullet$} ;
\draw (0,0) -- (1,0) ;
\draw (0,0) -- (0,1);
\draw (1,1) -- (1,0) ;
\draw (1,1) -- (0,1);
\draw (0,0.5) [fill=white] circle(0.15cm) ;
\draw (0,0.5)  node{$2$};
\draw (1,0.5) [fill=white] circle(0.15cm) ;
\draw (1,0.5)  node{$2$};
\end{scope}
\draw (5.3,0.9) node{$\displaystyle = \ \ \frac{ \tau_2^6 }{\pi^6} \sum_{(m,n) \neq (0,0)} \frac{ 1}{|m\tau+n|^4}  \sum_{(p,q),(r,s)  \neq (0,0)}   $};
\draw (5.3,0.4) node{$\times \, |p\tau+q|^{-2} \,|(p+m) \tau +(n+q)|^{-2} $};
\draw (5.3,0.1) node{$\times \, |r\tau+s|^{-2} \,|(r+m) \tau +(n+s)|^{-2} $};
\endtikzpicture


\section{Five-point integrals to weight $w=4$}
\label{app:w4}

The analytic string corrections to the five-particle closed-string amplitude at one loop were  given up to order $(\ap)^9$ in (\ref{fivegravcons}) and
(\ref{5ptviolating}). These results are based on the expansion of the five-point integrals (\ref{Kint}) as well as (\ref{int26}) to (\ref{int28}) to sixth
subleading order. The latter were explicitly expanded to weight $w\leq 3$ in (\ref{Jpexp}), and this appendix extends these results to weight $w=4$. Two
inequivalent five-point integrals are free of massless poles by themselves:
\begin{align}
J_{12|13} &\eq -\int \dd \mu_d(\tau)\ \Big( \, D_2 \, s_{23} \ + \ D_{111} \, (s_{24}s_{34} \, + \, s_{25} s_{35}) \ + \ \frac{D_3}{2} \,  s_{23} \, (s_{12} \, + \, s_{13} \, + \, s_{23}) 
\notag \\
& \ \ \ \ \  + \ \frac{D_4}{6} \, s_{23}\, \Big( \, s_{23}^2 \, + \, \tfrac{3}{2} \, s_{23} \, (s_{12}+s_{13}) \, + \, s_{12}^2+s_{13}^2 \, + \, \tfrac{3}{2} \, s_{12} s_{13}\, \Big) \notag \\
& \ \ \ \ \  - \ \frac{D_2^2}{4} \,  \Big( \, s_{23}^2 \, (s_{12}+s_{13}) \, + \, s_{12} s_{13} s_{23} - \, 2 s_{23}\, (s_{14}^2+s_{15}^2+s_{24}^2+s_{25}^2+s_{34}^2 + s_{35}^2+s_{45}^2)\notag \\
& \ \ \ \ \ \ \ \   \,    -  2 s_{23}\, (s_{14}(s_{24}+s_{34})+s_{15}(s_{25} +s_{35}))  +  2\, (s_{14}s_{24}s_{34} + s_{15}s_{25}s_{35})\Big)  \notag \\
& \ \ \ \ \ + \ \frac{D_{211}}{2} \, \Big( \, 2 s_{23} \, ( s_{24}s_{34} + s_{25}s_{35}) \, + \, s_{24} s_{34} \,( s_{24} + s_{34})  \, + \, s_{25} s_{35} \,( s_{25} + s_{35})  \notag \\
& \ \ \ \ \ \ \ \ \ \  \,   + \, (s_{12}+s_{13}) \, (s_{24}s_{34} + s_{25}s_{35}) \, + \, 2 \, (s_{14}s_{24}s_{34} + s_{15}s_{25}s_{35}) \, \Big) \notag \\
& \ \ \ \ \  - \ \frac{D_{1111}}{2} \, \Big( \, s_{23}\, (s_{14}s_{24}+s_{14}s_{34}+s_{15} s_{25} + s_{15}s_{35}) \, - \, s_{14}s_{24}s_{34} \, - \, s_{15}s_{25} s_{35} \notag \\
& \ \ \ \ \ \ \ \ \ \  \,   - \, 2 s_{45} \, (s_{24}s_{35} + s_{25}s_{34}) \, \Big) \, + \, \ldots \, \Big) \\
J'_{12|34} \eq &\int \dd \mu_d(\tau) \ \Big( \, D_{111} \, (s_{14}s_{23} - s_{13} s_{24}) \ + \ \frac{D_2^2}{2} \, \Big(  s_{13} s_{24} \, (s_{14}+s_{23}) - s_{14} s_{23} \, (s_{13}+s_{24}) \Big)
\notag \\
 & \ \ \ \ + \ \frac{D_{211}}{2} \, \Bigl( s_{14}s_{23} \, (s_{14}+s_{23}) - s_{13} s_{24} \, (s_{13}+s_{24})\, + \, (s_{12}+s_{34}) \, (  \, s_{14}s_{23} - s_{13} s_{24}) \notag \\
 & \ \ \ \ \ \ \ \ \ \ \  + \, 2s_{14} s_{23} \, (s_{13}+s_{24}) \, - \, 2s_{13} s_{24} \, (s_{14}+s_{23}) \, \Big) \notag\\
 & \ \ \ \ + \ \frac{D_{1111}}{2} \, \Big( \, s_{14} s_{23} \, (s_{13}+s_{24}) - s_{13} s_{24} \, (s_{14}+s_{23}) \ + \ 2 \,(s_{14}s_{25}s_{35} \, + \, s_{23} s_{15} s_{45}) \notag \\
 &  \ \ \ \ \ \ \ \ \ \ \   - \, 2 \, (s_{13} s_{25} s_{45} \, + \, s_{24} s_{15} s_{35})\, \Big) \, + \, \ldots \, \Big) 
 \end{align}
For the third type of holomorphically factored five-point integral (\ref{int26}), the residue of its massless pole is determined by the four-point integral (\ref{polea}). The residual task is to expand its regular part defined by (\ref{poleb}):
\begin{align}
I_{12}^{\te{reg}} &\eq  \, \int \dd \mu_d(\tau) \ \Big( \, \frac{D_2}{2} \, s_{12} \ + \ D_{111} \, (s_{13}s_{23} \, + \, s_{14}s_{24} \, + \, s_{15}s_{25} ) \notag \\
& \ \ + \ \frac{D_3}{6} \,  \Big( \, s_{12}^2 \, + \, 3 \, (s_{13}s_{23} + s_{14}s_{24} +s_{15}s_{25} )  \, \Big) \notag \\
& \ \ + \ \frac{D_4}{24}  \, \Big( \, s_{12}^3  \, + \, 4 s_{12} \, (s_{13}s_{23}+ s_{14}s_{24}+s_{15}s_{25} ) \notag \\
& \ \ \ \ \ \ \ \ \ \ + \ 6 \, ( s_{13}^2s_{23} + s_{13}s_{23}^2 + s_{14}^2s_{24} +  s_{14}s_{24}^2 +  s_{15}^2s_{25} +  s_{15}s_{25}^2 )\, \Big) \notag \\
& \ \ + \ \frac{D_2^2}{4} \, \Big( \, s_{12} \, (s_{13}^2+s_{14}^2+s_{15}^2 +s_{23}^2+s_{24}^2+s_{25}^2 + s_{34}^2+s_{35}^2+s_{45}^2)   \\
& \ \ \ \ \ \ \ \ \ \ - \ ( s_{13}^2s_{23} + s_{13}s_{23}^2 + s_{14}^2s_{24} +  s_{14}s_{24}^2 +  s_{15}^2s_{25} +  s_{15}s_{25}^2 ) \, \Big) \notag \\
& \ \ + \ \frac{D_{211}}{2} \,  \Big( \, s_{12} \, (s_{13}s_{23} + s_{14}s_{24} +s_{15}s_{25} )\, + \, s_{34} \, (s_{13}s_{24}+s_{14}s_{23}) \, + \, s_{35} \, (s_{13}s_{25}+s_{15}s_{23}) \notag \\
& \ \ \ \ \ \ \ \ \ \  + \, s_{45} \, (s_{14}s_{25} + s_{15}s_{24}) \, + \,
s_{13}^2s_{23} + s_{13}s_{23}^2 + s_{14}^2s_{24} +  s_{14}s_{24}^2 +  s_{15}^2s_{25} +  s_{15}s_{25}^2 \, \Big)
\notag \\
& \ \ + \ D_{1111} \, \Big( \, s_{34} \, (s_{13}s_{24}+s_{14}s_{23}) \, + \, s_{35} \, (s_{13}s_{25}+s_{15}s_{23}) \, + \, s_{45} \, (s_{14}s_{25} + s_{15}s_{24})  \, \Big) \, + \, \ldots \, \Big) \notag
\end{align} 
The weight $w=5,6$ analogues can be obtained from the auxiliary file included in the arXiv submission.

\section{Six-point integrals to weight $w=4$}
\label{anotherapp}

In this appendix, we provide the analytic part of the $\ap$ expansion to weight $w=4$ for a set of six-point world-sheet integrals appearing in the six closed-string amplitude at one-loop. As argued in section \ref{sixpoint}, these results are sufficient to infer weights $w \leq 4$ for any other world-sheet integral in the six-particle amplitude (\ref{6ptcorr}). 

\subsection{The singular part of six-point integrals}
\label{sixptw4a}

Among the seven topologies of six-point integrals with four propagator derivatives shown in figure \ref{sixpttop}, the integrals ${\cal I}^{1}_{ij,pqrs},
{\cal I}^{4}_{ij,pqr}$ and ${\cal I}^{7}_{ij,pq}$ have kinematic poles in $s_{ij}$ due to the integration region $z_i \rightarrow z_j$ where $\partial \ln
\chi_{ij} \tilde \partial \ln \chi_{ij}$ products in the integrand diverge as $\sim \frac{ 1}{|z_{ij}|^2}$. The single pole residues can be expressed through
five-point integrals $J_{qp|qr},J'_{pq|rs}$ and $I^{\te{reg}}_{pq}$ given by (\ref{int27}), (\ref{int28}) and (\ref{poleb}), promoted to functions of five
momenta $(k_p,k_q,k_r,k_s,k_t)$\footnote{Up to $(\ap)^3$ order, for instance, we have
\begin{align*}
J'_{pq|rs}(k_p,k_q,k_r,k_s,k_i+k_j) &\eq \int \dd\mu_d(\tau) \ D_{111}  \, (s_{ps}s_{qr} - s_{pr}s_{qs}) \\
J_{qp|qr}(k_q,k_p,k_r,k_i+k_j,k_s) &\eq - \int \dd \mu_d(\tau)\ \Big(\, D_2  \, s_{pr} \ + \  D_{111} \, ((s_{pi}+s_{pj})(s_{ri}+s_{rj})+s_{ps}s_{rs}) \ + \ \frac{ D_3}{2} \, s_{pr} \, s_{pqr} \Big) \\
I^{\te{reg}}_{pq}(k_p,k_q,k_i+k_j,k_r,k_s) &\eq  \int \dd \mu_d(\tau)\ \Big(\frac{ D_2 }{2} \, s_{pq}  \ + \ \frac{D_3}{6} \, s_{pq}^2
\notag \\
& \ \ \ \ \ \ \ \ + \ \left( \,  D_{111}   + \frac{ D_3 }{2} \, \right) \, ((s_{pi}+s_{pj})(s_{qi}+s_{qj})+s_{pr}s_{qr}+s_{ps}s_{qs})\, \Big)
\end{align*}}. Let ${\cal I}^{i,\te{reg}}$ denote the regular part of the integrals after subtracting off the poles, then
\begin{align}
{\cal I}^{1}_{ij,pqrs} &\eq {\cal I}^{1,\te{reg}}_{ij,pqrs} \ + \ \frac{1}{ s_{ij}} \, J'_{pq|rs}(k_p,k_q,k_r,k_s,k_i+k_j) \label{int101} \\
{\cal I}^{4}_{ij,pqr} &\eq {\cal I}^{4,\te{reg}}_{ij,pqr} \ - \ \frac{1}{  s_{ij}} \, J_{qp|qr}(k_q,k_p,k_r,k_i+k_j,k_s) \label{int102} \\
{\cal I}^{6}_{ij,pq} &\eq {\cal I}^{6,\te{reg}}_{ij,pq} \ + \ \frac{1}{  s_{ij}} \, I^{\te{reg}}_{pq}(k_p,k_q,k_i+k_j,k_r,k_s) \ + \ \frac{1}{ s_{pq}} \, I^{\te{reg}}_{ij}(k_i,k_j,k_p+k_q,k_r,k_s) \notag \\
& \ \  + \ \frac{1}{ s_{ij}s_{pq}} \, I(k_i+k_j,k_p+k_q,k_r,k_s) \ . \label{int103} 
\end{align}
The remaining ${\cal I}^{i}$ at $i=2,3,5,7$ do not have poles and coincide with their regular parts ${\cal I}^{i,\te{reg}}$.

\subsection{The regular part of six-point integrals}
\label{sixptw4}

In the following, we display the diagrammatic expansion of the six-point integrals defined in figure \ref{sixpttop}. Dashed lines represent derivatives $\partial \ln \chi$ or $\tilde \partial \ln \chi$, and a distinction between $\partial$ and $\tilde \partial$ is not needed since dashed lines with alike derivatives do not touch the same vertex for the integrals under consideration. The ellipses refers to contributions of higher weight $w \geq 5$.
\begin{align*}
{\cal I}^{1,\te{reg}}_{ij,pqrs} \eq& \ldots
\notag \\
{\cal I}^{2,\te{reg}}_{ijkl,pq} \eq &\int \dd \mu_d(\tau) \ D_{1111}  \, (s_{iq} s_{lp}-s_{ip}s_{lq}) \ + \ \ldots
\notag \\
{\cal I}^{3,\te{reg}}_{ijk,pqr} \eq &\int \dd \mu_d(\tau) \ \Big( \, D_2^2  \, s_{ik} s_{pr} \ + \ 
 D_{1111}   \, (s_{ip} s_{kr} + s_{ir} s_{kp})  \ + \ \ldots \, \Big)
 \notag \\
 {\cal I}^{4,\te{reg}}_{ij,pqr}  \eq &\int \dd \mu_d(\tau) \ \Big( \,\frac{ D_2^2 }{2} \, s_{i j} \, s_{pr} \ + \ \Big( 
  \,    D_{1111}   \ + \ \frac{ D_{211} }{2} \, \Big) \, (s_{ip}\, s_{j r} \, + \,s_{ir}\, s_{jp})  \ + \ \ldots \, \Big)
  \notag \\
  {\cal I}^{5,\te{reg}}_{ijklm} \eq &\int \dd \mu_d(\tau) \ \Big( \, D_{111} \, s_{i m} \ + \ \frac{ D_{211} }{2} \, s_{i m}^2 \ + \ 
 D_{1111}   \! \! \sum_{p\neq i,j,k,l,m} \! \! s_{ip} \,s_{mp} \ + \  D_{211}  \, s_{i m}  \, ( 
    s_{i k} +  s_{jl} +  s_{k m})
   \notag \\
    & \ \ \ + \ \frac{ D_{211} }{2} \, s_{i m} \, ( 
    s_{i j} + s_{j k} + s_{kl} + 
    s_{l m})  \ + \  \frac{ ( D_2^2 - D_{1111} )}{2} \,(s_{ik} s_{jm} + s_{il}s_{km})  \notag \\
   & \ \ \ + \ D_2^2 \, s_{ik}  s_{k m}  \ + \  \Big( \, \frac{ (  D_{1111}  - D_2^2)}{2} \ + \    D_{211}  \, \Big) \, \big( \, s_{im}\, ( 
   s_{jm} + s_{il}) \ - \ s_{il}\, s_{jm} \, \big) \ + \ \ldots \, \Big) \notag \\
   {\cal I}^{6,\te{reg}}_{ij,kl} \eq &\int \dd \mu_d(\tau) \ \Big( \,  \frac{ D_2^2 }{4} \, s_{i j} \, s_{kl} \ + \ \Big( \, (D_{1111} + D_{211} ) \ + \ \frac{ D_4 - D_2^2}{4} \, \Big) \, (s_{ik} s_{jl} \, + \, s_{il} s_{jk})  \ + \ \ldots  \, \Big) \notag \\
   {\cal I}^{7,\te{reg}}_{ijkl} \eq &\int \dd \mu_d(\tau) \ \Big( \, D_2 \ + \
\frac{ D_3}{2} \, ( 
   s_{ij} + s_{j k} + s_{k l} + s_{il} + 2 s_{i k} + 2 s_{j l} ) \ + \  \frac{ D_2^2}{2} \, \sum_{p<q \neq i,j,k,l} s_{pq}^2
    \notag \\
 & \ \ \ + \  \frac{ D_2^2 }{2} \,  \sum_{p \neq i,j,k,l} (s_{ip}^2+s_{jp}^2+s_{kp}^2+s_{lp}^2) \  + \  \frac{ D_4 - 
    D_2^2}{ 2} \, ( 
   s_{ij}+s_{jk}+s_{kl}+s_{il}) \, (s_{ik}+s_{jl}) \notag \\
    & \ \ \ + \ 
\frac{  D_4}{6} \, ( s_{ij}^2 + s_{jk}^2 + s_{kl}^2 + s_{il}^2 + 3 s_{ik}^2 + 
    3 s_{jl}^2 )  \ + \ \frac{ D_4 -2 D_2^2 +  D_{1111} }{2} \, s_{ik}s_{jl}  \notag \\
   & \ \ \ + \ \frac{ D_4 - D_2^2}{4} \, ( 
   s_{i j}  s_{k l} \, + \, s_{i l} s_{j k} \, +\, s_{i j} s_{j k} \, + \, 
    s_{j k} s_{k l}\, +\, s_{k l}s_{il}\, + \, s_{ij} s_{il})   \notag \\
   & \ \ \ + \  \frac{ D_2^2  -  D_{1111} }{2} \, \sum_{p \neq i,j,k,l} (s_{ip} s_{jp} +s_{jp} s_{kp} + s_{kp} s_{lp} + s_{ip} s_{lp})  \notag \\
   & \ \ \ + \ D_{211}   \, \sum_{p \neq i,j,k,l} (s_{ip} s_{kp} +s_{jp} s_{lp})  \ + \ \ldots \,  \Big)
\end{align*}  
We emphasise that all the above terms are spanned by the same set of $w\leq 4$ 1PI diagrams $D_0,D_2,D_{111},D_3,D_{1111},D_{211},D_2^2$ and $D_4$ appearing in the four-particle amplitude. This justifies the upper bounds on the six-point one-loop effective action in table \ref{operators} and \ref{uoneviolating}, also see the discussion in section \ref{sixpoint}.

\end{document}